\documentclass[prd,onecolumn,amsmath,amssymb,floatfix,superscriptaddress,notitlepage,nofootinbib,preprintnumbers]{revtex4-2}

\usepackage[T1]{fontenc}\usepackage[utf8]{inputenc}\usepackage{lmodern}
\usepackage{amsmath,amssymb,mathtools,bm}
\usepackage[margin=0.95in]{geometry}
\usepackage{xcolor}\usepackage{booktabs,longtable,array}\usepackage{enumitem}
\usepackage{graphicx}
\usepackage{placeins}

\usepackage{hyperref}
\hypersetup{
    colorlinks=true,       %
    linkcolor=blue!70!black,  %
    citecolor=blue!70!black,  %
    filecolor=blue,     %
    urlcolor=blue!70!black   %
}

\usepackage{natbib}
\providecommand{\aap}{Astron. Astrophys.}
\providecommand{\apj}{Astrophys. J.}
\providecommand{\baas}{Bull. Am. Astron. Soc.}
\providecommand{\mnras}{Mon. Not. Roy. Astron. Soc.}
\providecommand{\prd}{Phys. Rev. D}
\providecommand{\jcap}{JCAP}

\newcommand{\dd}{\mathrm d}
\newcommand{\Hc}{\mathcal H}
\newcommand{\Ox}{\widehat{\mathcal O}_x}

\begin{document}

\title{Spectral-Distortion Anisotropies:\\ Gauge-Covariant Formulation and Physical Interpretation}

\author{Wayne Hu}
\affiliation{Kavli Institute for Cosmological Physics, Enrico Fermi Institute, and Department of Astronomy \& Astrophysics, University of Chicago, Chicago IL 60637}
\author{Jens Chluba}
\affiliation{Jodrell Bank Centre for Astrophysics, University of Manchester, Manchester, UK}
\author{David Zegeye}
\affiliation{Kavli Institute for Cosmological Physics, Enrico Fermi Institute, and Department of Astronomy \& Astrophysics, University of Chicago, Chicago IL 60637}
\affiliation{Fermi National Accelerator Laboratory, MS209, P.O. Box 500, Batavia, IL, 60510, USA}
\affiliation{Department of Physics, University of Illinois Urbana-Champaign, 1110 W Green St, Urbana, IL 61821, USA}

\begin{abstract}
Spectral-distortion anisotropies in the cosmic microwave background provide
a unique window on the thermal history, particle content and inflationary
perturbations in the Universe.
Their sources are most simply specified in coordinates adapted to the
physical clock and emission frame, whereas numerical evolution benefits
from a common gauge choice.
We formulate a gauge-covariant frequency hierarchy for the linear scalar
Boltzmann equation, including Thomson and thermalization collision terms
in a general gauge. We provide Newtonian reconstructions from arbitrary
starting gauges, suitable as inputs to the cosmological thermalization code \texttt{CosmoTherm}.
These techniques allow sources to be specified in their most convenient
gauge and frame and spectral anisotropies to be evolved and transformed
between gauges, with photon frequency measured by observers normal to each time slice.
As an illustration, we consider the decay of a trace component of cold
dark matter directly into photons, emitted isotropically in its rest frame.
For adiabatic perturbations in the long-wavelength limit during a decay
in the chemical potential era (i.e.~when Compton equilibration is rapid), there is manifestly no fluctuation in the
chemical potential in synchronous gauge throughout the common {\it local} history.
In Newtonian gauge, four contributions to the chemical potential combine
to describe a temporal displacement along the evolving mean and cancel
once the average spectrum becomes static. 
The stationary spectral anisotropy that results has the spectrum of a
frequency-shifted background Bose--Einstein distribution rather than a
temperature-shifted blackbody. Since the orbital dipole shares this frequency dependence, fitting channel
gains to a Doppler shifted blackbody absorbs the stationary signal into
a single effective temperature map. The \texttt{CosmoTherm} frequency hierarchy reproduces this prediction
after we correct its effective low-frequency photon-production source to
include the required temperature and spectral-shift responses. Although small in amplitude, the correction removes a spurious intrinsic
chemical-potential mode in the long-wavelength limit and restores the
stationary frequency-shift response of the final mean. The formulation
allows distortion anisotropies to be treated in a gauge tailored to the
physical source.
\end{abstract}

\maketitle

\section{Introduction}
\label{sec:introduction}

The spectrum of the cosmic microwave background (CMB) provides a unique window into the
thermal history of the Universe \cite{SunyaevZeldovich:1970,Burigana1991,HuSilk:1993}, offering a broad science case for exploration with future CMB spectrometers \citep{Chluba2019BAAS, Chluba2021Voyage, Cyr2026SDscience}. Energy release and photon production
can leave departures from a blackbody that preserve information about
primordial fluctuations and particle interactions.
The anisotropy of the spectrum opens a spatial dimension to this probe, connecting through their causal generation the local processes to cosmological perturbations. Their prediction and interpretation involve the physics of production, the subsequent transport of photons, and the measurement of the resulting frequency-dependent signal.

Spectral distortions can constrain and potentially detect new particles and their interactions in the early universe.  
Even in the absence of new physics, 
acoustic dissipation provides a connection to the primordial perturbations:
the energy stored in short-wavelength fluctuations heats the photon
background and produces spectral distortions~\cite{Sunyaev:1970plh,Daly1991, Hu:1994bz,ChlubaKhatriSunyaev:2012}.
The transport of photons through  Compton energy
redistribution and photon production by double Compton (DC) scattering and
bremsstrahlung (BR) converts an injection into a temperature shift or a
surviving distortion~\cite{SunyaevZeldovich:1970,Burigana1991,HuSilk:1993,ChlubaSunyaev:2012}.
Correlations of distortion and temperature anisotropies can then probe
the coupling of these short fluctuations to long-wavelength modes,
providing a window onto primordial non-Gaussianity~\cite{PajerZaldarriaga:2012, Ganc2012}.

This transport is challenging to compute given that it involves the full radiative transport of the six-dimensional phase space distribution of the photons.   
The frequency hierarchy (FH) framework developed recently \citep{Chluba:2022xsd,Chluba:2022efq,kite_spectro-spatial_2023-III,Chluba:2026nmu} addresses this challenge and 
provides a common description of spectral evolution and angular
transport. Its spectral basis gives a finite representation of the
thermalization problem~\cite{Chluba:2022xsd}, while the generalized
Boltzmann hierarchy evolves temperature and distortion multipoles with
scattering, Doppler shifts, gravitational frequency shifts, and perturbed
heating~\cite{Chluba:2022efq}. Recent improvements to the collision treatment in \texttt{CosmoTherm} include stimulated scattering and
kinematic corrections~\cite{Chluba:2026nmu}. Here and below, the published
\texttt{CosmoTherm} treatment denotes the FH thermalization prescription
specified in that reference. Our calculations with photon production use
a corrected DC source of App.~\ref{app:dcbr-reduction} and, when BR is
retained, its extension in App.~\ref{app:br-extension}.
These developments supply
the spectral and transport framework for studying spatially varying
energy release and photon injection in the Newtonian gauge for cosmological perturbations.  

As with temperature and polarization anisotropy, being able to choose the gauge to suit the specific physics of a given source and evolution of spectral distortion anisotropy provides both a computational and interpretational advantage.  Gauge-covariant techniques further allow the sources and propagation to be described in different gauges. In particular the source description requires specifying
the clock that controls the source, the time slicing in which its spatial distribution is the simplest, the frame in which its dipole anisotropy vanishes,
and its emission spectrum in those coordinates.   This description is simplest in the coordinates picked out by the sources.  When the photon spectrum evolves in Newtonian gauge, its constant time surface samples a different stage of the source history while 
a change of the local frame shifts the photon frequency. Gravitational and scattering effects shift the frequency of any existing distortion, mixing terms in a fixed spectral basis.  Finally, how observational bands are calibrated changes the separation of propagation basis components into temperature and distortion maps.  In this paper we develop the methods to address all of these issues. 

Here, we reformulate the FH description in gauge-covariant form, which applies to any gauge choice, and provide the gauge-transformation properties of the photon distribution and collision sources that connect them.  These techniques can be used to promote any collision source defined originally in a given gauge, to its form in a different gauge.  For example, the thermalization sources 
of the \texttt{CosmoTherm} Newtonian gauge model~\cite{Chluba:2026nmu} can be promoted to any gauge through this method.  Conversely the gauge-covariant form for the Newtonian photon distribution and collision term can be used to reconstruct its gauge-invariant value from those quantities in the gauge for which its specification is the simplest, as inputs to \texttt{CosmoTherm}. This allows us to test whether the FH source and collision terms describe
the same local physics in different gauges, by comparing the transformed
Newtonian terms with independent local constructions on another slicing.
Such consistency checks were explicitly called for in
Ref.~\citep{Chluba:2022efq} and are carried out here for the retained isotropic thermalization
and unpolarized Thomson collision model.

We illustrate the use of these gauge-covariant FH techniques with an example of direct photon injection from the idealized decay of a collisionless dark matter component.   Here the source injection is most simply described in the source rest frame. The observers normal to source-comoving
synchronous slices share this frame.   For wavemodes that are above the horizon at the end of the decay, the injection is homogeneous and isotropic in these coordinates to leading order.   Recovering the common adiabatic history in Newtonian gauge requires
subtle cancellations among the primordial photon temperature perturbation
and three injection contributions: source density, the displaced decay
clock, and gravitational time dilation or lapse. Misspecifying any of
these terms can generate a spurious {\it intrinsic}\footnote{An intrinsic or local spectral distortion
anisotropy is a difference in spectral shape between patches compared at
the same local age, after removing the frequency shift of the common mean.
Stationary spectral anisotropy instead consists of frequency-shifted copies
of one common spectrum; for a Bose--Einstein spectrum, the physical
chemical potential is the same in every patch.} spectral distortion anisotropy.  

For this example, the only net spectral anisotropy is the usual adiabatic CMB anisotropy but in the so-called stationary form with the spectrum of a frequency-shifted background distortion, for example the Bose-Einstein distribution for decay during the chemical potential era, rather than a temperature-shifted blackbody~\cite{ChlubaKhatriSunyaev:2012,Chluba:2022efq,Evangelista:2026darkphoton}.  
With independently calibrated relative channel responses, fully correlated
temperature and distortion maps could recover the spectral information in
the common mean.
However, the kinematic orbital dipole shares the same frequency-shifted
background spectrum \citep{Danese1981dipole,Balashev:2015lla,Trombetti2021}. Calibrating observational frequency bands to this
dipole with a blackbody template places the entire stationary anisotropy
amplitude in an effective temperature map \cite{Zegeye:2023pbl}.
The gains fitted with that blackbody template absorb the common distortion
response of the orbital dipole and stationary sky anisotropy.

We demonstrate this
connection with a photon-injection source implemented in \texttt{CosmoTherm}:
its early-time evolution reproduces the simple moment calculation, while
continuous cosmological evolution preserves nearly stationary spectral anisotropy
through the acoustic peaks and damping tail, including photon-production
thermalization. Photon production continues to evolve the mean after
injection, while preserving the
common adiabatic history in the long-wavelength limit; the resulting
final anisotropy follows the frequency shift of the thermalized mean.

The thermalization example also exposes a consistency requirement for
the perturbation of the reduced low-frequency photon-production model.
The retained unpolarized Thomson and improved Compton terms agree with
the independent local construction under the stated approximations.
For DC, the emissivity must depend on the physical local photon
temperature, while the arbitrary reference temperature only labels
frequency. Restoring this distinction corrects an omitted temperature
response in the previous \texttt{CosmoTherm} FH setup.
The local photon temperature determines both how the distortion is
identified in the incident spectrum and the spectral shape of its
relaxation. Expressing this same local process in a fixed
reference-frequency basis requires both spectral responses, in addition
to the change in the thermalization rate.
Without these responses, patches that should share one thermalization
history acquire a spurious local chemical-potential fluctuation.
In the long-wavelength adiabatic limit, each patch follows the same
homogeneous injection and thermalization history. The separate-universe
test therefore requires the intrinsic chemical-potential fluctuation to
vanish. The corrected source preserves this common history even while
the mean relaxes. Appendix~\ref{app:dcbr-reduction} develops
the DC correction first, and App.~\ref{app:br-extension} gives the consistent BR
extension. The accuracy of these local models, especially the BR extension
remains a separate question \citep[see][for related discussions for average distortion evolution]{Chluba2014,Evangelista2025}.

Section~\ref{sec:gauge} presents the gauge-covariant FH with total collision
multipoles, its Newtonian
reduction, and the gauge-covariant forms of Newtonian variables. This formulation can be used to 
represent specified collision models in a general gauge or promote collision models from a different gauge with a Newtonian reconstruction.
Section~\ref{sec:source} provides the example of a photon source in the chemical potential era when observational modes are above the horizon.  We use gauge-covariant techniques to first derive the simple synchronous gauge solution and then show how the same result arises from cancellation of multiple Newtonian terms. In both treatments the result is a stationary spectral anisotropy which is
inseparable from an effective temperature anisotropy map after calibration with a blackbody orbital-dipole template.  In Section~\ref{sec:outlook}, we discuss these results and their application to other cases.  

Details of the method are provided in a series of Appendices.
Appendix~\ref{app:conventions} derives the gauge and observer-frame transformations
and the gauge-covariant transport equation.
Appendix~\ref{app:thermalization} derives the Thomson terms and constructs
the gauge-covariant representation of the Newtonian \texttt{CosmoTherm}
treatment~\cite{Chluba:2026nmu} and corrects its effective low-frequency
photon-production source to agree with the specified local model.
Appendix~\ref{app:numerics} relates the frequency-resolved photon source
of Sec.~\ref{sec:source} to the photon and source fluid equations
through the  energy moment hierarchy.  

\section{Gauge-covariant frequency hierarchy}
\label{sec:gauge}

Generalizing the treatment of Ref.~\cite{Chluba:2022efq}, we construct the frequency hierarchy (FH) for the phase-space distribution of CMB photons in a general scalar gauge. We call this hierarchy gauge covariant because its equations retain their form under gauge transformations of the metric, distribution, and collision terms. We also construct gauge-covariant forms that reconstruct the Newtonian variables from any starting gauge; their evaluated values are gauge invariant.   
In Sec.~\ref{sec:metric}, we define the linearized Boltzmann equation in a
general gauge using the Arnowitt--Deser--Misner (ADM) decomposition.
Section~\ref{sec:frequency} decomposes its frequency and angular dependence,
and Sec.~\ref{sec:transformation} gives the Newtonian reduction with
unpolarized Thomson collisions. Section~\ref{sec:gauge-reconstruction}
connects the gauge representations and constructs gauge-covariant forms
of the Newtonian variables.  These results hold for any set of collision terms. Appendix~\ref{app:thermalization} supplies the explicit unpolarized Thomson terms of Ref.~\cite{Chluba:2022efq} and the thermalization terms of Ref.~\cite{Chluba:2026nmu}; Sec.~\ref{sec:source} treats external photon injection.  We take the background to be spatially flat and set $c=\hbar=k_{\rm B}=1$ unless otherwise specified.

\subsection{Metric, observer frame, and photon distribution}
\label{sec:metric}

In a spatially flat background, the eigenmodes for the spatial decomposition of scalar quantities on a given time slice are ordinary plane waves, and vectors and tensors associated with them obey their azimuthal symmetry around the comoving wavevector $\bm k$
\begin{equation}
 {\mathbb Y}=e^{i\bm k\cdot\bm X},\qquad
 {\mathbb Y}_i=-k^{-1}\partial_i{\mathbb Y},\qquad
 {\mathbb Y}_{ij}=\bigl(k^{-2}\partial_i\partial_j
                     +\delta_{ij}/3\bigr){\mathbb Y}.
 \label{eq:harmonics}
\end{equation}
Here  $|\bm k|=k$, and $\bm X$ is the comoving position.

For a general gauge choice of time slicing and spatial threading, we follow the notational conventions of Kodama and Sasaki
(KS)~\cite{Kodama:1984ziu},
\begin{align}
 g_{00}&=-a^2(1+2A{\mathbb Y}), &%
 g_{0i}&=-a^2 B{\mathbb Y}_i,&%
 g_{ij}&=a^2\bigl[(1+2H_L{\mathbb Y})\delta_{ij}
                         +2H_T{\mathbb Y}_{ij}\bigr],
 \label{eq:metric}
\end{align}
where $A$, $B$, $H_L$, and $H_T$ depend on $\eta$ and $k$. Here the index $0$ represents conformal time $\eta= \int dt/a$.  Primes are $\partial_\eta$ whereas overdots are $\partial_t$ and the conformal Hubble rate $\Hc=a'/a = 
\dot a  = a H$.

The $3+1$ geometry of the gauge choice is provided by the correspondence of these variables to the Arnowitt--Deser--Misner (ADM) decomposition.  
In the metric
of Eq.~(\ref{eq:metric}), the unit normal to the slicing is, to first order,
\begin{equation}
 N^\mu=a^{-1}(1-A{\mathbb Y},B{\mathbb Y}^i).
 \label{eq:normal_vector}
\end{equation}
The normal frame is the local rest frame of observers with 4-velocity
$N^\mu$ normal to the chosen time slices such that $N_\mu N^\mu =-1$. Together with an orthonormal
spatial triad, $N^\mu$ defines a tetrad.
The lapse is
$\alpha=a(1+A{\mathbb Y})$ and the
shift is $\beta_i=g_{0i}=-a^2B{\mathbb Y}_i$ at first order in perturbations.
The induced metric on a constant-$\eta$ slice is
$h_{\mu\nu}=g_{\mu\nu}+N_\mu N_\nu$.
The scalar shear amplitude of the normal congruence is
$\sigma_{\rm g}=H_T'/k-B$.  For later notational convenience in defining shear-free time slicing we use
\begin{equation}
\label{eq:def_s}
s\equiv\sigma_{\rm g}/k=H_T'/k^2-B/k
\end{equation}
throughout.
For a fluid, with 4-velocity $u_f^\mu$, the 3+1 decomposition yields the scalar velocity amplitude $v_f$ as
\begin{equation}
 u_f^\mu=a^{-1}(1-A{\mathbb Y},v_f{\mathbb Y}^i).
 \label{eq:fluidvelocity_vector}
\end{equation}
Projecting $u_f^\mu$ into the 3-space of $N^\mu$ gives
$h^\mu{}_{\nu}u_f^\nu=a^{-1}(0,(v_f-B){\mathbb Y}^i)$.
Thus $V_f\equiv v_f-B$ is the scalar coefficient of the fluid velocity
measured by observers normal to the chosen slices.

The ADM geometry also defines the energy $q$ and  propagation direction $\hat{\bm\gamma}$ of the photon through the 3+1 decomposition of its 4-momentum $p^\mu$:
\begin{align}
 p^\mu &= -( p_\nu N^\nu)N^\mu  +h^\mu{}_{\nu}p^\nu \nonumber\\
 &=q(N^\mu+e^\mu),
  \label{eq:momentum-projection}
\end{align}
where
\begin{equation}
 q\equiv-p_\mu N^\mu,\qquad
 e^\mu\equiv q^{-1}h^\mu{}_{\nu}p^\nu , \qquad N_\mu e^\mu\equiv 0.
\end{equation}
Since $p^\mu p_\mu=0$, this implies the space-like vector $e^\mu$ is unit normalized $e_\mu e^\mu=1$.  This means that the components of $e^\mu$ in the normal observer's
orthonormal spatial triad form the direction $\hat{\bm\gamma}$.

For a given wavemode, symmetry requires that only 
$\chi=\hat{\bm k}\cdot\hat{\bm\gamma}$ enters into the scalar system.
We scale the energy $x=q/T_z$ to a reference temperature $T_z\propto a^{-1}$ that uniformly redshifts in spite of any injection of energy into the plasma.\footnote{While it is more conventional to label $\hat{\bm k}\cdot\hat{\bm\gamma}=\mu$, we follow Ref.~\cite{Chluba:2022efq} in defining it as $\chi$ to avoid confusion with the chemical potential. Likewise we choose $x$ for dimensionless frequency and distinguish spatial coordinates as $\bm X$.} To linear order, the phase-space occupation number for a single wavemode then becomes
\begin{equation}
 n(\eta,\bm X,x,\hat{\bm\gamma})
 =n^{(0)}(\eta,x)+n^{(1)}(\eta,k,x,\chi){\mathbb Y}(\bm k,\bm X),
 \label{eq:occupation}
\end{equation}
where the superscript $(0)$ denotes the homogeneous background and $(1)$ the linear fluctuation.  When extracting the relationships between first order quantities, the common $\mathbb Y$ factor is omitted below. 

The  Boltzmann equation defines the evolution of $n$ given arbitrary collisions 
\begin{equation}
 \frac{\mathrm D n}{\mathrm D\eta}=C .
 \label{eq:phase-space-boltzmann}
\end{equation}
Here $\mathrm D/\mathrm D\eta$ follows a photon geodesic, including its change
in position, frequency measured in the normal frame, and direction.  For multiple collision processes $j$, the total $C\equiv\sum_jC_j$, where each component defines a local collision or injection contribution to the occupation-number
equation per unit conformal time.  

We likewise can write
\begin{equation}
C_j(\eta,\bm X, x,\hat{\bm \gamma}) =  C_j^{(0)}(\eta,x)+ 
C_j^{(1)}(\eta,k,x,\chi){\mathbb Y}(\bm k,\bm X).
\end{equation}
Subtracting the background evolution $n^{(0)\prime} = C^{(0)}$ gives
the first-order scalar Boltzmann equation derived explicitly in
Appendix~\ref{app:hierarchy}:
\begin{equation}
 n^{(1)\prime}+ik\chi n^{(1)}-\mathcal G U
 =C^{(1)}.
 \label{eq:general-boltzmann}
\end{equation}
Here primes are taken at fixed $x$. Gravitational effects enter through\footnote{We note that $k^2s=H_T'-kB$ due to Eq.~\eqref{eq:def_s}. In Newtonian gauge, one furthermore has $s=0$.}
\begin{equation}
 \mathcal G\equiv-\left(H_L'+\frac{kB}{3}\right)-ik\chi A
   +\frac{2k^2s}{3}P_2(\chi),\qquad
 P_2(\chi)=\frac{3\chi^2-1}{2},
 \label{eq:general-redshift-source}
\end{equation}
where the three terms in $\mathcal G$ describe the expansion, acceleration,
and shear of the normal observers and generalize the usual gravitational
redshift from propagation across the gradient of the gravitational potential.
These effects change the spectrum through the frequency-shift response of
the unperturbed background
\begin{equation}
U(\eta,x)=\Ox n^{(0)}(\eta,x),
\end{equation}
where the frequency-shift operator is
$\Ox=-x\partial_x$ \citep[see Ref.][for broader discussion about the role of this operator in Lorentz boosts]{ChlubaBO25}.
Gravitational redshift and Thomson Doppler terms share the frequency-shift
response $U$, which sets the spectrum of stationary
anisotropy~\citep[e.g., Ref.][]{Evangelista:2026darkphoton}.  

\subsection{Frequency and angular hierarchy}
\label{sec:frequency}

The FH decomposes the phase space fluctuations $n^{(1)}$ into a basis that describes the frequency and angular propagation direction of the photons.    For the frequency dimension, Ref.~\cite{Chluba:2022xsd,Chluba:2022efq} introduced a discrete basis that compresses the information into a finite number of elements based on the reference blackbody distribution,
$n_{\rm bb}(x)\equiv(e^x-1)^{-1}$, and a hierarchy of successive frequency-shift operations
\begin{align}
 G&\equiv\Ox n_{\rm bb}, &
 Y&\equiv\Ox(\Ox-3)n_{\rm bb}, &
 Y_r&\equiv4^{-r}\Ox^rY\qquad(0\leq r\le N),
  \label{eq:fh-shapes}
 \end{align}
 complemented by
 \begin{align}
 M&\equiv (\beta_M^{-1}-x^{-1}) G,
   \label{eq:fh-M}
 \end{align}
 where
 \begin{align}
 \beta_M&\equiv\frac{\int_0^\infty x^2G\,\dd x}
                 {\int_0^\infty xG\,\dd x}=\frac{18}{\pi^2}
            \zeta(3) \approx 2.19229.
 \label{eq:betaM}
\end{align}
 The shape $M$ is related to the linear Bose--Einstein
chemical-potential response (see Sec.~\ref{sec:stationary-response} for the relation).   
However both $M$ and $Y_0=Y$ are defined to have vanishing photon-number moment by
construction.
For a spectral function of the dimensionless frequency $x$, we define
the photon-number and energy moments relative to the reference blackbody by
\begin{equation}
 \mathcal N[f]\equiv\frac{\int_0^\infty x^2f\,\dd x}
                   {\int_0^\infty x^2n_{\rm bb}\,\dd x},
 \qquad
 \mathcal E[f]\equiv\frac{\int_0^\infty x^3f\,\dd x}
                   {\int_0^\infty x^3n_{\rm bb}\,\dd x}.
 \label{eq:spectral-moments}
\end{equation}
Thus  $\mathcal N[Y]= \mathcal N[Y_r]=\mathcal N[M]=0$ so that the number moment is carried only by $\mathcal N[G]=3$.  For the energy moment $\mathcal E[Y]=\mathcal E[Y_r]=4$ and 
\begin{equation}
\mathcal E[M] = \frac{4}{\beta_M} - \frac{90}{\pi^4} \zeta(3) \approx \frac{1}{1.40066}.
\end{equation}
This is the $\epsilon_M$ normalization of Eq.~(2.3) in
Ref.~\cite{Chluba:2026nmu}.
Here we have used the integration by parts identities
$\mathcal N[\Ox f]=3\mathcal N[f]$ and
$\mathcal E[\Ox f]=4\mathcal E[f]$ under the assumption that boundary terms in the integrals vanish. 
The $Y_r$ truncation for a given $N$ can be extended to larger $N$ as needed to represent increasingly complicated spectral functions.   
For the background, the spectral basis representation becomes 
\begin{equation}
 \bm{\mathcal B}(x)=(G,Y_0,\ldots,Y_N,M)^T,\quad
 n^{(0)}-n_{\rm bb}\approx \bm{\mathcal B}\cdot\bm y^{(0)},
 \quad C^{(0)} \approx \bm{\mathcal B}\cdot\bm c^{(0)}.
 \end{equation}
 The approximation here represents the fact that the truncated basis is not complete for arbitrary spectra $n^{(0)}$ and this reconstruction can lose information, e.g.~for spectral lines.   More specifically, let $\mathcal B$ denote the $d\times(N+3)$ matrix whose columns sample the basis functions at $d$ frequencies. It constructs a discretized spectrum from $N+3$ basis coefficients. Let $\mathcal L$ be an $(N+3)\times d$ matrix of linear spectral weights. The projection onto the coefficients is the $(N+3)\times d$ matrix $\mathcal P$, defined for nonsingular $\mathcal L\mathcal B$ by
 \begin{equation}
 \mathcal P = (\mathcal L \mathcal B)^{-1} \mathcal L, \qquad
 \bm y^{(0)} = \mathcal P ( n^{(0)}-n_{\rm bb} ),
 \end{equation} 
 such that $\mathcal P \mathcal B = I$, the $(N+3)\times(N+3)$ identity matrix (see \cite{Chluba:2022xsd}, Appendix B for the specific linear weights $\mathcal L$ that preserve energy conservation). Loss of spectral information means that $\mathcal B \mathcal P \ne I$, the $d\times d$ identity matrix.  We hereafter assume that the retained basis represents the spectra to the required accuracy and write the approximate reconstruction as an equality. A discussion of limitations can be found in \citep{Chluba:2022xsd}.

 Since $n_{\rm bb}$ is static at fixed $x$, the background equation
$n^{(0)\prime}=C^{(0)}$ becomes
$\bm y^{(0)\prime}=\bm c^{(0)}$.
The perturbation $n^{(1)}$ and its collision term $C^{(1)}$ are additionally functions of the angle $\chi$.  
We decompose a general function $X(\chi)$ into multipole moments as
\begin{equation}
 X(\chi)=\sum_{\ell=0}^{\infty}(2\ell+1)(-i)^\ell
 X_\ell P_\ell(\chi),\qquad
 X_\ell=\frac{i^\ell}{2}\int_{-1}^{1}P_\ell(\chi)X(\chi)\,\dd\chi ,
 \label{eq:angular-multipoles}
\end{equation}
as usual.  Thus
\begin{align}
 n_\ell^{(1)}& =\bm{\mathcal B}\cdot\bm y_\ell ,\qquad
 C_\ell^{(1)}  = \bm{\mathcal B}\cdot\bm c_\ell ,
\end{align}
where for notational simplicity we omit the superscript $(1)$ on the first-order spectral basis coefficients.
Throughout, $n$ and $C$ denote frequency-dependent distributions and
collision terms, while bold vectors such as $\bm y_\ell$ and $\bm c_\ell$
denote their spectral-basis coefficients. The basis shapes $G$, $Y_r$ and
$M$ remain functions of frequency. Hatted frequency operators act on these
functions; matrices such as $M_{\rm B}$ below act on coefficient vectors.
The projection $\mathcal P$ maps functions to coefficients, and the basis
functions reconstruct the spectrum to the retained accuracy \citep[see][for discussions about convergence]{Chluba:2022xsd}.

Since the gravitational sources of the Boltzmann equation involve the
frequency shift of the background, it is useful to define its frequency-shift
coefficients in the spectral basis:
\begin{equation}
 \bm b^{(0)}\equiv\mathcal P[U]
 = \mathcal P[\Ox n^{(0)}]
 =\mathcal P[\Ox (\mathcal B \bm y^{(0)}+n_{\rm bb}) ].
 \end{equation}
Ref.~\cite{Chluba:2022efq} defines this composition in terms of the boost matrix $M_B$, which applies the frequency shift to $\bm y^{(0)}$
\begin{equation}
M_B = \mathcal P \Ox \mathcal B, \qquad 
\mathcal P[\Ox F]=M_{\rm B}\mathcal P[F],
\label{eq:boost-projection}
\end{equation}
for any spectrum $F$ represented in the retained basis, and so
\begin{equation}
 \bm b^{(0)}=\bm e_G+M_{\rm B}\bm y^{(0)},
 \label{eq:fh-background-response}
\end{equation}
where $\bm e_G$ is the unit vector
of the $G$ direction. Although the choice of closure in $\mathcal L$ is not unique, it is important to specify the same projection operation $\mathcal P$ throughout the source description and evolution as inconsistent choices would lead to residual inconsistencies in the predicted observables.

We can now assemble the gauge-covariant FH as\footnote{For nonzero background spatial
curvature $K$, the local spectral basis and collision kernels retain their
form, while plane waves and the flat-space propagation couplings are
replaced by curved-space harmonics and multipole couplings with factors
depending on $K/k^2$. Metric-source projections must use the same harmonic
normalization, as in Eqs.~(33) and (35)--(36) of Ref.~\cite{Hu:1997mn}.},
\begin{subequations}
\begin{align}
 \bm y_0'&=-k\bm y_1
   -\left(H_L'+\frac{kB}{3}\right)\bm b^{(0)}+\bm c_0,
 \label{eq:general-fh-monopole}\\
 \bm y_1'&=\frac{k}{3}(\bm y_0-2\bm y_2)
   +\frac{kA}{3}\bm b^{(0)}
   +\bm c_1,
 \label{eq:general-fh-dipole}\\
 \bm y_2'&=\frac{k}{5}(2\bm y_1-3\bm y_3)
   -\frac{2k^2s}{15}\bm b^{(0)}
   +\bm c_2,
 \label{eq:general-fh-quadrupole}\\
 \bm y_\ell'&=\frac{k}{2\ell+1}
   [\ell\bm y_{\ell-1}-(\ell+1)\bm y_{\ell+1}]
   +\bm c_\ell,\qquad\ell\geq3 .
 \label{eq:general-fh-higher}
\end{align}
\end{subequations}
For multiple collision sources 
$\bm c_\ell =\sum_j\bm c_{j,\ell}$.
We isolate the Thomson collision term next and relate the gauge-covariant FH to the Newtonian gauge FH given in 
Ref.~\cite{Chluba:2022efq}.
Appendix~\ref{app:thermal-fh}
gives the thermalization collision terms of
Ref.~\cite{Chluba:2026nmu} in a general gauge.  We treat the case of an external collision source in Sec. \ref{sec:source}.

\subsection{Newtonian reduction and Thomson collisions}
\label{sec:transformation}

With the gauge-covariant FH defined, we can proceed to reduce it to special cases for comparison to the literature, namely unpolarized Thomson scattering in Newtonian gauge with arbitrary additional collision terms.
The Newtonian gauge is defined by shear free slicing $s=0$ and shift-free threading and an isotropic spatial metric $H_T^{\rm N}=0$ which combined set the shift $B^{\rm N}=0$.
In our notation, a gauge label like '${\rm N}$' can appear
as a subscript or superscript chosen for clarity for each variable.
For notational compactness, we call the remaining metric variables 
$\Phi \equiv H_L^{\rm N}$ and $\Psi \equiv A^{\rm N}$.  

Expressing the transport equation in the Newtonian variables gives
the Newtonian form
\begin{equation}
 (n_{\rm N}^{(1)})'+ik\chi n_{\rm N}^{(1)}
 +(\Phi'+ik\chi\Psi)U
 =\sum_j C_{j{\rm N}}^{(1)}.
 \label{eq:invariant-boltzmann}
\end{equation}
For unpolarized Thomson scattering ($j={\rm T}$), the conformal scattering rate
is $\tau'=a n_e\sigma_T$, where $n_e$ is the background free-electron
number density and $\sigma_T$ the Thomson cross section.
Equation~\eqref{eq:app-thomson-projection} gives $C_{\rm T}^{(1)}$
in a general gauge. Its projection onto the spectral basis gives \citep[compare also Ref.][]{Chluba:2022efq}
\begin{align}
\bm c_{T,0} &= 0,\nonumber\\
\bm c_{T,1} &= - \tau'\left(\bm y_1 -\frac{V_b}{3}\bm b^{(0)}\right),\nonumber\\
\bm c_{T,2} &= -\frac{9}{10}\tau'\bm y_2, \nonumber\\
\bm c_{T,\ell} &= -\tau' \bm y_\ell,\qquad\ell\geq3 .
\end{align}
Notice that the dipole source takes the spectral basis form of a frequency shift.
This dipole source comes from the Doppler shift induced by the  baryon velocity, whose 4-velocity has the ADM components $u_b^0=a^{-1}(1-A{\mathbb Y})$, 
$u_b^i=a^{-1}v_b{\mathbb Y}^i$, and 
$V_b \equiv v_b-B$, the baryon velocity relative to the normal observers.  Specializing to Newtonian gauge one has $v_b^{\rm N}=V_b^{\rm N}$. 

In any gauge, define $\bm J_\ell\equiv\bm c_\ell-\bm c_{{\rm T},\ell}$
as the sum of collision terms other than Thomson scattering, with
$\bm J_{j,\ell}=\bm c_{j,\ell}$ for each non-Thomson sector $j$.
Under these assumptions the Newtonian gauge FH becomes
\begin{subequations}
\begin{align}
 (\bm y_0^{\rm N})'&=-k\bm y_1^{\rm N}-\Phi'\bm b^{(0)}+\bm J_0^{\rm N},
 \label{eq:fh-monopole}\\
 (\bm y_1^{\rm N})'&=\frac{k}{3}(\bm y_0^{\rm N}-2\bm y_2^{\rm N})
   +\frac{k\Psi}{3}\bm b^{(0)}
   -\tau'\left(\bm y_1^{\rm N}-\frac{V_b^{\rm N}}{3}\bm b^{(0)}\right)
   +\bm J_1^{\rm N},
 \label{eq:fh-dipole}\\
 (\bm y_2^{\rm N})'&=\frac{k}{5}(2\bm y_1^{\rm N}-3\bm y_3^{\rm N})
   -\frac{9}{10}\tau'\bm y_2^{\rm N}+\bm J_2^{\rm N},
 \label{eq:fh-quadrupole}\\
 (\bm y_\ell^{\rm N})'&=\frac{k}{2\ell+1}
   [\ell\bm y_{\ell-1}^{\rm N}-(\ell+1)\bm y_{\ell+1}^{\rm N}]
   -\tau'\bm y_\ell^{\rm N}+\bm J_\ell^{\rm N},\qquad \ell\geq3 .
 \label{eq:fh-higher}
\end{align}
\end{subequations}
The streaming, gravitational and unpolarized Thomson terms coincide
with Eq.~(2.30) of Ref.~\cite{Chluba:2022efq}.  The generalization to polarized Thomson scattering would introduce polarization sources to the Thomson collision terms and alter the $9/10$ factor for the $\ell=2$ equation, an extension that is left to future work.

\subsection{Gauge transformations and gauge-covariant forms}
\label{sec:gauge-reconstruction}

The gauge-covariant FH relates different gauges through transformations of the metric, phase space occupation function, and collision term. These transformations also supply gauge-covariant forms that reconstruct quantities in a specified target gauge from variables in another gauge.  
For example we illustrate the construction of Newtonian gauge collision sources from the synchronous gauge in Sec.~\ref{sec:source-gauges} and the general gauge thermalization term from Newtonian gauge in App.~\ref{app:thermal-fh}.

For a scalar coordinate change
$\bar\eta=\eta+\mathcal{T}{\mathbb Y}$ and
$\bar X^i=X^i+L{\mathbb Y}^i$, the metric fluctuations change as
\begin{align}
 \bar A&=A-\mathcal{T}'-\Hc \mathcal{T}, \nonumber\\
 \bar B&=B+L'+k\mathcal{T},\nonumber\\
 \bar H_L&=H_L-kL/3-\Hc \mathcal{T}, \nonumber\\
 \bar H_T&=H_T+kL,\nonumber\\
 \bar s&=s-\mathcal{T},
 \label{eq:metric-transform}
\end{align}
and a fluid velocity scalar relative to normal observers as
\begin{align}
\bar V_f \equiv \bar v_f-\bar B=V_f-k\mathcal{T}.
\end{align}
The combination implies that the fluid velocity relative to fixed spatial coordinates transforms as  $\bar v_f=v_f+L'$. 
Changing the slicing changes the normal observers and the frequency they
measure. As shown in Appendix~\ref{app:conventions}, the distribution
and collision terms expressed in the transformed normal frame obey
\begin{subequations}
\begin{align}
 \bar n^{(1)}-n^{(1)}
 &= (\Hc \mathcal{T}+ik\chi \mathcal{T})U-\mathcal{T}n^{(0)\prime},
 \label{eq:distribution-transform}\\
 \bar C_j^{(1)}-C_j^{(1)}
 &= \Hc \mathcal{T}\Ox C_j^{(0)}-(\mathcal{T}C_j^{(0)})'
   +ik\chi \mathcal{T}(\Ox-1)C_j^{(0)}.
 \label{eq:source-transform}
\end{align}
\end{subequations}
Note that the time surface change takes the background evolution of both quantities into a first order perturbation.
These transformations supply the gauge-covariant forms of FH variables by choosing $\mathcal{T}$ and $L$ to reach a specified target gauge.  

Finally for any local scalar $R=R^{(0)} +\delta R\,{\mathbb Y}$, comparing
the same spacetime event before and after the time shift gives
\begin{equation}
 \overline{\delta R}=\delta R-R^{(0)\prime}\mathcal{T} .
 \label{eq:app-scalar}
\end{equation}
If $R^{(0)}\ne0$, its fractional perturbation
$\varepsilon_R=\delta R/R^{(0)}$ consequently satisfies
$\bar\varepsilon_R=\varepsilon_R-(R^{(0)\prime}/R^{(0)})\mathcal{T}$.
These rules also apply to proper scalar rates defined in a fluid rest frame.
Their perturbations depend on the slicing. Choosing $\mathcal{T}=V_f/k$ reaches
fluid-comoving slices, where $\bar V_f=0$, and gives the gauge-covariant
reconstruction of the fractional rate perturbation on those slices:
\begin{equation}
 \varepsilon_{R,f}
 =\varepsilon_R-\frac{R^{(0)\prime}}{kR^{(0)}}V_f,
 \label{eq:app-rest-rate}
\end{equation}
which evaluates to a gauge-invariant value since the time-shift terms of $\varepsilon_R$ and $V_f$ cancel. 

For concreteness, let us apply this construction to Newtonian gauge variables from the previous section. 
For a target Newtonian gauge where $\bar s= s^{\rm N}=0$, 
Equation~\eqref{eq:metric-transform} 
fixes  $\mathcal{T}= s$ and $L = -H_T/k$.
Therefore the gauge-covariant forms
of the Newtonian occupation perturbation and collision terms become
\begin{subequations}
\begin{align}
 n_{\rm N}^{(1)}
 &=n^{(1)}+s
       (\Hc U-n^{(0)\prime})+ik\chi sU,
 \label{eq:invariant-distribution}\\
 C_{j{\rm N}}^{(1)}
 &=C_j^{(1)}+\Hc s\Ox C_j^{(0)}
   -\left(sC_j^{(0)}\right)'
   +ik\chi s(\Ox-1)C_j^{(0)},
 \label{eq:invariant-source}
\end{align}
\end{subequations}
and for the gravitational and Doppler sources
\begin{align}
 \Phi&=H_L+H_T/3-\Hc s,\nonumber\\
 \Psi&=A-s'-\Hc s,\nonumber\\
 V_b^{\rm N}&=V_b-ks.
 \label{eq:invariant-metric}
\end{align}
Given Eq.~\eqref{eq:invariant-distribution}, the spectral basis coefficients transform as
\begin{subequations}
\begin{align}
 \bm y_0^{\rm N}&=\bm y_0+s
       \left[\Hc\bm b^{(0)}-(\bm y^{(0)})'\right],
 \\
 \label{eq:dipoleN}
 \bm y_1^{\rm N}&=\bm y_1-\frac{ks}{3}\bm b^{(0)}, 
 \\
 \bm y_\ell^{\rm N}&=\bm y_\ell\qquad (\ell\geq2).
 \label{eq:invariant-fh-amplitudes}
\end{align}
\end{subequations}
The monopole includes a displacement along the evolving background and
a frequency rescaling; the dipole includes the change of normal frame.
The spectrally-projected collision term similarly follows from  Eq.~\eqref{eq:invariant-source} and involves the corresponding clock, frequency, and
boost terms.

We call these expressions gauge-covariant forms because the same expressions reconstruct the specified target-gauge quantities from variables in any starting gauge. Their evaluated values are gauge invariant: changing the starting gauge leaves the reconstructed quantities unchanged. Since reconstruction and direct evaluation in the target gauge give the same quantities, we use the same symbols for both. In the literature, the reconstructed combinations are often called gauge-invariant variables and distinguished symbolically from their directly evaluated counterparts.

In the inverse direction, these forms provide a check of the gauge
covariance of the FH for any specified collision term. Starting
from a Newtonian gauge FH, e.g.~Eqs.~(\ref{eq:fh-monopole})--(\ref{eq:fh-higher}),
substitute each variable's gauge-covariant form from
Eqs.~(\ref{eq:invariant-distribution})--(\ref{eq:invariant-fh-amplitudes}).
The resulting equations must reproduce the gauge-covariant FH,
Eqs.~(\ref{eq:general-fh-monopole})--(\ref{eq:general-fh-higher}),
as derived independently. An incorrectly derived FH can fail this
covariance check. Under an arbitrary time shift $\mathcal{T}$, the transformed
metric, distribution, and source must supply cancelling terms so that
the hierarchy retains its form. The reconstructed Newtonian quantities
must also retain their values.
For example, the dipole frame shift of Eq.~(\ref{eq:dipoleN}) provides the
 shear source in the gauge-covariant quadrupole evolution in Eq.~(\ref{eq:general-fh-quadrupole}) from the Newtonian gauge Eq.~(\ref{eq:fh-quadrupole}).  For the thermalization collision terms of the
\texttt{CosmoTherm} treatment~\cite{Chluba:2026nmu}, Appendix~\ref{app:thermal-checks}
provides the independent frequency-dependent comparison, and
Appendix~\ref{app:dcbr-reduction} corrects the effective low-frequency
photon-production source in the spectral basis with this technique.
These transformation techniques are also useful when collision sources for a given process are more easily specified in an alternate gauge from the Newtonian gauge  as we shall see next.

\section{Photon injection and the interpretation of spectral anisotropies}
\label{sec:source}

The gauge-covariant FH provides a method where the calculational gauge and representation of the collision terms can be chosen to best suit the physical problem at hand.   As an illustrative example, here we consider the direct injection of photons from an external and otherwise collisionless cold dark matter type source.   The photon injection term is simplest in the source rest frame, the
frame of observers normal to source-comoving synchronous slices.
The full computational machinery of \texttt{CosmoTherm}  has
instead been developed in Newtonian gauge.  
We first specify the source clock and energy transfer and provide their
gauge-transformation properties in Sec.~\ref{sec:source-clock}.
The emission spectrum and source collision term are specified in
Sec.~\ref{sec:source-collision}. We compare the simplicity of the collision term in synchronous gauge to its construction in Newtonian gauge in Sec.~\ref{sec:source-gauges}.
Section~\ref{sec:pulse} compares the moment and \texttt{CosmoTherm}
calculations of a prompt injection in the $\mu$-epoch, for which adiabatic
perturbations generate no leading intrinsic chemical-potential fluctuation
while injection changes the common background spectrum.
Section~\ref{sec:stationary-response} interprets stationary spectral
anisotropy in the basis, and Sec.~\ref{sec:cosmological-propagation} follows
its cosmological propagation. Finally,
Sec.~\ref{sec:calibration}
shows that dipole calibration of observational channels with the usual
blackbody template produces only effective ``temperature'' maps of CMB anisotropy.

\subsection{Source clock and energy transfer}
\label{sec:source-clock}

As a simple example that demonstrates gauge-covariant FH techniques, consider the direct photon injection
from a subdominant pressureless source component $X$. Apart from its
decay, we assume $X$ is collisionless and free falls like cold dark matter
(see Appendix~\ref{app:source-moments} for a detailed derivation of its
fluid equations). We give it adiabatic initial conditions and neglect its effect on the
background expansion and metric fluctuations. The analytic example takes
radiation domination; the cosmological calculation in
Sec.~\ref{sec:cosmological-propagation} includes matter and neutrinos.
Its stress tensor is $T_X^{\mu\nu}=\rho_Xu_X^\mu u_X^\nu$, where
$\rho_X$ is its proper energy density and $u_X^\mu$ its 4-velocity. The decay is assumed to be isotropic in the source rest frame defined
by $u_X^\mu$.  

The decay transfers energy and momentum to the photons.  
If $Q$ is the energy
transferred per unit  proper time and proper volume, the energy-momentum exchange in a general gauge is given by 
\begin{equation}
 Q_\gamma^\mu=Q u_X^\mu,
 \qquad
 \left.\nabla_\mu T_\gamma^{\mu\nu}\right|_Q=Q_\gamma^\nu,
 \qquad
 \nabla_\mu T_X^{\mu\nu}=-Q_\gamma^\nu ,
 \label{eq:source-ledger}
\end{equation}
where
$T_\gamma^{\mu\nu}$ is the photon stress tensor. 
The restriction $|_Q$ isolates the decay process; scattering and
thermalization contribute their own transfers to the photons.

The decay changes the abundance of $X$, which then continuously depletes the photon source, while the surviving particles
experience only gravity and continue to free fall. The injection rate therefore obeys 
\begin{equation}
 Q=\Gamma(\tau_X)\rho_X ,
 \label{eq:injectionrate}
\end{equation}
where $\Gamma$ is the prescribed decay probability per unit scalar age $\tau_X$; for example   in
Sec.~\ref{sec:pulse} we take a prompt pulse. The scalar age, or source clock, is the continuum analog of source proper time and obeys
$u_X^\mu\nabla_\mu\tau_X=1$ and vanishes on the
initial hypersurface chosen to be orthogonal to the initial $u_X^\mu$. Thus $\tau_X'=a$ in the background and we set $\tau_X(a=0)=0$.
 The constant-$\tau_X$ surfaces
then define a source-proper-time slicing, whose normal observers follow
the source. With spatial coordinates carried along the source flow from the initial surface where flow corresponds to the Hubble flow,
this is the source-comoving synchronous gauge, $A^{\rm S}=B^{\rm S}=0$; the specified
initial hypersurface fixes its residual time freedom of a general synchronous gauge. In what follows,
``synchronous slicing'' and ``synchronous gauge,'' labeled ${\rm S}$,
refer to this choice. 

For a given decay probability, the transfer $Q$ also responds to the depletion of $\rho_X$.
Thus in synchronous gauge, 
\begin{equation}
\delta Q^{\rm S} = \Gamma \delta\rho_X^{\rm S},
\end{equation}
making the source specification particularly simple in form.
In a general gauge, the conformal time and source clocks differ inhomogeneously due to the lapse perturbation.  In coordinates fixed by the latter, the scalar age picks up a linear fluctuation  $\tau_X=\tau_X^{(0)}+\delta\tau_X{\mathbb Y}$ and changes the proper rate
\begin{equation}
 \delta\tau_X'=aA,\qquad
 \delta Q = \Gamma  \delta\rho_X + 
\rho_X \frac{\dd\Gamma}{\dd\tau_X}\delta\tau_X .
 \label{eq:source-clock-rate}
\end{equation}
For linear perturbations unmarked densities denote their background values.
Since $\tau_X$, $\rho_X$, and $Q$ are spacetime scalars, the coordinate
change of Eq.~\eqref{eq:metric-transform} gives
\begin{align}
 \overline{\delta\tau_X}&=\delta\tau_X-a\mathcal{T}, \nonumber\\
 \overline{\delta\rho_X}&=\delta\rho_X-{\rho_X'}\mathcal{T},\nonumber\\
 \overline{\delta Q}&=\delta Q-Q'\mathcal{T}.
 \label{eq:source-scalar-transform}
\end{align}
Synchronous gauge is then reached by $\mathcal{T}=\delta \tau_X/a$ and utilizing this $\mathcal{T}$ in Eq.~(\ref{eq:metric-transform}) for the gauge-covariant form of the lapse with Eq.~(\ref{eq:source-clock-rate}) verifies $A^{\rm S}=0$ self-consistently.  

To complete the  system we need to evolve the source energy density through the injection. Eq.~(\ref{eq:source-ledger}) gives for the background
\begin{equation}
 \rho_X'+3\Hc\rho_X=-aQ.
 \label{eq:source-component-background}
 \end{equation}
The corresponding perturbed continuity and Euler equations are
\begin{align}
 \delta_X'&=-kv_X-3H_L'
   -a\left(\Gamma A+\Gamma_{,\tau_X}\delta\tau_X\right), \nonumber\\
 V_X'+\Hc V_X
 &=kA.
 \label{eq:sourceperturbation}
\end{align}
Here $\delta_X\equiv\delta\rho_X/\rho_X$, and the scalar source
velocity amplitude $v_X$ is defined by $u_X^i=a^{-1}v_X{\mathbb Y}^i$, and its
velocity amplitude relative to the normal observers is $V_X=v_X-B$.
Notice that under a coordinate transformation 
$\bar v_X=v_X+L'$ and together with
$\bar B=B+L'+k\mathcal{T}$, this gives
\begin{equation}
 \bar V_X=V_X-k\mathcal{T}.
 \label{eq:source-velocity-transform}
\end{equation}
Thus $\mathcal{T}=V_X/k$ reaches source-comoving slicing and equating this $\mathcal{T}$ with the synchronous $\mathcal{T}=\delta \tau_X/a$ verifies the Euler equation.  Hence synchronous gauge quantities remain self-consistently source-comoving when evaluated through their gauge-covariant forms from another gauge since $X$ is in free fall.
Again, the source specification in synchronous gauge is particularly simple in that the Euler equation for the source is automatically satisfied and with  $A^{\rm S}=\delta\tau_X^{\rm S}=0$, the energy equation takes a simple form.

\subsection{Source collision term}
\label{sec:source-collision}

The transfer 4-vector fixes the energy and momentum injected into the
photons, but does not determine their number or frequency distribution.
In the number-conserving Compton limit, the energy and number inputs together
determine the final chemical potential~\cite{HuSilk:1993, Chluba2015GreensII}. We therefore
specify the emission spectrum to complete the collision term.  The rate $Q$ defines the energy moment of the collision term relative to the uniformly redshifting reference blackbody defined in Eq.~(\ref{eq:spectral-moments})
\begin{equation}
\mathcal E[C_Q^{(0)}]=\frac{aQ}{\rho_z},
\end{equation}
where $\rho_z\equiv\pi^{-2}T_z^4\int x^3n_{\rm bb}\,\dd x$.  Let $F(\xi)$ be a non-negative occupation shape with a positive, finite
energy integral. For a characteristic source-frame photon energy scale $E_{\mathrm p}$, define
\begin{equation}
 q_X\equiv-p_\mu u_X^\mu,\qquad \xi\equiv q_X/E_{\mathrm p}.
 \label{eq:source-frequency}
\end{equation}
The background injection can then be written as
\begin{equation}
 C_Q^{(0)}(\eta,x)
 \equiv\frac{\pi^2 aQ}{E_{\mathrm p}^4}
    F\!\left(\frac{T_z}{E_{\mathrm p}}x \right),
 \label{eq:source-spectrum}
\end{equation}
where the energy-moment normalization is preserved by choosing
\begin{equation}
 \int_0^\infty \xi^3F(\xi)\,\dd\xi =1.
 \label{eq:source-shape-integral}
\end{equation}
The total photon-number injection is given by 
$\mathcal N[C_Q^{(0)}]$, which we assume is also finite.  

The perturbation in synchronous gauge takes the simple form
\begin{equation}
 C_{Q{\rm S}}^{(1)}=\frac{\delta Q^{\rm S}}{Q}C_Q^{(0)}
 =\frac{\Gamma \delta \rho_X^{\rm S}}{Q}C_Q^{(0)}
 = \Gamma \pi^2 a\frac{\rho_X}{E_{\mathrm p}^4}\delta_X^{\rm S} F\!\left(\frac{T_z}{E_{\mathrm p}}x\right),
 \label{eq:source-injection-sync}
\end{equation}
where we have defined the density fluctuation $\delta_X^{\rm S}\equiv \delta\rho_X^{\rm S}/\rho_X$.
We can then apply the  gauge transformation property 
Eq.~(\ref{eq:source-transform}) with $\mathcal{T}=V_X/k$ and $(a\mathcal{T})'=aA$ to find
the gauge-covariant expression
\begin{equation}
 C_{Q}^{(1)}=(A+\delta Q/Q)C_Q^{(0)}-i\chi V_X(\Ox-1)C_Q^{(0)},
 \label{eq:source-injection}
\end{equation}
where we have used
\begin{equation}
C_Q^{(0)\prime}=(\Hc+Q'/Q)C_Q^{(0)}+\Hc\Ox C_Q^{(0)},
\end{equation}
where the last term comes from the derivative of $F$ whose argument contains $T_z \propto a^{-1}$ while $E_{\mathrm p}$ is fixed.
Alternately we can derive Eq.~(\ref{eq:source-injection}) directly in the general gauge:
the lapse $A$ converts proper time to conformal time, while the $V_X$ term is
the boost of an isotropic source-frame injection, including the transformation
of the time rate. 

\subsection{Superhorizon injection in synchronous and Newtonian gauges}
\label{sec:source-gauges}

While the specific form of the FH spectral basis decomposition depends on the spectral shape $F(xT_z/E_{\mathrm p})$ of the injection, if the injection occurs exclusively in the chemical potential or $\mu$-era, the resultant spectral anisotropy takes a very simple form since the observable $k$-modes are on superhorizon scales at the time of injection.   In the long-wavelength limit, injection changes the common background
spectrum without generating an intrinsic spectral fluctuation on
source-comoving synchronous slices. After injection and Compton equilibration, the anisotropy takes
the stationary frequency-shift form. For this illustration, we assume prompt Compton equilibration with conserved photon number and energy. We neglect the small energy transferred to the thermal motion of the electrons and ions, including that required to offset their adiabatic cooling considered by \citep{Chluba2005, ChlubaSunyaev:2012}.  For this number-conserving example, we also neglect the low-frequency return to a blackbody at the electron temperature and its associated net photon production and chemical-potential relaxation over the interval considered. We derive the correction to \texttt{CosmoTherm}'s effective
photon-production source in App.~\ref{app:dcbr-reduction} and use the
new formulation in the cosmological calculation of
Sec.~\ref{sec:cosmological-propagation}.
Provided the injected moments admit the assumed Bose--Einstein equilibrium, the background result after number- and energy-conserving Compton equilibration depends only on the total photon number and energy injected, irrespective of $F$~\cite{Chluba:2015hma, Chluba2015GreensII}. This equilibrium approximation applies to injections completed at
$z\gtrsim3\times10^5$, while Compton redistribution is efficient.

Since the injection is particularly simple in synchronous gauge, we start with the synchronous FH.
The entire first-order injection collision term in Eq.~(\ref{eq:source-injection-sync}) is $C_{Q{\rm S}}^{(1)}\propto \delta_X^{\rm S}$, the density fluctuation in $X$. 
For superhorizon adiabatic fluctuations, $\delta_X^{\rm S}$ is suppressed relative to the initial comoving curvature fluctuation $\mathcal R$ by $(k\eta)^2\ll 1$.   
Using the KS metric and velocity conventions, the comoving curvature has the gauge-covariant form
\begin{equation}
 \mathcal R=H_L+\frac{H_T}{3}-\frac{\Hc}{k}(v-B),
 \label{eq:primordial-comoving-curvature}
\end{equation}
where $v$ is the total-fluid velocity amplitude. For adiabatic fluctuations,
$\mathcal R$ remains constant above the horizon. We fix the residual spatial threading in synchronous gauge by the initial condition $H_L^{\rm S}(\eta=0)=0$. For the regular growing adiabatic mode in radiation domination, $\mathcal R=H_T^{\rm S}/3$ to leading order. For this solution, the monopole and quadrupole equations of the
synchronous FH are
\begin{align}
 (\bm y_0^{\rm S})'&=-k\bm y_1^{\rm S}
       -(H_L^{\rm S})'\bm b^{(0)}+\bm J_0^{\rm S},
 \label{eq:synchronous-fh-monopole}\\
 (\bm y_2^{\rm S})'&=\frac{k}{5}
       (2\bm y_1^{\rm S}-3\bm y_3^{\rm S})
       -\frac{2}{15}(H_T^{\rm S})'\bm b^{(0)}
       -\frac9{10}\tau'\bm y_2^{\rm S}+\bm J_2^{\rm S}.
 \label{eq:synchronous-fh-hierarchy}
\end{align}
All of the sources on the right hand side are suppressed:  
The spatial metric drivers as $(H_L^{\rm S})'=\mathcal R\,O(k^2\eta)$ and $(H_T^{\rm S})'=\mathcal R\,O(k^2\eta)$; the injection monopole as $\bm J_{Q,0}^{\rm S}=\delta_X^{\rm S}\bm c_Q^{(0)}$, with $\delta_X^{\rm S}=\mathcal R\,O[(k\eta)^2]$; the higher injection multipoles vanish;  the inflationary adiabatic photon perturbation is suppressed as $\mathcal R\,O[(k\eta)^2]$.
Thus during the injection, which we assume to be confined to the $\mu$ epoch, only the background changes.   Within this approximation, rapid Comptonization leaves the background spectrum as a Bose--Einstein distribution.

After injection and Compton equilibration are complete, gravitational evolution,
streaming, and Thomson scattering preserve the frequency-shift form
$\bm y_\ell^{\rm S}=\Theta_\ell^{\rm S}\bm b^{(0)}$, or
$n_{\rm S}^{(1)}=\Theta^{\rm S}\Ox n^{(0)}$ in frequency space,
throughout horizon entry and subsequent evolution.
This frequency-shift form is the
``stationary solution'' of Evangelista et al.~\cite{Evangelista:2026darkphoton},
Sec.~2.3, Eq.~(2.10). Stationarity refers to the frequency-shift form of the anisotropy
relative to the mean spectrum, while its angular amplitude can still evolve. We call the average spectrum static when it is time
independent at fixed $x=q/T_z$. During evolution of the common adiabatic
mean, the Newtonian monopole also contains a displacement along that
history, as shown below in Eq.~\eqref{eq:two-gauge-fh-null}. This clock
term must be included when comparing slicings. A stationary final
anisotropy does not require the mean to have been static throughout its
formation and thermalization.
For a static mean, $\bm b^{(0)}$ factors out of the hierarchy: the spectral
shape is fixed while $\Theta_\ell^{\rm S}$ evolves in the same way as the
temperature multipoles in the absence of injection.
The Compton terms included in the \texttt{CosmoTherm} treatment vanish on
this solution when the equilibrium electron-temperature response is
included. The monopole cancels, and the dipolar collision source induced
by the electron bulk velocity vanishes with the background Compton
collision term [Sec.~4.7, Eqs.~(4.35)--(4.38), and Sec.~5, Eq.~(5.3), of
Ref.~\cite{Chluba:2026nmu}; see also Eq.~\eqref{eq:th-stationary-compton-null}]. Anisotropic thermal redistribution \citep[e.g., Ref.][for relevant collision terms]{ChlubaSunyaev:2012} in the
electron rest frame is neglected in the \texttt{CosmoTherm} treatment~\cite{Chluba:2026nmu}. Thus
$\bm J_{{\rm C},\ell}^{\rm S}=0$ at its retained order.

The same calculation in Newtonian gauge picks up multiple individual terms associated with the coordinate change that sum to the same simple result.  Omission of any of the terms leads to a false spectral anisotropy signal.  
Above the horizon during radiation domination when $a\propto\eta$, $\Hc=1/\eta$, and $a\eta = 2t$,
\begin{equation}
 \Phi=-\Psi=\frac{2}{3}\mathcal R 
 \quad (k\eta \ll 1).
 \label{eq:radiation-curvature-normalization}
\end{equation}
In this radiation-era example, we neglect the
anisotropic stress of free-streaming neutrinos for simplicity.  The full \texttt{CosmoTherm} calculation in
Sec.~\ref{sec:cosmological-propagation} includes massless neutrinos and matter.
Their effects on gravitational propagation preserve the stationary
frequency-shift solution under the retained collision approximations. 

Importantly $\delta_X^{\rm N}$ is leading order $\mathcal O(\mathcal R)$ for superhorizon modes.     To see this, note that the transformation in the time slicing to synchronous gauge is
\begin{equation}
 \mathcal{T}=\Psi \frac{t}{a} =
 \frac{V_X^{\rm N}}{k}=\frac{\delta \tau_X^{\rm N}}{a},\qquad
 \delta\tau_X^{\rm N}=a\mathcal{T}=\Psi t .
 \label{eq:adiabatic-source-clock}
\end{equation}
The gauge transformation formula for $\delta\rho_X$ in Eq.~(\ref{eq:source-scalar-transform}) gives a leading order contribution 
\begin{equation}
 \delta_X^{\rm N}=\mathcal R \left(1+\frac{2}{3}t\Gamma\right).
 \label{eq:adiabatic-source-density}
\end{equation}
The density fluctuation $\delta_X^{\rm N}$ and clock fluctuation $\delta \tau_X^{\rm N}$ then both modulate the proper rate $\delta Q^{\rm N}$ in Eq.~(\ref{eq:source-clock-rate}). 
Finally in the FH of Eq.~(\ref{eq:source-injection}), there is a third contribution to  $C_{Q{\rm N}}^{(1)}$ of the same order from the lapse entering the rate
\begin{equation}
 \delta Q^{\rm N}+\Psi Q
 =\Gamma\delta\rho_X^{\rm N}
  +\rho_X\dot\Gamma\delta\tau_X^{\rm N}+\Psi Q.
 \label{eq:pedagogical-newtonian-source}
\end{equation}
 We refer to these three Newtonian contributions to the source collision term as density, clock and lapse monopole terms respectively.   The dipole contribution to  $C_{Q{\rm N}}^{(1)}$ is higher order in $k\eta$.

We can alternately see this directly from the transformation in the spectral basis
of the monopole.  Since to leading order $\bm y_0^{\rm S}=0$, its  transformation to Newtonian gauge gives
\begin{equation}
 \bm y_0^{\rm N}
 =\mathcal{T}\left[ (\bm y^{(0)})'-\Hc \bm b^{(0)}\right] 
 = \frac{\Psi}{2} \left[ \frac{\partial \bm y^{(0)}}{\partial \ln a} -\bm b^{(0)}\right].
 \label{eq:two-gauge-fh-null}
\end{equation}
The first term describes the displacement along the evolving mean
spectrum; the second describes the rescaling of frequency relative to
$T_z$. With continuing photon production, the completed local source
preserves the common adiabatic history in the long-wavelength limit,
while the Newtonian time-displacement term can remain nonzero as the mean evolves
(App.~\ref{app:dcbr-reduction}). In the number-conserving example, the
background spectrum stops evolving at fixed $x$ after injection and
Compton equilibration are complete, and the displacement term vanishes.
The remaining term shifts the frequency of the background.
This is the radiation-era analogue of the local temperature contribution
to the Sachs--Wolfe effect. In matter domination, the adiabatic Newtonian
temperature perturbation is $-2\Psi/3$, which combines with the
gravitational redshift $+\Psi$ to give the familiar $\Psi/3$.
In radiation domination, the corresponding local temperature is
$\Theta_{\rm ad}=-\Psi/2=\mathcal R/3$.   Therefore in Newtonian gauge the net result is also just the usual gravitational and Thomson scattering induced anisotropy with the stationary spectrum of the shifted background that can now be spectrally distorted.  

\subsection{Prompt photon injection: moment and CosmoTherm calculations}
\label{sec:pulse}

To illustrate the stationary anisotropy that remains after a direct injection and Compton equilibration complete in the $\mu$ era while the modes are superhorizon, let us take the case of a prompt injection pulse with a fixed spectral shape.
Specifically we take the decay probability in
Eq.~\eqref{eq:injectionrate} to be
\begin{equation}
 \Gamma_\sigma(\tau_X)
=\frac{\lambda}{\sigma}
     g\!\left(\frac{\tau_X-\tau_{\mathrm p}}{\sigma}\right),
 \label{eq:pulse-profile}
\end{equation}
where the subscript ``$\mathrm p$'' labels the pulse, $\tau_{\mathrm p}$ is its
central proper time, and $\sigma$ its width. We use the standard unit-normalized Gaussian
$g(z)=e^{-z^2/2}/\sqrt{2\pi}$.
For the background $\tau_X^{(0)}=t$; hence the central background time is $t_{\mathrm p}=\tau_{\mathrm p}$, with reference temperature
$T_{\mathrm p}=T_z(t_{\mathrm p})$, scale factor $a_{\mathrm p}=a(t_{\mathrm p})$,
and conformal time $\eta_{\mathrm p}=\eta(t_{\mathrm p})$.
The dimensionless
$\lambda$ normalizes the total fraction of $X$ that has decayed during the injection as 
$1-e^{-\lambda}$.
Note that even in the limit of a prompt decay $H\sigma \ll 1$ near $\tau_{\mathrm p}$, it is important to keep terms that shift the decay surface by $\delta \tau_X$ from the synchronous source coordinates to obtain full gauge invariance.

Next for definiteness let us take the spectral shape of the decay to be
$F(\xi)\equiv e^{-\xi}/6$.  At fixed injected photon number and energy, rapid Compton equilibration
erases the detailed shape of $F$. Choosing $F$ fixes their ratio.   For this choice, the mean energy
of an emitted photon is $3E_{\mathrm p}$.

To expose the physical content of the result, we first simply
follow the energy and number changes through the moment diagnostic 
\begin{align}
 \mu_{\rm mom}&=\frac{1}{\mathcal E[M]}
  \ln\left[\frac{\rho_\gamma/\rho_z}
                 {(N_\gamma/N_z)^{4/3}}\right],
\nonumber\\
 \delta\mu_{\rm mom}
 &=\frac{1}{\mathcal E[M]}\left(\frac{\delta\rho_\gamma}{\rho_\gamma}
              -\frac43\frac{\delta N_\gamma}{N_\gamma}\right),
 \label{eq:mu-moment-diagnostic}
\end{align}
where $(\mathcal E[M])^{-1} \approx 1.4$.
Here $\rho_\gamma$ and $N_\gamma$ denote the photon energy and number
densities. Their reference blackbody values are $\rho_z$ and
$N_z=2\zeta(3)T_z^3/\pi^2$, respectively.
Before Comptonization $\delta\mu_{\rm mom}$
 is simply a diagnostic based on the number and energy moments of the photon distribution.  After Compton equilibration, this diagnostic equals the chemical potential to first order in the distortion~\cite{Chluba2015GreensII}.

For this injection spectrum, we can compute these quantities through the decay as
\begin{equation}
 \rho_\gamma'+4\Hc\rho_\gamma=aQ,\qquad
  N_\gamma' +3\Hc N_\gamma=a\frac{Q}{3E_{\mathrm p}},
 \label{eq:pulse-photon-moments}
\end{equation}
for the background and 
\begin{equation}
 \delta\rho'_\gamma+4 \Hc\delta\rho_\gamma=a (\delta Q+AQ),
 \qquad
 \delta N'_\gamma+3\Hc \delta N_\gamma
                         = a \frac{\delta Q+AQ}{3E_{\mathrm p}},
 \label{eq:pulse-photon-perturbations}
\end{equation}
for either synchronous or Newtonian gauge, with $A^{\rm N}=\Psi$ and $A^{\rm S}=0$ since on superhorizon scales $H_L'$ gravitational effects and $v_\gamma$ are negligible in radiation domination. 
Note that in the prompt limit $\sigma\rightarrow 0$ the final chemical potential in the background becomes
\begin{equation}
 \bar\mu_{\rm f}\simeq
 \frac{1}{\mathcal E[M]} \frac{\rho_X(\tau_{\mathrm p})}{\rho_z(\tau_{\mathrm p})}(1-e^{-\lambda})
 \left[1-\frac{2\pi^4}{135\zeta(3)}\frac{T_{\mathrm p}}{E_{\mathrm p}}\right],
 \label{eq:thin-pulse-mu}
\end{equation}
where the densities are understood to be evaluated just before the prompt injection at $\tau_{\mathrm p}$.
For finite $\sigma$, we integrate source depletion and the photon energy
and number through the pulse, starting from the early-time adiabatic
blackbody state. Time-step refinement confirms convergence of the
numerical moment evolution.

\begin{figure}[t]
\centering
\includegraphics[width=0.7\textwidth]{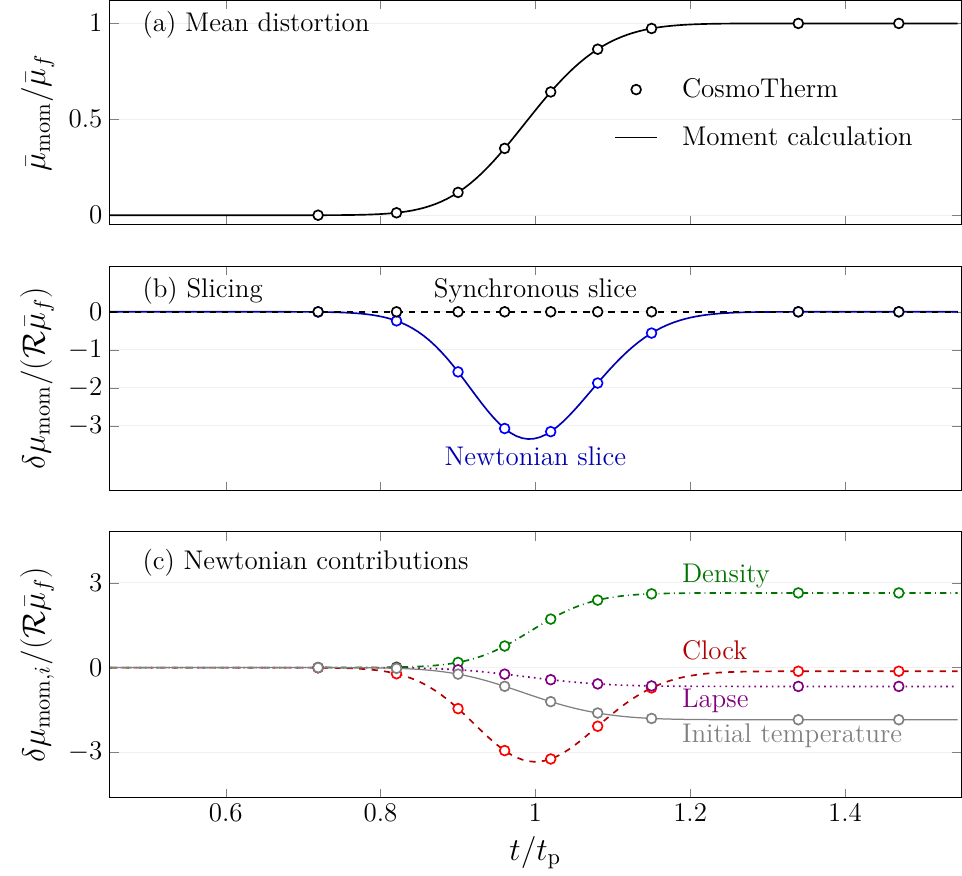}
\caption{Chemical potential evolution constructed from photon number and energy moments for the background and a superhorizon perturbation normalized to its inflationary initial curvature fluctuation $\mathcal R$.  The moment construction reflects the chemical potential after prompt Compton equilibration for injection in the radiation-dominated $\mu$-epoch. Curves show the moment calculation. Open circles show the corresponding moments from the independently evolved Newtonian \texttt{CosmoTherm} frequency hierarchy in the rapid-Comptonization limit, with DC/BR photon production and neutrinos omitted. In panel (b), black circles show the synchronous fluctuation reconstructed from that Newtonian evolution. Panel (a) shows the background growth as normalized to its final value $\bar \mu_f$.  
Panel (b) compares the fluctuation in Newtonian gauge $\delta\mu_{\rm mom}^{\rm N}$ (blue) with that of synchronous gauge
$\delta\mu_{\rm mom}^{\rm S}=0$ (black dashed). Notice that the  Newtonian moment fluctuation is transient and disappears once the injection completes.  Panel (c) decomposes this Newtonian fluctuation into density, clock,
lapse, and inherited-temperature contributions. The three individual source
terms in Eq.~\eqref{eq:pedagogical-newtonian-source} are integrated using the same source and background as panel (b) using the total photon energy and number in the background, including injection.  The gray curve represents the $\mu$ moment constructed from the adiabatic photon number and energy fluctuation before the injection. The four contributions  sum to panel (b) and cancel after the injection. The result after the injection and Comptonization is a stationary spectral anisotropy in both gauges with no local fluctuations in the chemical potential. 
}
\label{fig:two-gauge}
\end{figure}

After injection and Compton equilibration, a uniform chemical potential
coexists with the stationary anisotropy
$n_{\rm N}^{(1)}=\Theta_{\rm ad}\Ox n^{(0)}$.
The frequency shift of the distorted mean has nonzero $G$, $Y$ and $M$
components in the spectral basis, all sharing one angular amplitude.
Figure~\ref{fig:two-gauge} follows the chemical-potential moment fluctuation;
Fig.~\ref{fig:spectral-illustration} displays those spectral-basis components.

The example is thus described by 4 parameters: $\lambda$, $E_{\mathrm p}/T_{\mathrm p}$, $\sigma/\tau_{\mathrm p}$ and the overall scale of the injection $\bar\mu_{\rm f}$.
We denote the mean $G$ coefficient by $\bar\Theta$, whose photon-number
temperature interpretation is developed in Sec.~\ref{sec:stationary-response}.
We assume a small injection, with $|\bar\Theta|\ll1$ and
$|\bar\mu_{\rm f}|\lesssim10^{-4}\ll1$, consistent with observational
bounds~\cite{Fixsen:1996FIRAS, BF2022}. The amplitudes in
Figs.~\ref{fig:two-gauge} and \ref{fig:spectral-illustration} are normalized by
$\bar\mu_{\rm f}$ and are independent of the overall injection amplitude
at this order.
In Fig.~\ref{fig:two-gauge}, we illustrate the calculation for
$\lambda=0.7$, $E_{\mathrm p}/T_{\mathrm p}=2$, $\sigma/\tau_{\mathrm p}=0.08$, and show the result in units of $\bar\mu_{\rm f}$.
The top panel shows the growth of $\bar\mu_{\rm mom}$ through the injection.   
The middle panel shows the linear fluctuation for a superhorizon mode in units of the initial comoving curvature $\mathcal R$.  In synchronous gauge $\delta\mu_{\rm mom}$ vanishes in the long-wavelength limit.   In Newtonian gauge, during the injection there is a finite moment fluctuation but once the 
injection completes, the fluctuation vanishes, consistent with the general considerations of Eq.~(\ref{eq:two-gauge-fh-null}).   

The bottom panel resolves the density, clock, and lapse sources of
Eq.~\eqref{eq:pedagogical-newtonian-source}. Each contribution is
integrated through Eq.~\eqref{eq:pulse-photon-perturbations} with zero
initial photon perturbations, using the same consistently depleted source
abundance and proper-clock history, and is measured relative
to the total photon energy and number at that time, including the
injected photons.

The fourth curve retains the initial adiabatic temperature
$\Theta_i=-\Psi/2=\mathcal R/3$. Its freely redshifting energy and
number perturbations initially have $\delta\mu_{\rm mom}=0$, but their number and energy balance relative to the evolving background spectrum changes during injection. Writing
$\epsilon_\rho=\rho_\gamma/\rho_z-1$ and
$\epsilon_N=N_\gamma/N_z-1$, their contribution is
\begin{equation}
 \delta\mu_{{\rm mom},{\rm init}}(t)
 =\frac{4\Theta_i}{\mathcal E[M]}
 \left[\frac1{1+\epsilon_\rho(t)}-\frac1{1+\epsilon_N(t)}\right].
 \label{eq:initial-temperature-contribution}
\end{equation}
The four contributions satisfy
\begin{equation}
 \delta\mu_{{\rm mom},{\rm init}}
 +\sum_i\delta\mu_{{\rm mom},i}
 =\delta\mu_{\rm mom}^{\rm N}
 =\dot{\bar\mu}_{\rm mom}\delta\tau_X^{\rm N}.
 \label{eq:moment-contribution-sum}
\end{equation}
Thus the Newtonian moment fluctuation during injection is a temporal displacement along the background $\bar\mu_{\rm mom}(t)$.   We can also verify the gauge invariance of 
synchronous gauge: when constructed from Newtonian gauge the time shift exactly zeros this contribution.    Moreover after injection $\bar\mu_{\rm mom}=$\, const.\ and the four curves sum to zero. Omitting a transfer term
spoils this cancellation, even if source depletion and photon gain are
still balanced.  Our gauge-covariant methods provide an explicit check and guard against such omissions or misspecifications of the sources.

The Newtonian FH calculation reproduces these moment results while following
Compton redistribution explicitly. We implement the photon-injection source
of Sec.~\ref{sec:source-collision} in \texttt{CosmoTherm}, including the
source density, proper-clock and velocity perturbations and consistent
source depletion. To match the radiation-dominated, rapid-Comptonization
limit, the comparison in Figs.~\ref{fig:two-gauge} and
\ref{fig:spectral-illustration} uses $z_{\mathrm p}=10^7$ and
$k=10^{-4}\,{\rm Mpc}^{-1}$, with neutrinos and DC/BR photon production
switched off. The synchronous result is
reconstructed from the Newtonian evolution via gauge transformation. The open circles show selected
FH samples and are in excellent agreement with the moment calculation.
These comparisons test the implemented injection source together with
Compton redistribution; the photon-production correction of
App.~\ref{app:dcbr-reduction} is inactive with DC/BR switched off.

Both the moment calculation and the spectral evolution recover a homogeneous
chemical potential on the source-comoving synchronous slicing. The transient
Newtonian moment fluctuation represents a displacement along that common history.
A physical modulation of the injection instead changes the local history and
can leave a nonzero chemical-potential fluctuation. At fixed source age, let $\Gamma\to\Gamma(1+m_L{\mathbb Y})$, with
$m_L$ constant along the source trajectory. Hold the initial source abundance,
pulse time and width, and emitted spectrum fixed.
For the pulse of Eq.~\eqref{eq:pulse-profile}, this changes its normalization
as $\lambda\to\lambda(1+m_L{\mathbb Y})$. The final source-comoving chemical-potential fluctuation is
\begin{equation}
 \delta\mu_{\rm mom,f}^{\rm S}
 =m_L\lambda\frac{\partial\bar\mu_{\rm mom,f}}{\partial\lambda}.
 \label{eq:modulated-mu-control}
\end{equation}
This parameter variation changes the photon-number and energy additions,
whereas the adiabatic Newtonian fluctuation above displaces their common clock
history. The modulated moment fluctuation agrees with independent homogeneous
calculations in which $\lambda$ is varied, including the consequent change
in source depletion. A separate superhorizon calculation using
\texttt{CosmoTherm}'s spectral evolution verifies this parameter-response
relation for the full photon spectrum, including finite Compton redistribution.
Here the initial source abundance, background expansion, and other physical
parameters are held fixed. These separate-universe comparisons test the
source prescription for both a displaced adiabatic history and a physical
change of injection strength.

\subsection{Stationary anisotropy in the spectral basis}
\label{sec:stationary-response}

The adiabatic injection example also clarifies the difference between stationary anisotropy with a single angular amplitude and its representation by multiple components in the spectral basis.  
After injection and Compton equilibration are complete, the anisotropy spectrum takes the form of a frequency shifted background with no intrinsic variation in that spectrum across the sky.  This common frequency shift is described by one angular amplitude even
though it projects onto several spectral-basis functions.

First recall that the spectral basis is defined against a blackbody with a uniform reference temperature $T_z\propto a^{-1}$ which defines the
fixed frequency coordinate $x=q/T_z$.
Through the injection, the actual background photon spectrum $n^{(0)}$ changes relative to this reference, including
its mean temperature. After injection and Compton equilibration, the background is represented in the spectral basis by
\begin{equation}
 n^{(0)}-n_{\rm bb}=\bar\Theta G+\bar\mu M,
 \label{eq:mean-equilibrium-fh}
\end{equation}
to first order in the change due to the injection. Since $M$ conserves photon number,
the background additions $\Delta N_\gamma\equiv N_\gamma-N_z$ and
$\Delta\rho_\gamma\equiv\rho_\gamma-\rho_z$ determine the two coefficients:
\begin{equation}
 \frac{\Delta N_\gamma}{N_z}=3\bar\Theta,\qquad
 \frac{\Delta\rho_\gamma}{\rho_z}
       =4\bar\Theta+\mathcal E[M] {\bar\mu}.
 \label{eq:mean-equilibrium-fh-moments}
\end{equation}
The coefficient $\bar\Theta$ gives the photon-number temperature shift.
The Bose--Einstein temperature $\bar T_{\rm BE}$ defines the spectrum
\begin{equation}
 n^{(0)}=\left[\exp\left(\frac{h\nu}
              {k_{\rm B}\bar T_{\rm BE}}+\bar\mu\right)-1\right]^{-1},
 \label{eq:constant-mu-temperature}
\end{equation}
where for clarity we have restored $h$ and $k_{\rm B}$.
As a fluctuation to the reference temperature it becomes
\begin{equation}
 \frac{\bar T_{\rm BE}}{T_z}-1=\bar\Theta+\frac{\bar\mu}{\beta_M},
 \label{eq:mean-be-temperature-shift}
\end{equation}
as follows by expanding the Bose--Einstein spectrum with
$\theta_{\rm BE}\equiv\bar T_{\rm BE}/T_z-1$, thus yielding:
\begin{equation}
 n^{(0)}-n_{\rm bb}
 =\theta_{\rm BE}G-\bar\mu G/x
 =\left(\theta_{\rm BE}-\frac{\bar\mu}{\beta_M}\right)G+\bar\mu M.
 \label{eq:app-be-fh-temperature}
\end{equation}
The compensating temperature term in $M$ therefore makes $\bar\Theta$
the photon-number temperature shift rather than $\theta_{\rm BE}$.
Substitution of Eq.~\eqref{eq:mean-equilibrium-fh-moments} into
Eq.~\eqref{eq:mu-moment-diagnostic} also gives
$\bar\mu_{\rm mom}=\bar\mu$ to first order in the injection.
Both the injected temperature change and
chemical potential enter the background spectrum used to construct its
adiabatic frequency shift.

This change in the background enters into the spectral anisotropy basis description as well.   In both the synchronous and Newtonian calculations, the spectral anisotropy is stationary after injection and Compton equilibration are complete.
Thus it can be characterized as a generalization of the usual temperature anisotropy $\Theta_{\rm ad}$ 
for adiabatic fluctuations to the spectrum of a frequency-shifted Bose-Einstein distribution instead of that of the reference blackbody
\begin{equation}
 n_{\rm N}^{(1)}=\Theta_{\rm ad}\Ox n^{(0)},\qquad
 \Ox n^{(0)}=G+\bar\Theta(3G+Y)+\bar\mu\Ox M,
 \label{eq:stationary-mu-response}
\end{equation}
to first order in the mean distortion, using $\Ox G=3G+Y$ from
Eq.~\eqref{eq:fh-shapes}. 
The consequence is that in the spectral basis, a single anisotropy map of $\Theta_{\rm ad}$ is represented as multiple spectral basis maps: $G$, $Y$ and the frequency shifted $M$.  For the latter, we follow Ref.~\cite{Chluba:2022efq}, Sec.~2.3.1 in approximating its projection onto the spectral basis with the lowest-order $Y_r$ representation 
\begin{equation}
 \Ox M\simeq c_Y Y+c_M M,\qquad
 c_M=1.9069,\qquad c_Y=\frac{4-c_M}{4}\mathcal E[M]\simeq0.3736,
 \label{eq:illustrative-mu-boost}
\end{equation}
which preserves the number and energy moments of $\Ox M$.  The approximate spectral basis decomposition of the injection example then carries $G$, $Y$, and $M$ anisotropy maps, all of which share a common fluctuating amplitude 
$\Theta_{\rm ad}$.

\begin{figure}[htbp]
\centering
\includegraphics[width=0.7\textwidth]{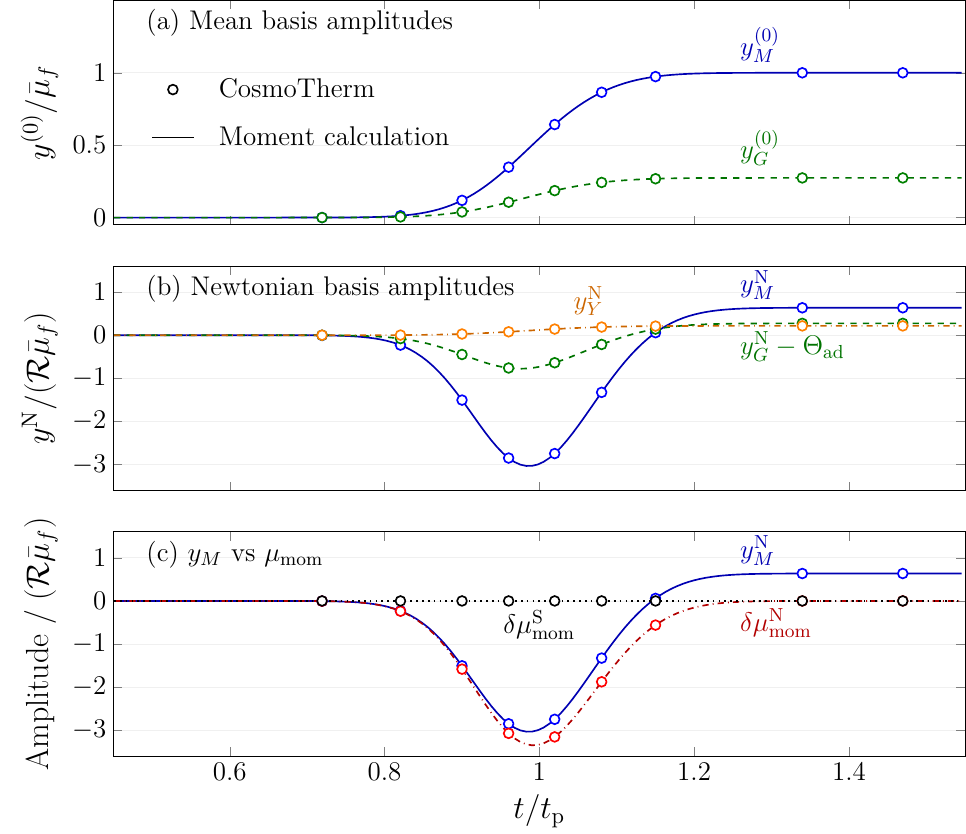}
\caption{Spectral basis amplitudes for the photon injection of Fig.~\ref{fig:two-gauge} (see caption for details),
with curves assuming instantaneous Compton equilibrium and the basis approximation to frequency shifts
in Eq.~\eqref{eq:illustrative-mu-boost}. Open circles show the corresponding lowest-order \texttt{CosmoTherm} frequency hierarchy, with the same source as Fig.~\ref{fig:two-gauge}. Panel (a) shows the mean
$G$ and $M$ coefficients, $y_G^{(0)}=\bar\Theta$ and
$y_M^{(0)}=\bar\mu$, relative to the uniformly redshifting reference $T_z$.
Panel (b) shows the $G$, $Y$, and $M$ Newtonian monopole amplitudes; the initial
adiabatic fluctuation  $\Theta_{\rm ad}=\mathcal R/3$ is subtracted
from $y_G^{\rm N}$ to expose its injection-induced change.
Panel (c) compares the Newtonian $y_M^{\rm N}$ with
$\delta\mu_{\rm mom}^{\rm N}$ and
$\delta\mu_{\rm mom}^{\rm S}$ as in Fig.~\ref{fig:two-gauge}.  
After injection and Compton equilibration, $y_M^{\rm N}$ remains nonzero
while both chemical-potential moment perturbations vanish. The stationary
$G$ and $M$ amplitudes are fully correlated parts of a frequency-shifted
spectrum with no {\it local} chemical-potential fluctuation.
}
\label{fig:spectral-illustration}
\end{figure}

We illustrate the spectral basis decomposition through the injection in Fig.~\ref{fig:spectral-illustration} for the same case as shown in Fig.~\ref{fig:two-gauge}.  Here we assume
 instantaneous Compton equilibration during the pulse, as if $\mu=\mu_{\rm mom}$.  Nonetheless during the injection, the Newtonian monopole still includes its clock displacement as well as its frequency shift. In the radiation-era superhorizon limit, the local Sachs--Wolfe temperature is $\Theta_{\rm ad}=-\Psi/2=\mathcal R/3$.  The resulting
$G$,  $M$ and $Y$ entries of the Newtonian monopole vector
$\bm y_0^{\rm N}$
\begin{align}
 y_G^{\rm N}&=\delta\tau_X^{\rm N}\dot{\bar\Theta}
                 +\Theta_{\rm ad}(1+3\bar\Theta),\nonumber\\
 y_M^{\rm N}&=\delta\tau_X^{\rm N}\dot{\bar\mu}
                 +c_M\Theta_{\rm ad}\bar\mu, \nonumber\\
  y_Y^{\rm N}&=\Theta_{\rm ad}(\bar\Theta+c_Y\bar\mu).
 \label{eq:illustrative-g-m-y-amplitude}
\end{align}
are shown in the middle panel, with the corresponding \texttt{CosmoTherm}
amplitudes overlaid as circles. Here $\bar\Theta = \mathcal O(\bar\mu_f) \ll 1$ so that 
$y_G^{\rm N} -\Theta_{\rm ad} = 3 \Theta_{\rm ad} \bar \Theta$ after $\dot{\bar\Theta}$ vanishes after injection and displays the net impact from the injection \citep[see also][]{Evangelista:2026darkphoton}. 

Together these coefficients have moments
$\delta N_\gamma/N_z=3y_G^{\rm N}$ and
$\delta\rho_\gamma/\rho_z=4y_G^{\rm N}+4y_Y^{\rm N}
+\mathcal E[M]y_M^{\rm N}$. The moment diagnostic of
Eq.~\eqref{eq:mu-moment-diagnostic} therefore becomes
\begin{equation}
 \delta\mu_{\rm mom}^{\rm N}=\frac{1}{\mathcal E[M]}
 \left[\frac{4y_G^{\rm N}+4y_Y^{\rm N}+\mathcal E[M]y_M^{\rm N}}
                  {1+\epsilon_\rho}
       -\frac{4y_G^{\rm N}}{1+\epsilon_N}\right],
 \label{eq:illustrative-basis-moment}
\end{equation}
where $\epsilon_\rho=\rho_\gamma/\rho_z-1$ and
$\epsilon_N=N_\gamma/N_z-1$. The boost constraint
$4c_Y+c_M\mathcal E[M]=4\mathcal E[M]$ ensures that after injection
and Compton equilibration they reduce to
\begin{equation}
 \frac{\delta\rho_\gamma^{\rm N}}{\rho_\gamma}=4\Theta_{\rm ad},\qquad
 \frac{\delta N_\gamma^{\rm N}}{N_\gamma}=3\Theta_{\rm ad},\qquad
 \delta\mu_{\rm mom}^{\rm N}=0.
 \label{eq:stationary-basis-moment-null}
\end{equation}
as in our direct analysis above and the bottom panel. In the spectral representation,
the $Y$ coefficient and the injected background denominators are both
needed for this cancellation. The nonzero
$y_M^{\rm N}=c_M\Theta_{\rm ad}\bar\mu$ describes part of the shifted
mean spectrum, while the chemical potential itself is uniform without new spectro-spatial information \citep{Evangelista:2026darkphoton}.
Both $y_G^{\rm N}$ and $y_M^{\rm N}$ reach finite values after injection
and Compton equilibration. Their common frequency-shift origin predicts
correlated angular maps after propagation. 

More generally, let $\Theta_\ell^{\rm N}$ denote the ordinary blackbody
temperature multipoles evolved with the same metric and scattering inputs
and adiabatic initial conditions, omitting injection and distortion
corrections. We define departures from the associated stationary
frequency-shift response by
\begin{equation}
 \bm r_\ell^{\rm N}
 \equiv\bm y_\ell^{\rm N}-\Theta_\ell^{\rm N}\bm b^{(0)},
 \label{eq:stationary-residual}
\end{equation}
The reference amplitude $\Theta_\ell^{\rm N}$ differs from the full $G$
coefficient of the spectral vector. The stationary reference solution has
$\bm r_\ell^{\rm N}=0$ even though its individual spectral-basis
components are nonzero. Away from this solution, the residual can include
a change in the common temperature amplitude as well as a change in
spectral shape.
A nonzero residual alone therefore does not establish an intrinsic
spectral distortion. These variables provide a useful starting point for a
tight-coupling expansion that separates the common frequency-shift response
from additional spectral structure. During evolution of the mean, the
clock displacement must also be accounted for, as described below.

Evolution of the mean does not by itself imply an intrinsic departure.
In the long-wavelength adiabatic limit,
$\Theta_0^{\rm N}=-H\delta\tau_X^{\rm N}$. During injection or subsequent
thermalization, Eq.~\eqref{eq:two-gauge-fh-null} therefore gives
\begin{equation}
 \bm r_0^{\rm N}
 =\delta\tau_X^{\rm N}\dot{\bm y}^{(0)},
 \label{eq:adiabatic-residual-clock}
\end{equation}
This residual describes a displacement along the common local spectral
history; subtracting the
clock displacement as well leaves no intrinsic fluctuation at equal
source age. Once the mean ceases evolving, the clock term vanishes and
the anisotropy takes the stationary frequency-shift form. The DC-on
calculation of Sec.~\ref{sec:cosmological-propagation} illustrates this
distinction: photon production changes the mean after injection while
preserving the common adiabatic history in the long-wavelength limit,
leaving nearly stationary final anisotropy.

\subsection{Cosmological propagation of stationary anisotropy}
\label{sec:cosmological-propagation}

We next follow the same pulse family through injection and cosmological
propagation with the more complete spectral evolution of \texttt{CosmoTherm}.
For this angular calculation we change the normalization of the uniformly
redshifting reference $T_z$. The earlier moment example sets it by the
initial Planck spectrum; here its blackbody energy density $\rho_z$
matches the final mean photon energy density. Define $\epsilon_{\rm inj}$
as the fractional increase in $a^4\rho_\gamma$ from injection, relative
to its pre-injection value. The initial Planck temperature then obeys
$T_{\gamma,{\rm i}}/T_{z,{\rm i}}=1-\epsilon_T$, with
$\epsilon_T=\epsilon_{\rm inj}/4$ to first order in the injection.
Thus the initial spectrum has $\bm y_{\rm i}^{(0)}=-\epsilon_T\bm e_G$.
Its initial adiabatic spectral multipoles include the matching correction
$-\epsilon_T\Theta_{\ell,{\rm i}}M_{\rm B}\bm e_G$,
where $\Theta_{\ell,{\rm i}}$ is the ordinary temperature multipole \citep{Chluba:2026nmu, Evangelista:2026darkphoton}.
We express the mean, anisotropy and orbital dipole in this same reference,
retaining first order in the injected mean and its product with the linear
perturbation.

Figure~\ref{fig:cosmotherm-stationary}(a) shows the spectral anisotropy
power spectra in temperature-equivalent $\mu{\rm K}^2$ units in a flat
Planck-2018-based cosmology~\cite{Planck:2018params}
with matter, recombination and massless neutrinos.\footnote{We use
$h=0.6732$, $\omega_b=0.022383$, $\omega_c=0.12011$,
$T_0=2.7255\,{\rm K}$, $Y_p=0.2454$, $N_{\rm eff}=3.046$,
$\ln(10^{10}A_s)=3.0448$ and $n_s=0.96605$ at
$k_0=0.05\,{\rm Mpc}^{-1}$. The calculation follows unpolarized, unlensed
transport through recombination, with reionization and the trace source's
backreaction on the leading cosmology omitted.}
Injection at $z_{\mathrm p}=10^6$ leaves the relevant modes well outside the horizon
while the mean approaches a chemical-potential spectrum. 
At this injection redshift, DC suppresses the surviving chemical
potential relative to the same pulse with photon production switched off:
$\bar\mu_{\rm f}^{\rm DC}\simeq
\mathcal J_{\rm bb}(z_{\rm p})\bar\mu_{\rm f}^{\rm noDC}$.
The no-DC value is determined by both the injected energy and photon number
through Eq.~\eqref{eq:mu-moment-diagnostic}. The approximate visibility
$\mathcal J_{\rm bb}(z_{\rm p})\approx
\exp[-(z_{\rm p}/z_{\rm dc})^{5/2}]\simeq0.84$, with
$z_{\rm dc}\approx2\times10^6$, agrees with the calculated ratio
$\bar\mu_{\rm f}^{\rm DC}/\bar\mu_{\rm f}^{\rm noDC}=0.8382$.
The corresponding mean chemical potentials are $2.31\times10^{-5}$ and
$2.76\times10^{-5}$, respectively.
We use the
emitted spectrum of Sec.~\ref{sec:source-collision} with
$E_{\mathrm p}/T_{\mathrm p}\simeq2$. The $N=15$ hierarchy retains $G$, $Y_0,\ldots,Y_{15}$ and $M$, with DC photon
production on or off. BR is omitted from both the mean and perturbed
equations in both cases. The on-case uses the local DC-only closure and
corrected source of App.~\ref{app:dcbr-reduction}; the optional BR
extension is given in App.~\ref{app:br-extension} and we have verified that it gives qualitatively the same results with  a slightly smaller $\bar\mu_{\rm f}$. For this illustration,
we switch off effective DC thermalization on a surface of constant source
proper age whose background redshift is $z=2\times10^4$, while continuing
Compton redistribution and cosmological transport.

To test stationarity, we compare the directly calculated $GM$ and $GY_0$
cross power spectra with their stationary predictions obtained by rescaling $GG$:
\begin{equation}
 C_\ell^{GX}\simeq\frac{b_X^{(0)}}{b_G^{(0)}}C_\ell^{GG},
 \qquad X=M,Y_0,
 \label{eq:stationary-angular-cross}
\end{equation}
where the spectral bias factors $b_X^{(0)}/b_G^{(0)}$ are fixed by the
frequency shift of the final mean spectrum through
Eq.~\eqref{eq:fh-background-response}.
These predictions are shown as curves in Fig.~\ref{fig:cosmotherm-stationary}(a),
while circles show the directly evolved cross-spectra. Their agreement across the acoustic
peaks and damping tail demonstrates nearly stationary anisotropy with
DC thermalization as well as in the number-conserving idealization of
Sec.~\ref{sec:source-gauges}. In the DC-on case, the mean continues
relaxing after injection, yet the final anisotropy follows the frequency
shift of the resulting mean. The mean need not remain static throughout
the history for this final relation to hold.
The full band spectra use all eighteen components: the selected $M$ and
$Y_0$ amplitudes describe parts of one frequency-shifted spectrum. The small
departures that become increasingly visible toward high $\ell$ are consistent
with finite-gradient corrections to the superhorizon stationary solution.
The leading injection contribution is suppressed by $(k\eta_{\mathrm p})^2$.
With DC active, local temperature and density fluctuations can also
modulate relaxation during the subsequent evolution of the mean. This
contribution is weighted over the relaxation history and is suppressed
by $(k\eta)^2$ while the mode remains superhorizon; after horizon entry,
the injection-time suppression alone does not bound it.
The largest $GM$ departure through $\ell=2000$ is about $0.03\%$
in this effective model. Selected source-grid, tolerance, angular and
radial refinements change the full plotted powers by less than $0.001\%$.

The effect of correcting the perturbative source is tested separately
in App.~\ref{app:dcbr-adiabatic}, with the same pulse parameters and a
DC-only mean held fixed between the original and corrected prescriptions.
BR is omitted from both the mean and perturbed equations in that test.
The original source generates a spurious local chemical-potential
fluctuation $\delta\mu_{\rm intr}\simeq0.08\,\bar\mu_{\rm f}\mathcal R$
after prompt Compton equilibration. Here $\bar\mu_{\rm f}$ is the
surviving mean chemical potential and $\mathcal R$ is the primordial
curvature fluctuation. The coefficient remains finite as $k \rightarrow 0$;
the corrected source passes the adiabatic separate-universe test and
leaves only gradient-suppressed departures.

The original \texttt{CosmoTherm} FH photon-production source incorrectly
assigns different chemical-potential relaxation to patches
that should share the same proper-time history. The resulting intrinsic
$\mu$ fluctuation survives after photon production freezes out and can
source a small intrinsic $\mu T$ correlation under cosmological
transport from a chemical-potential mode that is not gradient suppressed.
This mode is absent in the adiabatic separate-universe limit.
This differs from the stationary distortion-basis correlations produced
by a frequency shift of a common mean.  In particular we shall see that the dipole calibration absorbs that
common response into the temperature map, but does not erase an intrinsic
chemical-potential fluctuation. Figure~\ref{fig:cosmotherm-stationary}
compares photon production on and off with the corrected source; the
original-versus-corrected test instead fixes the mean relaxation.

\FloatBarrier
\subsection{Dipole calibration of stationary anisotropy}
\label{sec:calibration}

For an observer, stationary spectral anisotropy means that every sky pixel
contains a frequency-shifted copy of the same mean photon spectrum.
The fractional shift, $\Theta_{\rm eff}(\hat{\boldsymbol n})$, varies
across the sky, while intrinsic distortion parameters such as the chemical
potential remain common across patches. All frequency channels therefore trace the same
angular pattern, with frequency-dependent amplitudes fixed by the shift
response of the mean spectrum. The orbital dipole measures this same
spectral response. Calibrating each channel to that dipole with a blackbody
template therefore absorbs the common distortion response into the fitted gains,
leaving a single effective temperature map.

\begin{figure}[t]
\centering
\includegraphics[width=0.7\textwidth]{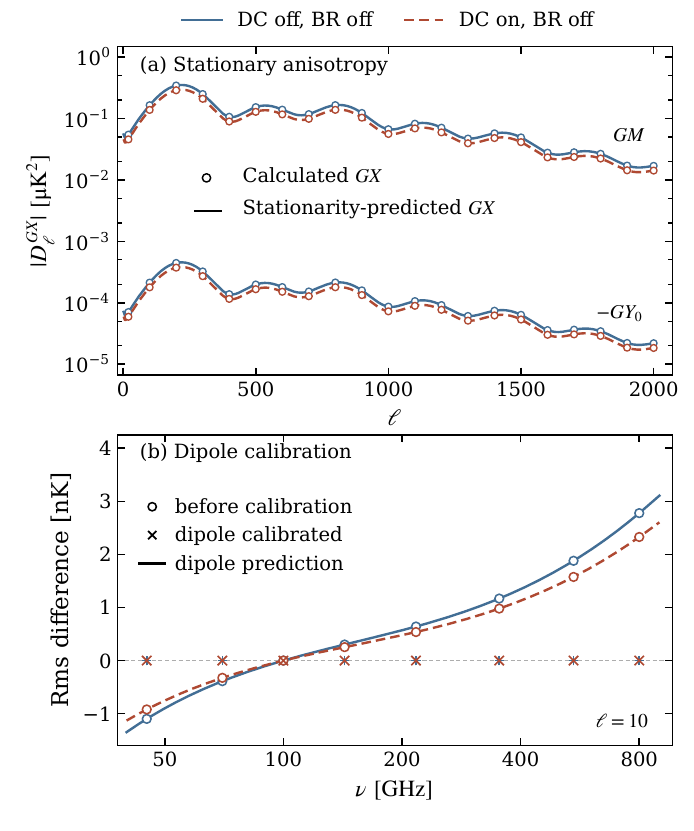}
\caption{Top: nearly stationary anisotropy in temperature-equivalent $\mu{\rm K}^2$ units, $D_\ell^{GX}=T_0^2\ell(\ell+1)C_\ell^{GX}/(2\pi)$ with $T_0=2.7255\,{\rm K}$, from cosmologically
evolved photon injection with \texttt{CosmoTherm} in an $N=15$ basis.
DC photon production is off (blue solid) or on (red dashed); BR is omitted
from both the mean and perturbed equations in both cases. The on-case
uses the corrected effective source of App.~\ref{app:dcbr-reduction}.
The optional BR extension is given in App.~\ref{app:br-extension}.
DC continues to evolve the mean after injection. The calculated $GM$ and $-GY_0$ cross-powers (circles) agree
with the prediction of Eq.~\eqref{eq:stationary-angular-cross} (curves)
from the frequency shift of the final mean spectrum and $GG$. Thus the
final anisotropy is nearly stationary even when the mean evolves during
thermalization.
 Here $GY_0$ is negative.
 The pulse is centered at $z_{\mathrm p}=10^6$ and has
$\lambda=0.7$, $\sigma/\tau_{\mathrm p}=0.08$ and injected energy fraction
$5.0\times10^{-5}$. Bottom: open circles show the
band rms before dipole calibration; lines predict it from the physical dipole of
the mean spectrum, and crosses show the result after that calibration.
The difference is relative to $100$ GHz; zero means a common rms,
$\sqrt{D_{10}^{TT}}=29.8\,\mu{\rm K}$, rather than vanishing anisotropy.
The bands are Gaussian with width $0.08\nu_a$. The small residual below the
plot resolution is mainly due to the finite-basis representation of the
dipole derivative.}
\label{fig:cosmotherm-stationary}
\end{figure}

Here $\Theta_{\rm eff}$ is the common fractional frequency-shift field
after cosmological propagation, including its gravitational and scattering
contributions; it replaces the local radiation-era amplitude
$\Theta_{\rm ad}=\mathcal R/3$. The correlated components in
Fig.~\ref{fig:cosmotherm-stationary}(a) represent this common angular field
in the spectral basis.

CMB anisotropy experiments calibrate their frequency channels using the orbital dipole of a satellite or the primary temperature anisotropy of such a calibration measured independently from the ground. For the orbital dipole, the calibrator is a Doppler shift of the background \cite{Danese1981dipole,Balashev:2015lla} and therefore shares the stationary anisotropy spectrum \cite{Evangelista:2026darkphoton}. Applying the calibration argument of Ref.~\cite{Zegeye:2023pbl} (see their  Eqs.~(8)--(10) for the Bose-Einstein case) to this common spectrum recovers this same amplitude as a single effective temperature map
$\Theta_{\rm eff}$. This logic applies to any background spectrum whose frequency-shift
response is shared by the sky anisotropy and orbital dipole at observation;
it does not require a static background throughout formation and
thermalization. The assumption that the orbital dipole is the Doppler boost of a blackbody means that the obtained map will be called a temperature map, rather than a spectrally distorted map where gain factors absorb the distortion. See Refs.~\cite{Planck:2018bsf,Planck:2018lkk, Trombetti2021} for foreground and instrumental complications in practical dipole calibration, and Ref.~\cite{Remazeilles:2018kqd} for calibration uncertainties in distortion reconstruction.

To be more explicit, the background specific intensity 
is $\bar I_\nu=2h\nu^3 n^{(0)}/c^2$ and the boost or shifted form is
\begin{equation}
 D_\nu=(3-\partial_{\ln\nu})\bar I_\nu
       =\frac{2h\nu^3}{c^2}\Ox n^{(0)}.
\end{equation}       
The orbital dipole is $\delta I_\nu^{\rm orb}=(\hat{\bf n} \cdot {\bf v}_{\rm orb}/c)D_\nu$,
where ${\bf v}_{\rm orb}$ is the orbital velocity and $\hat{\bf n}$ is the line of sight.  The stationary anisotropy spectrum is $\delta I_\nu^{\rm stat}=\Theta_{\rm eff} D_\nu$.

Now assume that the instrumental channels have their gains calibrated to the orbital dipole by instead assuming the shift of a blackbody  whose specific intensity is $B_\nu(T_{\rm ref})$
\begin{equation}
 G_\nu=T_{\rm ref}\left.\frac{\partial B_\nu(T)}{\partial T}
                                     \right|_{T_{\rm ref}},
\end{equation}
where $T_{\rm ref}$ is the adopted reference blackbody temperature,
constrained observationally by FIRAS. In our illustration we use
$T_{\rm ref}=2.7255\,{\rm K}$ for the energy-defined present reference
of Sec.~\ref{sec:cosmological-propagation}. We express the mean spectrum,
anisotropy coefficients and orbital-dipole template relative to this same
reference blackbody, preserving the physical intensity and dipole under
the reference change.
For a frequency channel labeled $a$ with fixed bandpass $R_a(\nu)$, let
$D_a=\int\dd\nu R_aD_\nu$ and $G_a=\int\dd\nu R_aG_\nu$.
Let $g_a$ be the true conversion from band-integrated intensity to
detector signal and $\widehat g_a$ the gain fitted by assigning the
blackbody template $G_a$ to the orbital dipole
\begin{equation}
 \widehat g_a=g_a\frac{D_a}{G_a}.
\end{equation} 
Applying this gain
to the stationary anisotropy gives the calibrated intensity as 
\begin{equation}
 \frac{g_a\Theta_{\rm eff} D_a}{\widehat g_a}=\Theta_{\rm eff} G_a ,
 \label{eq:dipole-calibration-cancellation}
\end{equation}
which is then extracted as an effective ``temperature'' map of the shifted blackbody.   
In our example, and indeed for any stationary anisotropy,
the stationary $M$ and $Y$ components of the spectral basis cannot be separated from the $G$ component if the bands are calibrated only by the dipole.   Separation requires an external calibration of the relative frequency responses and in that case, the background spectral distortion can be measured instead from a stationary spectral anisotropy \cite{Zegeye:2023pbl}.
For modes near or inside the horizon during injection, spectral
anisotropies can depart from the common dipole response and consequently retain
independent information after calibration~\citep{kite_spectro-spatial_2023-III}.

Figure~\ref{fig:cosmotherm-stationary}(b) demonstrates this cancellation for
the cosmologically evolved spectra. We reconstruct the intensity in ideal
frequency bands and compare their $\ell=10$ rms with the $100$-GHz band.
The lines predict the frequency dependence directly from the analytic
derivative $\Ox n^{(0)}$ of the final $G/M$ mean, independently of the
finite-basis boost matrix. Calibration by this physical dipole brings the
calculated band rms to a common temperature amplitude, with small residual
differences dominated by the finite spectral basis. Together with the
stationary cross-spectra in Fig.~\ref{fig:cosmotherm-stationary}(a), this
demonstrates that dipole calibration recovers, to good approximation, a
single effective temperature map for this example.

\section{Discussion}
\label{sec:outlook}

We have formulated the general gauge-covariant FH
for an arbitrary intensity
collision term and photon frequencies measured by observers
normal to the time slices. A gauge transformation that changes the slicing spatially modulates an evolving
mean spectrum in time and changes the local frequency and conformal
collision rate. Combining these transformations with those of the spatial threading preserves the
form of the hierarchy and supplies gauge-covariant expressions.
It also supplies gauge-covariant equations for 
Newtonian variables for their gauge-invariant evaluation.  

These gauge-covariant methods are especially useful when collision sources take their simplest form in a different gauge.   We construct an explicit example where an isotropic photon injection is specified by its own clock,
emission frame, and rest-frame spectrum.  For adiabatic fluctuations and a cold dark matter source, this specifies a synchronous gauge description where a superhorizon 
injection in the $\mu$-epoch leaves the usual adiabatic anisotropy with stationary frequency dependence.  
We show how the Newtonian reconstruction obtains the same through canceling initial adiabatic perturbations, source density perturbations, shifted clocks and lapse-distorted rates.  This stationary spectral anisotropy becomes inseparable from an effective temperature
anisotropy when channels are calibrated to the orbital dipole with a
blackbody template, since the dipole shares the stationary spectral form and the mismatch in the spectral template is absorbed in the gain factors.

The moment example isolates the injection physics through prompt,
number- and energy-conserving Compton equilibration (see Fig.~\ref{fig:two-gauge}). Its agreement with the
Newtonian \texttt{CosmoTherm} calculation demonstrates the source construction
and the cancellation of the transient Newtonian moment fluctuation. Continuous $N=15$ evolution
then follows injection, thermalization and cosmological transport with
matter and massless neutrinos. DC changes the common mean spectrum
after injection while preserving the common adiabatic history in the
long-wavelength limit. The resulting final spectral cross-correlations follow the
mean-spectrum prediction across the acoustic peaks and damping tail:
evolution of the mean need not generate an intrinsic spectral fluctuation
or spoil the final stationary frequency-shift form. Calibration by the physical dipole of that mean returns a common
angular temperature map, with the small remaining rms difference
dominated by the finite-basis representation of its frequency derivative (Fig.~\ref{fig:cosmotherm-stationary}).
These conclusions apply to the retained isotropic rest-frame thermalization
model. Intrinsic anisotropic Compton redistribution supplies additional
spectral sources, which can act even when the mean spectrum is Planckian.

For the thermalization collision terms, our gauge-covariant techniques provide a simple prescription for turning established Newtonian representations into those of a general gauge.   We construct the gauge-covariant forms of the unpolarized Thomson, Kompaneets, double-Compton and bremsstrahlung terms of the \texttt{CosmoTherm} treatment~\cite{Chluba:2026nmu} in App.~\ref{app:thermalization}.
The independent local construction in a general gauge in
Secs.~\ref{app:Thomson}, \ref{app:thermal-checks} and
\ref{app:dcbr-reduction} tests that correspondence.
The retained Thomson terms and the improved Compton terms, including
the stimulated and kinematic terms of Ref.~\cite{Chluba:2026nmu}, agree
with the independent local construction under the stated approximations.
Photon production instead requires a correction to the DC emissivity's
temperature dependence and its effective perturbative source:
covariance of the frequency-dependent emission kernel and consistency of
its perturbed $G/M$ reduction are separate checks. The gauge-covariant reconstruction and independent local construction together identify the
spectral-shift and local-rate variations which previously were not included consistently.
Including these variations corrects the effective photon-production
source for the specified mean thermalization model. The corrected source
agrees with the local construction on arbitrary slicing and preserves
the common adiabatic thermalization history in the long-wavelength limit.

The DC-only comparison isolates the physical consequence of the omitted
responses at a fixed mean relaxation. The original source creates a
false inhomogeneity in chemical-potential relaxation and leaves a local
chemical-potential fluctuation
$\delta\mu_{\rm intr}\simeq0.08\,\bar\mu_{\rm f}\mathcal R$ in the long-wavelength limit; this fluctuation survives after
spectral transients decay. Although small in amplitude, its intrinsic $\mu T$ correlation is
spurious in the adiabatic separate-universe limit and is distinct from the expected
stationary mean-spectrum response.
The corrected source removes that leading intrinsic mode while retaining
the evolution of the common mean and finite-gradient effects.
Our cosmological calculation uses this corrected prescription. The separate-universe comparisons test the common adiabatic history and
physical parameter responses of the specified local model. Its scalar
relaxation-rate consistency is further assessed by independent homogeneous
full-frequency evolution.   Consistency of the local model does not guarantee its accuracy: in this case, the  low-frequency reduction after correction is consistent but remains approximate.

Indeed, future work could improve the accuracy of  FH predictions.  In addition to more accurately modeling low-frequency emission,
extending the hierarchy to polarized Thomson transport and anisotropic
Compton redistribution would distinguish transport of the stationary
mean-spectrum response from additional spectral anisotropies generated by
scattering. Their consistent treatment requires the corresponding intensity
and polarization sources, electron-temperature response, and momentum
exchange.

Our techniques should also be useful for refining resonant photon--dark-photon
conversion calculations~\cite{Chluba2024DP,Evangelista:2026darkphoton}. Its local plasma-frame
description specifies both the conversion strength and the resonance
condition. Transforming this source into Newtonian gauge requires the
displacement of the resonance profile and the accompanying conversion of
its differential rate, in addition to amplitude and spectral responses.
The adiabatic separate-universe response provides a consistency test
of this source construction. Conversion after horizon entry can instead
generate a spectral departure from the common frequency shift of the mean.
Following that departure requires coupled photon and dark-sector transport,
thermalization, and metric evolution. 
A synchronous-gauge treatment may also be useful for extending these
calculations to conversions during recombination and reionization, or to
multiple conversions \citep[e.g.,][]{Cyr2024Axions}. We leave the choice
and implementation of such a treatment to future work.

Another example is injection from acoustic dissipation of short wavelength modes.  
 The FH treatment
of Ref.~\cite{Chluba:2026acoustic} follows the
distortion sourced by short acoustic modes and its subsequent transport.
For inflationary initial conditions obeying the single-field consistency
relation, a constant long adiabatic mode dilates the coordinate ruler
used to describe those short modes. The short modes must be consistently evolved in the presence of long wavelength modes to consistently preserve the single-field consistency relation at dissipation~\cite{Inomata:2023faq}.  The dissipated short-scale power 
must be compared with the same local ruler and at the same fluid clock.  This would generalize the 
$\mu T$ cancellation in the long mode established by Ref.~\cite{Cabass:2018singlefield} for the FH and allow a direct computation of the subleading residual \cite{Zegeye:2021yml}.
For scenarios with primordial non-Gaussianity, joint predictions for
mean spectral distortions and distortion--temperature correlations could also
assess their combined sensitivity to enhanced small-scale power and its
coupling to long-wavelength modes.
Applications to multi-field inflation models
\citep[see][for discussion in this context]{Dimastrogiovanni2016} and to
more general primordial bispectrum shapes are useful directions for
future work.

Dipole calibration removes the component sharing the dipole's spectral
response. When a perturbation's wavelength approaches or falls below the
horizon scale at injection or conversion, the source can generate spectral
departures that retain independent information even after
dipole calibration~\citep{kite_spectro-spatial_2023-III, Evangelista:2026darkphoton}. Anisotropic Compton
scattering supplies additional spectral sources. Disentangling these
contributions requires polarized transport and consistently specified local
source and collision models. We leave these and other considerations to future work.

Independent spectrometry with the sub-orbital CMB spectrometers
BISOU~\citep{BISOU}, COSMO~\citep{Masi2021} and TMS~\citep{Jose2020TMS}
could supply the mean spectrum needed to construct the physical dipole
template. With sufficiently accurate relative channel calibration, CMB imaging
could provide an independent measurement of the mean distortion through
its stationary anisotropy signal. In the longer term, {\it FOSSIL} \citep{FOSSIL2026}, a proposed CMB spectrometer that has now advanced to Phase 0 of the ESA M8 selection process, could deliver low-resolution absolutely calibrated CMB maps for studying primordial distortion anisotropies.

\section*{Acknowledgments}
We thank Tom Crawford and Austin Joyce for useful discussions.
WH was supported by U.S.\ Department of Energy contract
DE-SC0009924 and the Simons Foundation.
  DZ was partially supported by a
Grainger Distinguished Postdoctoral Fellowship and the South Pole Telescope program supported by the National Science Foundation (NSF) through award OPP-2332483. Work at the Fermi National Accelerator Laboratory (Fermilab), a U.S. Department of Energy, Office of Science, Office of High Energy Physics HEP User Facility, is managed by Fermi Forward Discovery Group, LLC, acting under Contract No. 89243024CSC000002.   JC acknowledges support from the UK Space Agency LiteBIRD UK grant
ST/Y005945/1.

The authors used OpenAI Codex (accessed October--November 2026) as an interactive assistant
for literature organization, derivation checks, symbolic and numerical
coding, manuscript and \LaTeX\ preparation, and figure production.
They directed and revised the outputs used in this work. All scientific
judgments, derivations, calculations, interpretation, and responsibility
for the manuscript remain with the authors.

\appendix

\section{Gauge transformations and covariance of spectral quantities}
\label{app:conventions}

In App.~\ref{app:slicing_normal}, we derive the gauge transformation properties of the photon distribution function and collision term, introduced in the main text as Eqs.~\eqref{eq:distribution-transform} and
\eqref{eq:source-transform}, by combining the scalar time shift with the
change of frequency measured by the normal observer.  We then construct the gauge-covariant forms of
the Newtonian distribution and source  in App.~\ref{app:gauge-covariance-invariance} and verify their gauge-invariant evaluation. Finally, App.~\ref{app:hierarchy} derives
the normal-observer frequency shift along photon geodesics and the
gauge-covariant transport equation.

\subsection{Time slicing and its normal observers}
\label{app:slicing_normal}

In the ADM 3+1 decomposition of Sec.~\ref{sec:metric}, the chosen time slices define their unit
normal and the observers who follow it from slice to slice. Changing
the slicing changes this observer frame, in addition to the passive
coordinate changes of clocks and rulers discussed in
Sec.~\ref{sec:gauge-reconstruction}.

The occupation number $n$ has the same value for a given spacetime event
and photon 4-momentum in either coordinate system.
Its arguments
$(\eta,\bm X,x,\hat{\bm\gamma})$, however, use a frequency and
direction defined in the normal frame, so their values change
with the slicing. Compare the same photon 4-momentum at the same
spacetime event in the two slicings.
Since the normal is fixed by $N_\mu=-\alpha\partial_\mu\eta$, the time shift
$\bar\eta=\eta+\mathcal{T}{\mathbb Y}$ tilts the new normal by
$\bar N_i-N_i=-a\partial_i(\mathcal{T}{\mathbb Y})$ when both are expressed
in the original coordinates, to first order. Using the normal projection
in Eq.~(\ref{eq:momentum-projection}), the resulting energy shift is
\begin{equation}
 \frac{\bar q-q}{q}
 =-\frac{p^\mu}{q}(\bar N_\mu-N_\mu)
 =\hat\gamma^i\partial_i(\mathcal{T}{\mathbb Y})
 =ik\chi \mathcal{T}{\mathbb Y}.
 \label{eq:normal-shift-energy}
\end{equation}
The homogeneous reference temperature is
evaluated at the new time coordinate of that event, so
$(\bar T_z-T_z)/T_z=(T_z'/T_z)\mathcal{T}{\mathbb Y}
=-\Hc \mathcal{T}{\mathbb Y}$. Combining $x=q/T_z$ with these two changes gives
\begin{equation}
 \frac{\bar x-x}{x}
   =(\Hc \mathcal{T}+ik\chi \mathcal{T}){\mathbb Y}
 \label{eq:app-frequency-map}
\end{equation}
at first order. At fixed new phase-space coordinates, the old arguments
of the homogeneous spectrum are
$\eta=\bar\eta-\mathcal{T}{\mathbb Y}$ and
$x=\bar x[1-(\Hc+ik\chi)\mathcal{T}{\mathbb Y}]$. Its first-order Taylor
expansion is therefore\footnote{For a static Planck background, the frequency shift measured by the
normal observers and
distribution transformation agree with Eqs.~(4.15) and (4.21) of
Naruko et al.~\cite{Naruko:2013aaa}. An evolving mean spectrum also
requires the time-displacement term $-\mathcal{T}n^{(0)\prime}$ retained here.
}
\begin{equation*}
 n^{(0)}(\eta,x)-n^{(0)}(\bar\eta,\bar x)
 =\left[-\mathcal{T}n^{(0)\prime}+(\Hc \mathcal{T}+ik\chi \mathcal{T})\Ox n^{(0)}\right]{\mathbb Y},
\end{equation*}
where we recall that $\Ox=-x\partial_x$ and $\Ox n^{(0)}=U$. Shifting the arguments of the already
first-order $n^{(1)}$ by $\mathcal{T}$ contributes only at second order.  Thus we obtain Eq.~(\ref{eq:distribution-transform}) for the first-order transformation
\begin{align}
 \bar n^{(1)}-n^{(1)}
 &= (\Hc \mathcal{T}+ik\chi \mathcal{T})U-\mathcal{T}n^{(0)\prime}.
 \label{eq:app-distribution-transform}
\end{align} 
The spatial shift $L$
 acts only on the first-order perturbation: $n^{(0)}$ has no
spatial dependence, and $L$ does not change the normal to the time
slices. Thus it supplies no term at this order.

Now let us consider the gauge transformation properties of collision terms.
Each individual contribution $C_j$ is an occupation-number source per unit conformal time
along the photon trajectory contributed by a process, so its increment satisfies
$\bar C_j\,\dd\bar\eta=C_j\,\dd\eta$ at the same phase-space event.
Along the background trajectory, $\dd X^i/\dd\eta=\hat\gamma^i$;
therefore the time-coordinate shift gives
\begin{equation*}
 \frac{\dd\bar\eta}{\dd\eta}
 =1+\left(\mathcal{T}'+ik\chi \mathcal{T}\right){\mathbb Y}.%
\end{equation*}
The $\mathcal{T}'$ term comes from the explicit time dependence of the shift,
while $ik\chi \mathcal{T}$ comes from the photon crossing its spatially varying
time slices. At fixed new phase-space coordinates, the background
source has the same argument shift as the mean distribution,
\begin{equation*}
 C_j^{(0)}(\eta,x)-C_j^{(0)}(\bar\eta,\bar x)
 =\left[-\mathcal{T}C_j^{(0)\prime}
   +(\Hc \mathcal{T}+ik\chi \mathcal{T})\Ox C_j^{(0)}\right]{\mathbb Y}.
\end{equation*}
Multiplying it by $\dd\eta/\dd\bar\eta$ and retaining first-order
terms gives
\begin{align}
 \bar C_j^{(1)}-C_j^{(1)}
 ={}&-\mathcal{T}C_j^{(0)\prime}-\mathcal{T}'C_j^{(0)}
     +\Hc \mathcal{T}\Ox C_j^{(0)}\nonumber\\
 &+ik\chi \mathcal{T}\Ox C_j^{(0)}-ik\chi \mathcal{T} C_j^{(0)} .
 \label{eq:app-source-pieces}
\end{align}
Physically,
 $-\mathcal{T}C_j^{(0)\prime}$ is the displaced background source,
$-\mathcal{T}'C_j^{(0)}$ is the time Jacobian $\partial \bar\eta/\partial \eta$, and
 $-ik\chi \mathcal{T} C_j^{(0)}$ comes from the photon-path
part of the time Jacobian $\hat\gamma^i \partial_{X^i} \bar\eta$.  
Here
$\Hc \mathcal{T}\Ox C_j^{(0)}$ is the isotropic frequency shift and the dipole
$ik\chi \mathcal{T}\Ox C_j^{(0)}$ term comes from the change of normal
frame. The $\chi$-dependent pieces  combine to form the total dipole term,
$(\Ox-1)C_j^{(0)}$ in Eq.~(\ref{eq:source-transform}).   These operations supply gauge-covariant forms
in Sec.~\ref{app:gauge-covariance-invariance} and are applied to the thermalization collision
terms  in Sec.~\ref{app:thermal-fh}.

\subsection{Gauge-covariant forms of the distribution}
\label{app:gauge-covariance-invariance}

The gauge transformation properties of the distribution and collision terms provide gauge-covariant forms for their (gauge-invariant) evaluation from a general gauge rather than directly in the target gauge.  

This construction applies to any target gauge, but Newtonian gauge supplies an important example given its use in \texttt{CosmoTherm}.  Newtonian gauge utilizes shear-free slicing and photon frequencies are specified in its normal frame.   
Applying \eqref{eq:app-distribution-transform} to the Newtonian case where $\mathcal{T}=s$ gives the covariant form of 
Eq.~(\ref{eq:invariant-distribution}) for $n_{\rm N}^{(1)}$. 
Its reconstruction from a general gauge involves the
displacement along the evolving mean spectrum, the frequency rescaling,
and the change of normal frame. The same choice in
Eq.~\eqref{eq:app-source-pieces} gives the collision source form
\eqref{eq:invariant-source}, including the photon-path time Jacobian.

For the collision term this reconstruction provides a means of deriving a general gauge collision term directly from Newtonian gauge expressions. 
  Operationally, the multipole moments of any collision term satisfy
\begin{align}
 C_{j{\rm N},0}^{(1)}
 &=C_{j,0}^{(1)}
   +\Hc s\Ox C_j^{(0)}
   -\left(sC_j^{(0)}\right)',
 \nonumber\\
 C_{j{\rm N},1}^{(1)}
 &=C_{j,1}^{(1)}
   -\frac{ks}{3}(\Ox-1)C_j^{(0)} \nonumber\\
    C_{j{\rm N}, \ell}^{(1)}
 &=C_{j,\ell }^{(1)} \qquad (\ell\ge 2),
 \label{eq:app-source-multipoles}
\end{align}
at first order,
or in terms of the spectral basis
\begin{align}
 \bm c_{j,0}^{\rm N}
 &=\bm c_{j,0}
   +\Hc sM_{\rm B}\bm c_j^{(0)}
   -\left(s\bm c_j^{(0)}\right)',
 \nonumber\\
 \bm c_{j,1}^{\rm N}
 &=\bm c_{j,1}
   -\frac{ks}{3}(M_{\rm B}-I)\bm c_j^{(0)},\nonumber\\
    \bm c_{j,\ell}^{\rm N}
 &=\bm c_{j,\ell} \qquad (\ell\ge 2).
 \label{eq:app-fh-source-completion}
\end{align}
Substituting the gauge-covariant representation of the Newtonian variables on the left hand sides gives a closed form expression for the collision terms in the variables of a general gauge.   
Comparing this Newtonian construction of the general gauge collision term with the direct construction in a general gauge provides a check of the consistency of the original Newtonian gauge result.   We call this the gauge invariance check. 
While this check would seem like a tautology since the time shift of the pieces in Eq.~\eqref{eq:invariant-source} cancel by definition, it can identify inconsistencies in the construction.

Finally it is worth noting that other gauge-covariant distribution functions can be constructed. The observer that defines the photon frequency need not be the normal observer.  For collisionless photons,
Durrer and Straumann's~\cite{DurrerStraumann:1988}
invariant is related to ours by
\begin{equation}
 {\cal F}_{\rm DS}\equiv n_{\rm N}^{(1)}+\Phi U
 = n_{\rm flat}^{(1)}-ik\chi\frac{\Phi}{\Hc}U,
 \qquad n^{(0)\prime}=0.
 \label{eq:app-ds-frame}
\end{equation}
Here $n_{\rm flat}^{(1)}$ uses spatially flat slices and photon
frequencies measured by their normal observers.
The shift $\mathcal{T}=\Phi/\Hc$ from shear-free to flat slicing gives this
relation by Eq.~\eqref{eq:distribution-transform}.  On the other hand  ${\cal F}_{\rm DS}$ omits the energy shift of Eq.~(\ref{eq:normal-shift-energy}) that would result from specifying observers by the normal to the slicing.
Thus the monopole of ${\cal F}_{\rm DS}$ refers to flat slicing,
while its dipole retains the normal frame of shear-free slices.  
Their collisionless transport equation agrees with ours under this
variable change.
The restriction $n^{(0)\prime}=0$ is their collisionless assumption. When collisions change the mean spectrum, the distribution
function at fixed $x$ is no longer conserved even in the background and $\mathcal F_{\rm DS}$ would require a generalization from the $n^{(0)\prime}$ term in 
Eq.~\eqref{eq:distribution-transform} for the shift to flat slicing.

\subsection{Gauge-covariant transport}
\label{app:hierarchy}

To obtain the gauge-covariant transport equation, we begin with the left hand side or Liouville term ${\rm D}
 n(\eta,\bm X,x,\hat{\bm\gamma})
/{\rm D}\eta$.  Embedded in the total derivative is the term  $\dd\ln x/\dd \eta$, the change in the frequency relative to the uniformly redshifting reference along the free photon trajectory.  Here we derive this term for a general metric fluctuation to linear order. 

Along a null geodesic with affine parameter $\lambda_{\rm aff}$,
$p^\nu=\dd X^\nu/\dd\lambda_{\rm aff}$ and the null-geodesic equation gives
$p^\nu\nabla_\nu p^\mu=0$. Differentiating the normal-frame energy
$q=-N_\mu p^\mu$ along the ray and using the geodesic equation gives
\begin{equation*}
 \frac{\dd q}{\dd\lambda_{\rm aff}}
 =-p^\nu\nabla_\nu(N_\mu p^\mu)
 =-p^\nu p^\mu\nabla_\nu N_\mu .
\end{equation*}
Here the covariant derivative of the normal can be separated into
the expansion $\vartheta$, acceleration $a_i$ and shear $\sigma_{ij}$ as
\begin{equation*}
 \nabla_\nu N_\mu=-N_\nu a_\mu
       +\frac{\vartheta}{3}h_{\nu\mu}+\sigma_{\nu\mu}.
\end{equation*}
Because $N^\mu$ is normal to the time slices, its vorticity vanishes. 
Substituting $p^\mu=q(N^\mu+e^\mu)$ and using
$N_\mu e^\mu=0$ gives
\begin{equation}
\frac{\dd q}{\dd\lambda_{\rm aff}}=-q^2\left(\frac{\vartheta}{3}+a_\mu e^\mu
+\sigma_{\mu\nu}e^\mu e^\nu\right).
\label{eq:energy-affine}
\end{equation}
Next we can convert derivatives with respect to 
$\lambda_{\rm aff}$ to those with respect to $\eta$.  Given the ADM lapse $\alpha\equiv(-g^{00})^{-1/2}$, $N_\mu=-\alpha\partial_\mu\eta$ and $\dd\eta/\dd\lambda_{\rm aff}=p^0=q/\alpha$.   Dividing Eq.~(\ref{eq:energy-affine}) by
$q\,\dd\eta/\dd\lambda_{\rm aff}$ yields
\begin{equation}
 \frac{\dd\ln q}{\dd\eta}
 =-\alpha\left(\frac{\vartheta}{3}
                 +a_\mu e^\mu+\sigma_{\mu\nu}e^\mu e^\nu\right).
 \label{eq:app-geodesic}
\end{equation}
To linear order, following KS,
\begin{align}
\alpha& =a(1+A{\mathbb Y}), \nonumber\\ 
 \frac{\vartheta}{3}
 &=\frac1a\left[\Hc+
  \left(H_L'+\frac{kB}{3}-\Hc A\right){\mathbb Y}\right],
 \nonumber\\
 a_i&=A\partial_i{\mathbb Y}, 
 \nonumber\\
 \sigma_{ij}&=ak^2s{\mathbb Y}_{ij},
 \label{eq:app-normal-kinematics}
\end{align}
so that
\begin{align*}
 \frac{\alpha\vartheta}{3}
 &=\Hc+\left(H_L'+\frac{kB}{3}\right){\mathbb Y},\\
 \alpha a_\mu e^\mu&=ik\chi A{\mathbb Y},\\
 \alpha\sigma_{\mu\nu}e^\mu e^\nu
 &=-\frac{2k^2s}{3}P_2(\chi){\mathbb Y},
\end{align*}
where we have used $k_i e^i = (k/a)\chi$ at background order and 
$\hat\gamma^i\hat\gamma^j{\mathbb Y}_{ij}
=-2P_2(\chi){\mathbb Y}/3$.
Finally, using $x = q/T_z$ where 
$\dd\ln T_z/\dd\eta=-\Hc$ we obtain the metric-induced frequency shift
\begin{align}
 \frac{\dd\ln x}{\dd\eta}
 &=\mathcal G\,{\mathbb Y},\nonumber\\
 \mathcal G
 &=-\left(H_L'+\frac{kB}{3}\right)P_0(\chi)
   -ikA P_1(\chi)
   +\frac{2k^2s}{3}P_2(\chi).
 \label{eq:app-frequency-shift}
\end{align}
Notice that this term is first order in fluctuations and therefore its corresponding Liouville term is
\begin{equation}
 \frac{\dd\ln x}{\dd\eta} \frac{\partial n^{(0)}}{\partial\ln
x} = - \frac{\dd\ln x}{\dd\eta} U  = -\mathcal G U\,{\mathbb Y},
\end{equation}
where $U = -x \partial_x n^{(0)}$. 
Combined with the linearized free streaming term
\begin{equation}
\frac{\dd X^i}{\dd\eta}  \partial_{i}  n=
ik\chi n^{(1)}  {\mathbb Y},
\end{equation} 
and the collision term,
we obtain the full first order transport equation
\begin{equation}
 n^{(1)\prime}+ik\chi n^{(1)}-\mathcal G U
 =C^{(1)},
 \label{eq:app-raw-boltzmann}
\end{equation}
in gauge-covariant form, completing the derivation of 
\eqref{eq:general-boltzmann}. Its metric frequency shift $\mathcal G$
contains the expansion, acceleration, and scalar shear of the normal
congruence, while $C^{(1)}=\sum_jC_j^{(1)}$ contains all collisions.

The angular hierarchy follows  from the addition of angular momentum in the $\chi n^{(1)}$ term \cite{Hu:1997hp}
\begin{equation*}
 (ik\chi n^{(1)})_\ell
 =\frac{k}{2\ell+1}
   \left[(\ell+1)n_{\ell+1}^{(1)}-\ell n_{\ell-1}^{(1)}\right].
\end{equation*}
Once further decomposed into the spectral basis, this gives the gauge-covariant FH equations (\ref{eq:general-fh-higher}). 

\section{Thomson and Thermalization Collision Terms}
\label{app:thermalization}

We use the construction of App.~\ref{app:gauge-covariance-invariance}
to obtain collision terms on a general slicing. We first treat unpolarized
Thomson scattering (Sec.~\ref{app:Thomson}), reconstruct the Newtonian
thermal variables (Sec.~\ref{app:thermal-variables}), and express the
Newtonian Kompaneets and effective photon-production terms of
Ref.~\cite{Chluba:2026nmu} in a general gauge and spectral basis
(Sec.~\ref{app:thermal-fh}).

Section~\ref{app:thermal-checks} compares the reconstructed Compton and
frequency-dependent DC sources with independent local constructions.
The original DC emissivity fails this comparison because parts of the temperature
response were omitted in its reference-frequency normalization. Restoring
the physical local normalization makes the direct and reconstructed
kernels agree. Section~\ref{app:dcbr-reduction} then derives the effective
DC relaxation and its complete perturbed source, including the frequency
conversion of the operator's input and output. The same mean emission
matrix is used to compare the original and corrected perturbations; only
the corrected source preserves the common adiabatic history in the
long-wavelength limit.

Section~\ref{app:br-extension} gives the optional BR kernel and its
effective completion. At the later epochs when BR becomes important for
low-frequency emission, Compton redistribution no longer efficiently
couples these photons to the spectrum near its peak. The single-profile
conversion between $\mu$ and temperature then becomes
inaccurate~\citep{Chluba2014,Evangelista2025}. We therefore omit BR from
Fig.~\ref{fig:cosmotherm-stationary}. A DC-only model omits BR from both the background
and perturbed equations; a model retaining BR must include its rate and
local responses in both. The gauge and separate-universe comparisons
establish consistency within the specified local model. The accuracy of
the low-frequency reduction, especially outside the $\mu$ era, is a
separate question assessed against frequency-resolved evolution.

For thermalization we retain first order separately in the mean departure
from a blackbody and the cosmological perturbation, including their
product. We assume fully ionized matter in the analytic thermalization
comparisons below. The cosmological calculation of
Sec.~\ref{sec:cosmological-propagation} includes recombination in the electron
abundance and Thomson transport. We adopt a quasi-stationary electron
temperature and isotropic rest-frame thermalization~\cite{Chluba:2026nmu}.
The general FH construction and Thomson terms allow an arbitrary isotropic
mean spectrum. Absorption of pre-existing anisotropic occupations and
intrinsic anisotropic Compton scattering lie beyond the isotropic
thermalization approximation, and have been neglected as before \citep{Chluba:2022efq}. Polarized Thomson scattering admits a
corresponding comparison with its local kernel.

\subsection{Thomson scattering}
\label{app:Thomson}

Retaining terms linear in anisotropic intensity and setting
polarization to zero in the Thomson kernel of
Refs.~\cite{Pitrou:2008hy,Naruko:2013aaa} provides the 
Thomson kernel adopted in the \texttt{CosmoTherm} treatment~\cite{Chluba:2026nmu}.
In this limit, the Thomson source in the normal
frame is
\begin{equation}
 C_{\rm T}^{(1)}=\tau'
 \left[n_0^{(1)}-\frac12n_2^{(1)}P_2(\chi)-n^{(1)}
       -i\chi(v_b-B)U\right].
 \label{eq:app-thomson}
\end{equation}
Here the $v_b-B$ term expresses the Doppler shift from the relative velocity of the normal observers and the baryons.  It takes the spectral form of a frequency shift $U$.  Aside from this generalization of a temperature shift to a frequency shift on an arbitrary background $n^{(0)}$, this is the usual unpolarized Thomson term of the blackbody Boltzmann hierarchy \citep[e.g.,][]{Ma1995, Hu:1997mn}. 

Decomposing this expression into multipole moments, the monopole term vanishes and 
\begin{equation}
 C_{{\rm T},1}^{(1)}
 =-\tau'\left[n_1^{(1)}-\frac{V_b}{3}U\right],
 \qquad
 C_{{\rm T},2}^{(1)}=-\frac9{10}\tau'n_2^{(1)},
 \qquad
 C_{{\rm T},\ell}^{(1)}=-\tau'n_\ell^{(1)}
 \quad(\ell\geq3).
 \label{eq:app-thomson-projection}
\end{equation}
In both Newtonian gauge and synchronous gauge $V_b=v_b$, the coordinate velocity amplitude of the baryons. The monopole term $(n_0^{(1)}-n^{(1)})_{\ell=0}=0$ vanishes because Thomson scattering conserves photon number and redistributes photon directions at fixed frequency in the baryon rest frame.

We can perform the gauge invariance check of App.~\ref{app:gauge-covariance-invariance} for Thomson scattering.  The first step is 
to 
 express the Newtonian terms in a general gauge using their gauge-covariant forms.
For the dipole, substituting $n_{{\rm N},1}^{(1)}=n_1^{(1)}-{ksU}/{3}$
and $V^{\rm N}=V-ks$ gives
\begin{align}
C_{{\rm TN},1}^{(1)}
 &= -\tau'\left[n_{{\rm N},1}^{(1)}-\frac{V_b^{\rm N}}{3}U\right]\nonumber\\
 &= -\tau'\left[n_1^{(1)}-\frac{ks}{3}U
                    -\frac{V_b-ks}{3}U\right]
  =-\tau'\left[n_1^{(1)}-\frac{v_b-B}{3}U\right].
 \label{eq:app-thomson-reconstruction}
\end{align}
The dipole source  involves the relative velocity of the photon and baryon fluids and hence is likewise gauge invariant on its own.  

For the check to pass, the transformed collision term must also be gauge invariant on its own.
Using \eqref{eq:app-source-multipoles} gives
\begin{align}
 C_{{\rm TN},0}^{(1)}
 &=C_{T,0}^{(1)}
   +\Hc s\Ox C_{\rm T}^{(0)}
   -\left(sC_{\rm T}^{(0)}\right)' =C_{T,0}^{(1)}=0,
 \nonumber\\
 C_{{\rm TN},1}^{(1)}
 &=C_{T,1}^{(1)}
   -\frac{ks}{3}(\Ox-1)C_{\rm T}^{(0)}=C_{T,1}^{(1)},
 \end{align}
  which, combined with Eq.~\eqref{eq:app-thomson-reconstruction} and the gauge invariance of $\ell \ge 2$ moments, verifies this and implies
 \begin{equation}
  C_{\rm T}^{(1)}=\tau'
 \left[n_0^{(1)}-\frac12n_2^{(1)}P_2(\chi)-n^{(1)}
       -i\chi(v_b-B)U\right],
 \end{equation}      
where $C_{\rm T}^{(0)}=0$ because Thomson scattering of an isotropic
background redistributes photon directions at fixed frequency.  This check returns the same form as the direct construction in Eq.~\eqref{eq:app-thomson-projection}.

The Thomson scattering check is particularly simple given that relative velocities are already gauge invariant.   To promote the Newtonian thermalization variables to their covariant forms is less trivial and requires the
temperature, density, and heating reconstructions given next.

\subsection{Reconstructing Newtonian thermal variables}
\label{app:thermal-variables}

The \texttt{CosmoTherm} treatment~\cite{Chluba:2026nmu} requires electron temperature,
electron density, and heating inputs in addition to the metric, photon,
and baryon variables. Here we give their gauge-covariant Newtonian
forms for their reconstruction from a general gauge. The metric and baryon-velocity reconstructions are supplied by
Eq.~\eqref{eq:invariant-metric}, and the photon coefficients by
Eq.~\eqref{eq:invariant-fh-amplitudes}. These inputs will be used to evaluate the Newtonian \texttt{CosmoTherm} treatment and
then construct its collision terms in a general gauge in App.~\ref{app:thermal-fh}.

The \texttt{CosmoTherm} treatment eliminates the electron temperature through
quasi-stationary balance between electron heating and Compton energy
exchange [Sec.~4.4 of Ref.~\cite{Chluba:2026nmu}]. Its reconstruction
below also supplies the general gauge temperature input for the direct thermalization construction in App.~\ref{app:thermal-checks}.

Write the physical electron temperature as
$T_e=T_{e0}+\delta T_e{\mathbb Y}$, with homogeneous value
$T_{e0}=T_z[1+\vartheta_e^{(0)}]$ and perturbation
$\delta T_e=T_z\vartheta_e^{(1)}$.
Since $T_e$ is a scalar,
$\overline{\delta T_e}=\delta T_e-T_{e0}'\mathcal{T}$ under a change of slicing,
so the normalized perturbation transforms as
\begin{equation}
 \bar\vartheta_e^{(1)}-\vartheta_e^{(1)}
 =\left[\Hc(1+\vartheta_e^{(0)})
              -\vartheta_e^{(0)\prime}\right]\mathcal{T},
 \label{eq:th-temperature-transform}
\end{equation}
where the $\Hc \mathcal{T}$ contribution comes from the redshifting of $T_z$ and $T_{e0}'$ gives the evolution of the background. 
The Newtonian reconstruction is therefore
\begin{align}
 \vartheta_{e{\rm N}}^{(1)}
 &=\vartheta_e^{(1)}
 +\left[\Hc(1+\vartheta_e^{(0)})
                    -\vartheta_e^{(0)\prime}\right]s,
 \label{eq:th-electron-temperature-reconstruction}
\end{align}
where recall $s^{\rm N}=0$.
Similarly, for the electron number density fluctuation $\delta_e = \delta n_e/n_{e0}$, define $n_e=n_{e0}(1+\delta_e{\mathbb Y})$. Its Newtonian reconstruction is
\begin{align}
 \delta_e^{\rm N}&=\delta_e-\frac{n_{e0}'}{n_{e0}}s
          =\delta_e+3\Hc s.
 \label{eq:th-thermal-invariants}
\end{align}
Here conserved baryon number and full ionization give
$n_{e0}\propto a^{-3}$ and $\delta_e^{\rm N}=\delta_b^{\rm N}$, where   $\delta_b = \delta \rho_b/\rho_{b0}$ is the baryon energy density fluctuation.
We also assume that electrons and baryons share the same 4-velocity,
$u_e^\mu=u_b^\mu$, as usually ensured by Coulomb interactions.

Next let $Q_h$ be the energy deposited into electrons per unit proper volume
and proper time in their rest frame, with homogeneous value $Q_h^{(0)}$.
Its scalar-rate reconstruction is
\begin{equation}
 \delta Q_h^{\rm N}=\delta Q_h-(Q_h^{(0)})'s.
 \label{eq:th-heating-invariant}
\end{equation}
This input describes electron heating; direct photon injection is
specified separately in Sec.~\ref{sec:source}.

Finally, the thermalization prescription requires the temperature perturbation
of the reference blackbody radiation. Let $\Theta_0^{(1)}$ denote the photon
temperature monopole evolved with the standard blackbody Boltzmann hierarchy,
neglecting injection and distortion corrections. Its Newtonian reconstruction
is $\Theta_{\rm N}=\Theta_0^{(1)}+\Hc s$, the monopole
$\Theta_0^{\rm N}$ of the ordinary temperature solution used in
Eq.~\eqref{eq:stationary-residual}. Just as $T_z$ supplies the homogeneous
reference temperature, this solution supplies its perturbed counterpart.
Its products with the mean distortion describe how the ordinary temperature
anisotropy changes thermalization in App.~\ref{app:thermal-fh}.
The full $G$ coefficient also contains distortion-dependent corrections and
transforms according to Eq.~\eqref{eq:invariant-fh-amplitudes}.

\subsection{Gauge-covariant thermalization in the spectral basis}
\label{app:thermal-fh}

To supply the collision terms of the \texttt{CosmoTherm} treatment~\cite{Chluba:2026nmu} in a general gauge, first
apply the prescription of App.~\ref{app:gauge-covariance-invariance}: replace Newtonian quantities in the Newtonian collision term with gauge-covariant forms of the photon, electron,
and heating variables. This constructs the transformation of the prescribed
Newtonian source; its agreement with independently specified local physics
is tested separately for the frequency-dependent collision terms in
Sec.~\ref{app:thermal-checks} and the effective photon-production prescription
in Sec.~\ref{app:dcbr-reduction}. Then add the terms
that transform the collision term itself which involve the evolving mean
collision spectrum and its conformal-time rate. Here the sector label
``th'' collects Compton and DC/BR collisions and the prescribed
electron heating after eliminating the quasi-stationary electron temperature.

The thermal inputs are reconstructed with
Eqs.~\eqref{eq:th-electron-temperature-reconstruction},
\eqref{eq:th-thermal-invariants}, and \eqref{eq:th-heating-invariant}.
The photon coefficients follow from Eq.~\eqref{eq:invariant-fh-amplitudes},
while the metric and baryon velocity follow
Eq.~\eqref{eq:invariant-metric}.
The reference blackbody temperature perturbation is $\Theta_{\rm N}$ as defined in
Sec.~\ref{app:thermal-variables}. Define $\theta_z=T_z/m_e$, where
$m_e$ is the electron mass. The adopted matrices for spectral basis vector conversions are those of
Eqs.~(4.34)--(4.37) of Ref.~\cite{Chluba:2026nmu}:
$M_{\rm K}$ describes number-conserving Compton redistribution with its
equilibrium temperature response. $M_{\rm T}$ additionally includes effective
DC/BR relaxation of the chemical potential toward zero, accompanied by a
change in the $G$ coefficient that conserves photon energy. Their actions on
the equilibrium directions and on a general coefficient vector are \citep[see][]{Chluba:2026nmu}
\begin{align}
 M_{\rm K}\bm e_G&=M_{\rm K}\bm e_M=0,\nonumber\\
 (M_{\rm T}-M_{\rm K})\bm y
 &=x_c\mu(\gamma_T\bm e_G-\gamma_N\bm e_M),
 \label{eq:th-effective-emission-matrix}
\end{align}
where $\bm e_G$ and $\bm e_M$ are unit vectors along $G$ and $M$, and
$\mu$ is the $M$ coefficient of $\bm y$. The critical dimensionless
frequency $x_c$ characterizes the balance of low-frequency DC/BR and Compton
redistribution. The adopted coefficients $\gamma_T\simeq0.1387$ and
$\gamma_N\simeq0.7769$ satisfy $4\gamma_T=\gamma_N\mathcal E[M]$.
Thus positive $\mu$ decays while the $G$ coefficient increases: photon number
increases and the two energy changes cancel [Sec.~4.5, Eq.~(4.27a), and
Sec.~4.7, Eq.~(4.36f), of Ref.~\cite{Chluba:2026nmu}]. This effective rate
describes the evolution of the broad spectrum without resolving the
low-frequency turnover back to thermal equilibrium. Improvements to this effective treatment (i.e., including more detailed frequency-dependent corrections and FH responses) could be constructed analytically following \cite{Chluba2014} or by using the matching to the numerical solutions \citep{Evangelista2025}.

The boost matrix $M_{\rm B}$ represents the frequency-shift generator $\Ox$ in spectral basis space \cite[see][]{Chluba:2022efq}.
The commutator $M_{\rm BK}=M_{\rm B}M_{\rm K}-M_{\rm K}M_{\rm B}$
compares shifting the Compton collision spectrum with applying Compton
redistribution to the shifted spectrum. It supplies the corresponding
temperature-dependent correction to the monopole below. The
stimulated-scattering rearrangement and physical approximations are given in
Ref.~\cite{Chluba:2026nmu}.

Let $\bm e_Y$ and $\bm e_{Y_1}$ be the spectral-basis unit vectors along $Y_0$ and $Y_1$.
The heating and heat-redistribution vectors per unit conformal time,
normalized by the reference photon energy density $\rho_z$, are
\begin{align}
 \bm h^{(0)}&=\frac{aQ_h^{(0)}}{\rho_z}\bm e_Y, \nonumber\\
 \bm h_{\rm N}^{(1)}
   &=\frac{a}{\rho_z}\left[\delta Q_h^{\rm N}+\Psi Q_h^{(0)}\right]\bm e_Y,\nonumber\\
 \bm r_h^{(0)}&=\frac{aQ_h^{(0)}}{\rho_z}(\bm e_{Y_1}-\bm e_Y).
 \label{eq:th-heating-vectors}
\end{align}
The factor $a$ converts the background proper rate to conformal time and $\Psi$ accounts for the lapse of proper time in Newtonian gauge \cite{Senatore2009}. 
These correspond to the heating and heat-redistribution vectors of
Eqs.~(4.36f,g) of Ref.~\cite{Chluba:2026nmu}. Its Eqs.~(4.37a,b) then give
\begin{subequations}
\begin{align}
 \bm c_{\rm th}^{(0)}
 &=\tau'\theta_zM_{\rm T}\bm y^{(0)}+\frac{\bm h^{(0)}}4,
 \label{eq:th-fh-background}\\
 \bm c_{{\rm th},0}^{\rm N}
 &=\tau'\theta_zM_{\rm T}\bm y_0^{\rm N}
   +\frac{\bm h_{\rm N}^{(1)}}4+\Theta_{\rm N}\bm r_h^{(0)}
   +\tau'\theta_z
   \left[(\delta_e^{\rm N}+\Theta_{\rm N}+\Psi)M_{\rm T}
                 +\Theta_{\rm N}M_{\rm BK}\right]\bm y^{(0)},
 \label{eq:th-fh-monopole} \\
  \bm c_{{\rm th},1}^{\rm N}
 &=\frac{V_b^{\rm N}}{3}(M_{\rm B}-I)\bm c_{\rm th}^{(0)},
 \label{eq:th-fh-dipole}
\end{align}
\end{subequations}
where we implicitly replace direct Newtonian variables with their
gauge-covariant forms on the right-hand sides.  
The $1/4$ in the heating term converts the deposited energy to the amplitude of $Y$, whose
normalized energy moment is four. The density and lapse modulate the
collision rate; the temperature terms describe redistribution and the
spectral response to heating.  The dipole term includes the Compton and effective DC/BR boosts. These
equations state the baseline prescription of the reference.
Section~\ref{app:dcbr-reduction} derives the addition to
Eq.~\eqref{eq:th-fh-monopole} needed for DC relaxation to agree with the
direct local thermalization history; Sec.~\ref{app:br-extension} extends
it to BR. The general reconstruction below applies to the completed
Newtonian monopole.

Solving Eq.~\eqref{eq:app-fh-source-completion} for the collision terms in a general gauge, with the reconstructed Newtonian inputs, gives:
\begin{align}
 \bm c_{{\rm th},0}
 &=\bm c_{{\rm th},0}^{\rm N}
   -\Hc s M_{\rm B}\bm c_{\rm th}^{(0)}
   +(s\bm c_{\rm th}^{(0)})',\nonumber\\
 \bm c_{{\rm th},1}
 &=\frac{V_b}{3}(M_{\rm B}-I)\bm c_{\rm th}^{(0)},
 \label{eq:th-fh-general}
\end{align}
where we have left the gauge-covariant substitution of
$\bm c_{{\rm th},0}^{\rm N}$ implicit and have explicitly written the additional terms:  the reference-frequency shift and $(s\bm c_{\rm th}^{(0)})'$.   The latter contains
$(s\bm c_{\rm th}^{(0)})'=s'\bm c_{\rm th}^{(0)}+
 s\bm c_{\rm th}^{(0)\prime}$, the time-coordinate Jacobian and 
 the displacement of the mean collision history.
 For the dipole we have  used the cancelling contributions from the transformation of $V_b^{\rm N}$ and  $c_{{\rm th},1}$ to show its explicit form in a general gauge.

The stationary solution of Sec.~\ref{sec:source-gauges} illustrates these
matrix operations. After injection, the example neglects electron heating
and net DC/BR relaxation over the interval considered. The collision matrix
therefore reduces to $M_{\rm K}$. For this static Compton-equilibrium
mean, preservation of the stationary anisotropy requires the Compton
collision term to vanish. The evolving-mean DC case is treated
in App.~\ref{app:dcbr-reduction}, with its BR extension in
App.~\ref{app:br-extension}. 

 A change in temperature or chemical
potential moves along the family of Compton-equilibrium spectra, so both
directions are annihilated by $M_{\rm K}$ in
Eq.~\eqref{eq:th-effective-emission-matrix}. The equilibrium mean
$\bm y^{(0)}=\bar\Theta\bm e_G+\bar\mu\bm e_M$ consequently obeys
$M_{\rm K}\bm y^{(0)}=0$ and $\bm c_{\rm C}^{(0)}=0$.
A blackbody temperature shift is also annihilated by $M_{\rm T}$ because
it has no chemical potential to relax. A nonzero-$\mu$ equilibrium mean is
static only when the DC/BR relaxation is omitted; retaining $M_{\rm T}$
would evolve its $G$ and $M$ coefficients together leading to the conversion of $\mu\rightarrow \Theta$ under energy conservation \citep{Chluba:2022xsd}.

The stationary anisotropy is a local frequency shift of that mean spectrum.
Let $\Theta_0$ denote the monopole amplitude of this shift, so that
$\bm y_0=\Theta_0\bm b^{(0)}$. In the Newtonian superhorizon example of
Sec.~\ref{sec:source-gauges}, $\Theta_0^{\rm N}=\Theta_{\rm ad}$; the
subscript labels the angular monopole. The $G$ coefficient of $\bm y_0$
also includes the frequency shift of the mean distortion. The shift vector is $\bm b^{(0)}=\bm e_G+M_{\rm B}\bm y^{(0)}$
from Eq.~\eqref{eq:fh-background-response}.
The first term shifts the reference blackbody; the second shifts the mean
distortion. Applying $M_{\rm K}$ to this vector alone need not give zero.
The accompanying temperature perturbation of the Compton operator supplies
$M_{\rm BK}\bm y^{(0)}$, whose $-M_{\rm K}M_{\rm B}\bm y^{(0)}$ term
cancels redistribution of the shifted distortion. The remaining
$M_{\rm B}M_{\rm K}\bm y^{(0)}$ shifts the vanishing equilibrium collision
spectrum. For this stationary monopole,
Eq.~\eqref{eq:th-fh-monopole} therefore gives
\begin{align}
 \bm c_{{\rm C},0}
 &=\tau'\theta_z\Theta_0
   [M_{\rm K}\bm b^{(0)}+M_{\rm BK}\bm y^{(0)}]\nonumber\\
 &=\tau'\theta_z\Theta_0 M_{\rm B}M_{\rm K}\bm y^{(0)}=0.
 \label{eq:th-stationary-compton-null}
\end{align}
Since $\bm c_{\rm C}^{(0)}=0$, the slicing completion
in Eq.~\eqref{eq:th-fh-general} also vanishes. The kinematic dipole is
proportional to $\bm c_{\rm C}^{(0)}$ and vanishes as well. These
statements include the equilibrium electron-temperature response in
$M_{\rm K}$ and the retained stimulated terms. 
Under these \texttt{CosmoTherm} assumptions, the stationary anisotropy spectrum remains stationary under Comptonization, which is a result we used in Sec.~\ref{sec:source-gauges}.
Compton redistribution and photon-production effects acting on pre-existing
dipole and higher multipoles in the electron rest frame are neglected as
in the \texttt{CosmoTherm} treatment~\cite{Chluba:2026nmu}. The kinematic dipole
generated by boosting the isotropic collision source is retained.

\subsection{Gauge invariance of the frequency-dependent collision terms}
\label{app:thermal-checks}

We now work with frequency-dependent distributions $n$ and collision
terms $C$, before projection onto the retained spectral basis. Their
independent local construction tests the source transformations used in
App.~\ref{app:thermal-fh}. We retain the thermalization assumptions of the
\texttt{CosmoTherm} treatment~\cite{Chluba:2026nmu} and vary the local frequency,
temperature, rate and boost inputs. The Compton monopole check uses the
fully ionized, unheated quasi-stationary limit. For DC we check the
frequency-dependent emission and absorption kernel; Sec.~\ref{app:br-extension}
gives the corresponding BR kernel and its check. A common spectral
projection supplies the corresponding coefficient relations; the effective
photon-production prescription obtained by eliminating the low-frequency
layer requires the separate comparison in Sec.~\ref{app:dcbr-reduction}. 

\subsubsection{Compton scattering}
\label{app:compton}

Expanding the local Compton operator allows an independent comparison
with the collision terms constructed in Sec.~\ref{app:thermal-fh}.
Its monopole includes electron-density and lapse perturbations and the
temperature response of the evolving mean spectrum. We use the frequency operators of Ref.~\cite{Chluba:2026nmu} and in this section we provide the specific equations from their work that correspond to our equations.  The defining operators are given in their
Eqs.~(4.2) and (A.1a)--(A.1c):
\begin{align}
 \widehat{\mathcal D}_x&=\Ox(\Ox-3)=x^{-2}\partial_x x^4\partial_x,&
 \widehat{\mathcal D}_x^*F&=x^{-2}\partial_x(x^4F),\nonumber\\
 \widehat{\mathcal K}_x F
 &=\widehat{\mathcal D}_xF+\widehat{\mathcal D}_x^*(A_xF),&
 A_x&=1+2n_{\rm bb},\qquad Y=\widehat{\mathcal D}_x n_{\rm bb}.
 \label{eq:th-operators}
\end{align}
With $f=n^{(0)}-n_{\rm bb}$, their Eqs.~(4.1a,b) give the background source
$C_{\rm C}^{(0)}=\tau'\theta_z[\vartheta_e^{(0)}Y+
\widehat{\mathcal K}_x f]$ and the intrinsic monopole.
Adding the normal-frame rate perturbation gives
\begin{align}
 C_{{\rm C},0}^{(1)}=\tau'\theta_z\big\{&
 \vartheta_e^{(1)}[Y+\widehat{\mathcal D}_x f]
 +\widehat{\mathcal K}_x n_0^{(1)}
 +\vartheta_e^{(0)}\widehat{\mathcal D}_x n_0^{(1)}\nonumber\\
 &+2\widehat{\mathcal D}_x^*[f n_0^{(1)}]
 +(\delta_e+A)[\vartheta_e^{(0)}Y+\widehat{\mathcal K}_x f]\big\}.
 \label{eq:th-compton-monopole}
\end{align}
We retain first order separately in the mean departure and the
cosmological perturbation, including their product. In products with
$f$ or $\vartheta_e^{(0)}$, only the leading blackbody part of the
perturbed distribution or temperature is retained. The stimulated term
$2\widehat{\mathcal D}_x^*[f n_0^{(1)}]$ is already present in their Eq.~(4.1b); its rearrangement into the FH matrices is given
in their Eqs.~(4.3)--(4.5) and (4.34)--(4.37).

This rearrangement also displays the physical frequency response of
Compton scattering. In the second line of their Eq.~(4.5),
$\widehat{\mathcal K}_x[n_0^{(1)}-\Theta_{\rm N}\Ox n^{(0)}]$
subtracts the common frequency shift from the incident perturbation, while
$\Theta_{\rm N}\Ox[C_{\rm C}^{(0)}/(\tau'\theta_z)]$ supplies the
frequency shift of the mean collision output. The accompanying term
$\Theta_{\rm N}C_{\rm C}^{(0)}/(\tau'\theta_z)$ accounts for the
Compton clock's temperature dependence. Combining the input and output
responses with the quasi-stationary electron-temperature response gives
$M_{\rm BK}=[M_{\rm B},M_{\rm K}]$, as in their Eq.~(4.35) and
Eq.~\eqref{eq:th-fh-monopole} here. The commutator is therefore part of
the local Compton temperature response, including stimulated scattering.
Its frequency-coordinate interpretation will also apply to the reduced
DC operator in Sec.~\ref{app:dcbr-source-check}. The velocity boost that
produces the dipole below is a separate observer-frame transformation.

To compare with App.~\ref{app:thermal-fh}, set electron heating and the DC/BR
terms to zero, so that $M_{\rm T}$ reduces to $M_{\rm K}$ in
Eq.~\eqref{eq:th-fh-monopole}. Its Compton monopole is the spectral projection
of Eq.~\eqref{eq:th-compton-monopole} in Newtonian gauge after eliminating
the electron temperature with the same quasi-stationary balance.
Equation~\eqref{eq:th-fh-general} then reconstructs that monopole in a
general gauge. We compare its slicing dependence with the direct changes
of the temperature, density, distribution, and lapse inside the local operator.

In the fully ionized, unheated quasi-stationary limit,
$\tau'\theta_z\propto a^{-3}$. Substituting these changes into
Eq.~\eqref{eq:th-compton-monopole} and subtracting the source transformation
of Eq.~\eqref{eq:source-transform} reduces the remaining frequency-dependent
residual to the following operator identity for $f=n^{(0)}-n_{\rm bb}$:
\begin{align}
 &\widehat{\mathcal K}_x\Ox f
       -\Ox\widehat{\mathcal K}_x f
       +\widehat{\mathcal D}_xf
       +2\widehat{\mathcal D}_x^*(Gf)
       -\widehat{\mathcal K}_xf\nonumber\\
 ={}&\widehat{\mathcal D}_x^*[(A_x-2G)f]
       -\widehat{\mathcal D}_x^*(A_xf)
       +2\widehat{\mathcal D}_x^*(Gf)=0.
 \label{eq:th-compton-identity}
\end{align}
Here $[\Ox,\widehat{\mathcal K}_x]f
=-\widehat{\mathcal D}_x^*[(A_x-2G)f]$ follows from the operator
definitions. Thus the direct local operator has the same source transformation
used in Eq.~\eqref{eq:th-fh-general}. With the same spectral projection and
electron-temperature closure, it agrees with the Compton sector of
App.~\ref{app:thermal-fh} in this limit, including for an evolving mean distortion.
This cancellation requires both the full scalar-temperature shift of
Eq.~\eqref{eq:th-temperature-transform} and stimulated scattering. Omitting
the latter leaves the nonzero residual $-2\widehat{\mathcal D}_x^*(Gf)$.
This contribution was
neglected in the original FH monopole, Eq.~(C.11) of
Ref.~\cite{Chluba:2022efq}, and retained in the improved treatment,
Eqs.~(4.1b) and (4.3)--(4.5) of Ref.~\cite{Chluba:2026nmu}.
The improved treatment recovers the stationary response of a blackbody whose temperature differs from $T_z$~\cite{Chluba:2026nmu}. Its numerical comparison with the earlier treatment~\cite{kite_spectro-spatial_2023-III} found only small changes in the decaying-particle power spectra from the combined refinements.
The independent
comparison therefore fails with that original truncation and passes
when the stimulated term is retained at the order considered here.
A dynamical electron
bath requires its heat equation and the opposite Compton energy transfer.

The boost of an isotropic electron-frame collision supplies a shared
dipole check for Compton scattering and DC/BR emission. Its frequency
and photon-path time transformations give
\begin{equation}
 C_{j,\rm dip}^{(1)}
 =-i\chi V_b (\Ox-1)C_j^{(0)}.
 \label{eq:th-local-boost}
\end{equation}
This local boost supplies the dipole comparison with
Eq.~\eqref{eq:th-fh-general}: its
change under $V_b \to V_b-k\mathcal{T}$ is exactly the dipole part of
Eq.~\eqref{eq:source-transform}. Omitting this dipole for a general
evolving thermalization source leaves that transformation unmatched. The
Compton kinematic terms are given in Eq.~(A.8) of
Ref.~\cite{Chluba:2022efq} but were omitted from its final hierarchy;
the improved treatment retains them in Eq.~(5.5) of Ref.~\cite{Chluba:2026nmu}, where the background-temperature-shift tests also demonstrated the importance of these terms for preserving the full stationary solution.
This completes the independent check of the Compton terms in agreement with \citep{Chluba:2026nmu}.

\subsubsection{Frequency-dependent double Compton}
\label{app:dcbr}

The DC collision term has been treated in detail \citep{Thorne1981, Lightman1981}, including various relativistic temperature and frequency corrections \citep{Gould1984, Chluba2007a, Ravenni2020DC}.
Our strategy is to specify a local collision model in the soft-photon limit and low-temperature limit \citep[see][for detailed comparisons]{Chluba2007a,Ravenni2020DC} and map it
self-consistently to its gauge and spectral-basis representations.
Emission and absorption are specified in the electron rest frame using
physical photon energies and bath temperatures. Their expressions on
another slicing and in reference-frequency coordinates must follow from
those same local inputs. Here we identify the emissivity inconsistency
and construct the corrected frequency-dependent kernel;
Sec.~\ref{app:dcbr-reduction} then reduces that kernel and derives its
complete response in the fixed spectral basis.

We apply the gauge-invariance check of
App.~\ref{app:gauge-covariance-invariance} to the frequency-dependent
emission kernel. We first take the stated soft-DC scaling of
Ref.~\cite{Chluba:2026nmu} [Sec.~4.5, Eq.~(4.24) and the discussion below it]
as a trial local prescription and compare the direct general-gauge
source with the transformation of the Newtonian source for that same
prescription.

The local emission and absorption model is specified in the baryon
rest frame, which we identify with the electron rest frame,
$u_b^\mu=u_e^\mu$. Let $E_b=-p\cdot u_b$ be the photon energy measured
there and $x_b=E_b/T_z$ its reference-frequency coordinate. The subscript
$b$ labels the measuring frame; the temperature unit remains the
reference $T_z$. In contrast, $u_\gamma=E_b/T_\gamma$ and
$u_e=E_b/T_e$ use the physical photon and electron temperatures.
Write the electron-frame kernel per Thomson depth as
$\mathcal L_{\rm em}=P_{\rm em}D_{\rm em}$, with
$P_{\rm em}=\Lambda/x^3_b$ and $D_{\rm em}=r_e-n(1-r_e)$, where
$r_e=e^{-E_b/T_e}$ [Eq.~(B.1) of Ref.~\cite{Chluba:2026nmu}].
The background quantities are
\begin{equation}
 D_{\rm em}^{(0)}=r_e^{(0)}-n^{(0)}(1-r_e^{(0)}),\qquad
 r_e^{(0)}=e^{-x/(1+\vartheta_e^{(0)})},\qquad
 \mathcal L_{\rm em}^{(0)}=P_{\rm em}^{(0)}D_{\rm em}^{(0)}.
 \label{eq:th-emission-balance}
\end{equation}
First consider the perturbed monopole, for which $x_b=x$ at this
order. At fixed $x$, the balance factor varies as
\begin{equation*}
 D_{{\rm em},0}^{(1)}
 =-(1-r_e^{(0)})n_0^{(1)}
  +(1+n^{(0)})r_e^{(0)}
     \frac{x\vartheta_e^{(1)}}{(1+\vartheta_e^{(0)})^2}.
\end{equation*}
We retain the first order separately in the mean departure and the
cosmological perturbation, including their product.

For the trial prefactor $P_{\rm trial}=K\theta_\gamma^2/x^3_b$, let
$\theta_I=T_I/m_e$ and $\Theta_\gamma=\delta T_\gamma/T_{\gamma0}$.
Here $K$ is constant for leading soft, nonrelativistic DC with Planck
incident photons. Varying the kernel at fixed $x$ and including the
electron-density and lapse perturbations of the conformal-time rate gives
the direct general-gauge monopole
\begin{equation*}
 \left.C_{{\rm trial},0}^{(1)}\right|_{\rm dir}
 =\tau'\left[
   P_{\rm trial}^{(0)}D_{{\rm em},0}^{(1)}
   +(2\Theta_\gamma+\delta_e+A)\mathcal L_{\rm trial}^{(0)}
 \right],
\end{equation*}
where $\tau'=a n_{e0}\sigma_{\rm T}$ and
$\mathcal L_{\rm trial}^{(0)}=P_{\rm trial}^{(0)}D_{\rm em}^{(0)}$.

In transforming the Newtonian source, the detailed-balance and rate
terms give their direct local variations. In particular, the occupation
and electron-temperature transformations give
$\Delta D_{{\rm em},0}^{(1)}
 =\Hc \mathcal{T}\Ox D_{\rm em}^{(0)}-\mathcal{T} D_{\rm em}^{(0)\prime}$, and the rate
modulation obeys $\Delta(\delta_e+A)=-\mathcal{T}'-(\tau''/\tau')\mathcal{T}$.
The remaining frequency shift acts on the prefactor according to
\begin{equation}
 \Ox\!\left(\frac{\Lambda}{x^3}\right)
  =[3+\Ox\ln\Lambda]\frac{\Lambda}{x^3}.
 \label{eq:th-emission-prefactor}
\end{equation}
Since $\Lambda_{\rm trial}$ is independent of $x$, the inverse
transformation from Newtonian gauge gives
\begin{equation}
 \left.C_{{\rm trial},0}^{(1)}\right|_{\rm rec}
 =\tau'\left[
   P_{\rm trial}^{(0)}D_{{\rm em},0}^{(1)}
   +(2\Theta_\gamma+\delta_e+A-3\Hc s)\mathcal L_{\rm trial}^{(0)}
 \right].
 \label{eq:th-emission-raw}
\end{equation}
The direct source minus this reconstructed source is
$3\Hc s\, \tau' \mathcal L_{\rm trial}^{(0)}$ and  the consistency check of App.~\ref{app:gauge-covariance-invariance} fails.
The mismatch comes from treating the arbitrary reference 
$T_z$ as a physical input to the emissivity prefactor $P_{\rm trial}$ through $x_b$. At fixed $E_b$, occupation
and local bath, rescaling the arbitrary reference temperature unphysically changes the result.

The correction is therefore made at the level of the local kernel,
before spectral projection. We use the physical photon bath to set the
frequency scale of the soft-DC emissivity and then rewrite that kernel
in reference-frequency coordinates, retaining every normalization factor
required by the change of variables. This specifies the same local
collision process for the background and its perturbations.

For soft DC the frequency in the local emissivity is
$u_\gamma=E_b/T_\gamma$. Expressing this kernel in reference units gives
\begin{equation}
 P_{\rm DC}=K\frac{\theta_\gamma^2}{u_\gamma^3}
    =K\frac{\theta_\gamma^5}{\theta_z^3 x^3_b},
 \label{eq:th-emission-local-dc}
\end{equation}
whereas Refs.~\citep{Chluba:2022efq,Chluba:2026nmu} incorrectly used
$P_{\rm DC}\propto\theta_\gamma^2$ at fixed $x_b$ and $T_z$, omitting
the normalization factor $(T_\gamma/T_z)^3$.\footnote{One of the authors is embarrassed to admit this mistake given their experience with the DC process \citep{Chluba2007a, Ravenni2020DC}.}
The two prescriptions agree on the leading background
$T_\gamma=T_z$. At fixed $x$ and $T_z$, the local kernel gives
$5\Theta_\gamma$ for its prefactor response. In the inverse transformation
from Newtonian gauge, $T_{z,{\rm N}}=T_z(1-\Hc s)$ and
$x_{b,{\rm N}}=x_b(1+\Hc s)$ for the monopole. The 
$K\theta_\gamma^5/\theta_{z,{\rm N}}^3$ factor then supplies
$5\Theta_\gamma+3\Hc s$, and $x^{-3}_b$ supplies the counter term $-3\Hc s$, leaving the physical result independent of $T_z$.
The balance factor continues
to use the same physical ratio $E_b/T_e$. Relative to the reconstruction
from the original Newtonian prescription, the corrected source therefore
adds $3(\Theta_\gamma+\Hc s)C_{\rm trial}^{(0)}$ at the matched background.
The quantity in parentheses is the photon-temperature perturbation on the
Newtonian slicing at this order. The difference is gauge invariant, so a
comparison of transformation laws alone would miss it.

For the corrected DC prefactor, the background dependence at fixed $x$
comes from the scalar bath variables and $E_b=xT_z$. The latter has
$E_b'=-\Hc E_b$, so varying the scalar inputs at fixed $x$ gives
$\Delta P_{{\rm em},0}^{(1)}
 =-\mathcal{T} P_{\rm em}^{(0)\prime}+\Hc \mathcal{T}\Ox P_{\rm em}^{(0)}$.
Combining this with the balance and rate variations gives
\begin{equation}
 \Delta C_{{\rm em},0}^{(1)}
 -\left[\Hc \mathcal{T}\Ox C_{\rm em}^{(0)}
              -(\mathcal{T}C_{\rm em}^{(0)})'\right]=0.
 \label{eq:th-emission-covariance}
\end{equation}
The local boost in Eq.~\eqref{eq:th-local-boost} supplies
$\Delta C_{{\rm em},1}^{(1)}=-k\mathcal{T}(\Ox-1)C_{\rm em}^{(0)}/3$.
Thus the corrected DC kernel reproduces the monopole and dipole
transformations of App.~\ref{app:thermal-fh} before spectral projection,
whereas the trial prescription fails the independent local comparison.
The following section reduces this kernel to the spectral basis;
Sec.~\ref{app:br-extension} gives the corresponding BR extension.
As in Ref.~\cite{Chluba:2026nmu}, the reduced model neglects the small
net energy exchange with electrons. If that energy moment is retained,
matter or a specified reservoir must carry the opposite transfer.

\subsection{Double-Compton relaxation in the spectral basis}
\label{app:dcbr-reduction}

DC emission and absorption maintain a nearly blackbody spectrum at low
frequencies, while Compton scattering redistributes the produced photons
toward the spectral peak. Their balance determines the effective
chemical-potential relaxation \citep[see][for a detailed discussion of
corrections to this treatment]{Chluba2014}. Rather than resolving the
low-frequency layer with additional spectral functions, \texttt{CosmoTherm}
incorporates its influence through energy-conserving evolution of the
retained $G$ and $M$ coefficients. We first reduce the corrected local DC
kernel of Sec.~\ref{app:dcbr} and perturb that same model. The extension
including BR is given in Sec.~\ref{app:br-extension}.

In this DC-only model, BR is omitted from both the mean relaxation and
its perturbation. Its emissivity is set to zero before solving for the
low-frequency profile: removing a perturbed BR term while retaining its
contribution to the mean rate would describe different local models in
the two equations. The gauge check requires the same retained physics in
both; it also permits BR to be included consistently in both.

The gauge-invariance comparison of
App.~\ref{app:gauge-covariance-invariance} tests whether the Newtonian
source reconstructs the independent local source on arbitrary slicing.
A separate-universe comparison tests whether a long-wavelength perturbation
follows the homogeneous history or parameter response of that same model.
Comparison with independently evolved, more complete kinetic physics
assesses the physical accuracy of the reduction. These tests distinguish
the consistency of a specified model from the accuracy of its approximations.

\subsubsection{Low-frequency reduction and relaxation rate}
\label{app:dcbr-rate}

The original basis prescription in App.~\ref{app:thermal-fh} represents
photon production through $M_{\rm T}-M_{\rm K}$. To compare the
perturbative sources at a fixed mean rate, we evaluate the original
prescription with the emission matrix matched below. Its contributions
to Eqs.~\eqref{eq:th-fh-background} and \eqref{eq:th-fh-monopole} are
\begin{align}
 \bm c_{\rm em}^{(0)}
 &=\tau'\theta_z(M_{\rm T}-M_{\rm K})\bm y^{(0)},\nonumber\\
 \bm c_{{\rm em},0}^{\rm N}
 &=\tau'\theta_z(M_{\rm T}-M_{\rm K})\bm y_0^{\rm N}
   +(\delta_e^{\rm N}+\Theta_{\rm N}+\Psi)\bm c_{\rm em}^{(0)}.
 \label{eq:th-effective-dcbr-sector}
\end{align}
The first term in the monopole relaxes the chemical-potential perturbation;
the second modulates the mean relaxation by density, temperature and lapse.
This effective form follows from the low-frequency emission--Compton
balance and its matching to energy-conserving $G$ and $M$ evolution
[Eqs.~(4.24)--(4.32) of Ref.~\cite{Chluba:2026nmu}]. In that reduction,
the critical frequency is treated as unperturbed when the photon and
electron temperatures track one another, while the Compton clock supplies
the explicit $\Theta_{\rm N}$ prefactor [their Eq.~(4.32) and the
discussion below it]. The absorption response and the ratio of time
scales remain essential inputs to the corrected reduction.

The local frequency normalization of Sec.~\ref{app:dcbr} changes the
temperature response of the turnover. With $T_e=T_\gamma=T$, the leading
DC turnover in local frequency units scales as $T^{1/2}$. Its frequency
in reference units is larger by $T/T_z$ and scales as $T^{3/2}$ at fixed
$T_z$. A local perturbation changes both the effective rate and the
frequency coordinates of the emission operator's input and output.
These responses must be retained together.
The absorption and Compton-matching treatment of
Ref.~\citep{Chluba:2022efq} provides the starting point for this reduction.
Here we complete its fixed-reference spectral representation using the
corrected local emissivity.

The low-frequency profile is assumed to adjust rapidly compared
with relaxation of the outer chemical potential. DC emission and absorption
drive the soft spectrum toward a blackbody at the local (electron) temperature,
while Compton redistribution connects it to the outer distortion. We
therefore solve their quasi-stationary balance in the Rayleigh--Jeans
region and integrate the net photon production. Energy conservation fixes
the accompanying temperature change as the chemical potential relaxes.
This supplies a local constitutive model for the retained $G/M$ evolution.
Time-dependent and relativistic temperature corrections \citep[for detailed treatment see][]{Chluba2014} are omitted, but can have noticeable effects deep into the thermalization era \citep{Chluba2005,ChlubaSunyaev:2012, Khatri2012b, Chluba2014}.

To determine the effective rate, let $u=E_b/T$ and write the small
chemical-potential profile in the unresolved layer as
$\mu(u)=\mu_\infty m(u)$. The leading Rayleigh--Jeans balance requires
\begin{equation}
 \frac{\dd}{\dd u}\left(u^2\frac{\dd m}{\dd u}\right)
       -\frac{\mathcal A(u)}{u^2}m=0,
 \qquad \mathcal A(u)=\frac{\Lambda_{\rm DC}(u)}{\theta},
 \qquad \theta=\frac{T}{m_e}.
 \label{eq:th-reduced-profile}
\end{equation}
Here $\Lambda_{\rm DC}$ is the numerator of the local-frequency prefactor
in Eq.~\eqref{eq:th-emission-local-dc}, evaluated for $u\ll1$.
For constant $\mathcal A=x_c^2$, $m=e^{-x_c/u}$ connects the inner
blackbody, $m\rightarrow0$, to the outer chemical potential,
$m\rightarrow1$. Integrating the net photon production gives
$\int_0^\infty x_c^2m(u)\,\dd u/u^2=x_c$. For the adopted profile we
define $x_{c,\rm eff}=\int\mathcal A(u)m(u)\,\dd u/u^2$ by matching this
integrated production.\footnote{For the numerical DC-only example we use
leading soft DC emission with a Planck incident moment and no BR emissivity.
We solve on $10^{-5}\leq u\leq0.2$, with $m=0$ at the lower boundary and
$m+u\,\dd m/\dd u=1$ at the upper boundary. The rate and its temperature
derivative use this same matching problem.}

Photon production relaxes the mean chemical potential toward zero while
increasing the temperature coefficient at fixed photon energy. This is
the $M$-to-$G$ conversion of Eqs.~(4.27a) and (4.36f) of
Ref.~\cite{Chluba:2026nmu}, represented by
Eq.~\eqref{eq:th-effective-emission-matrix} with $x_c$ replaced by the
matched $x_{c,\rm eff}$. Eliminating the profile defines the rate for the
retained coefficients rather than additional functions resolving the layer.
The effective emission matrix per unit proper time is
\begin{equation*}
 \mathsf D\bm y
 =Cx_{c,\rm eff}y_M(\gamma_T\bm e_G-\gamma_N\bm e_M),
 \qquad C=\theta n_e\sigma_{\rm T}.
\end{equation*}
Here $y_M$ is the $M$ coefficient of $\bm y$. Dots denote proper-time
derivatives, so $\dot{\bm y}^{(0)}|_{\rm em}=\mathsf D\bm y^{(0)}$ and
$\dot y_M^{(0)}|_{\rm em}=-\Gamma_\mu y_M^{(0)}$, where
$\Gamma_\mu=\gamma_N Cx_{c,\rm eff}$.

The local temperature response is
\begin{equation}
 a_T=\left.\frac{\partial\ln\Gamma_\mu}{\partial\ln T}\right|_{n_e}.
 \label{eq:th-dc-rate-response}
\end{equation}
The electron-density response is unity. There is no independent ion-density
response in this DC-only closure. In the leading nonrelativistic limit
with Planck incident photons, $\Lambda_{\rm DC}\propto T^2$ and
$x_c=\sqrt{\Lambda_{\rm DC}/\theta}$, giving
\begin{equation*}
 \Gamma_\mu\propto n_eT^{3/2},\qquad a_T=\frac32.
\end{equation*}
The unit contribution to $a_T$ comes from the Compton clock $C\propto T$;
the remaining $1/2$ comes from the turnover set by emission--Compton
balance. The numerical example
retains the small correction from its finite matching boundary by
differentiating that same profile. The exponent $5$ of the fixed-reference
emissivity prefactor in Eq.~\eqref{eq:th-emission-local-dc} is therefore
not the logarithmic derivative of $\Gamma_\mu$: the rate also includes
absorption, redistribution and photon-number normalization.

\subsubsection{Perturbed source in reference-frequency coordinates}
\label{app:dcbr-source-check}

The reduced law $\dot\mu|_{\rm em}=-\Gamma_\mu\mu$ evolves the
physical outer chemical potential $\mu=\mu_\infty$ and its accompanying
energy-conserving temperature change. The matched low-frequency profile
enters through $\Gamma_\mu$; the retained input and output spectra are
specified relative to the local temperature. Their representation in a
basis tied to $T_z$ therefore {\it also} changes when that physical temperature changes.
Section~\ref{sec:stationary-response} already illustrates the distinction:
a common frequency shift of a distorted mean has a nonzero $M$-basis
anisotropy even though the chemical potential is uniform. Relaxing that
coefficient as if it were an intrinsic chemical-potential fluctuation would
mistake a frequency shift for a change in the local distortion.

The input and output conversions follow directly from this physical
model. For a given local $\mu$, its effective broad-spectrum collision
source per unit proper time is
$C_{\rm em,loc}^{\rm eff}(u)=\Gamma_\mu\mu[(\gamma_T/\gamma_N)G(u)-M(u)]$.
The relation $4\gamma_T=\gamma_N\mathcal E[M]$ ensures that this source
changes photon number while conserving photon energy. Both the chemical
potential inferred from the incident spectrum and the $G/M$ shapes of
this relaxation refer to the local temperature. A reference-basis input
must first be expressed in those local coordinates to identify the
chemical potential on which the operator acts. Its output must then be
expressed in reference coordinates to give the collision spectrum used
by the FH. An evolution written directly for physical $\mu$ and temperature
contains these responses when the spectrum is reconstructed.

At fixed photon energy, a fractional temperature perturbation
$\Theta=\delta T/T$ rescales the local frequency argument by
$1/(1+\Theta)$. Holding the observer frame and time slicing fixed, this
is a frequency-coordinate dilation. To first order and at the retained
spectral accuracy, $S_\Theta=I+\Theta M_{\rm B}$ maps local-temperature
coefficients to the fixed reference-frequency basis. The ordinary
blackbody temperature shift is annihilated by $\mathsf D\bm e_G=0$;
for the distortion component, $S_\Theta^{-1}$ converts the input,
$\mathsf D$ acts locally, and $S_\Theta$ converts its output back.
Including the scalar rate response therefore gives
\begin{equation}
 S_\Theta[(1+a_T\Theta)\mathsf D]S_\Theta^{-1}
 \approx\mathsf D+\Theta[M_{\rm B},\mathsf D]+a_T\Theta\mathsf D.
 \label{eq:th-reduced-operator-response}
\end{equation}
The commutator is $M_{\rm B}\mathsf D-\mathsf D M_{\rm B}$: its two
terms shift the output collision spectrum and correct the input on which
the operator acts. This input and output conversion applies to any linear
spectral operator defined in local-temperature coordinates. A prescribed
additive source vector requires its output conversion.

This spectral response is not specific to eliminating the low-frequency
layer. The Compton construction in Sec.~\ref{app:compton} already includes
it in $M_{\rm BK}$ after the electron-temperature response is incorporated.
Elastic Thomson angular redistribution is frequency independent and
commutes with this dilation. A frequency-resolved collision calculation
can include the same response through its explicit temperature-dependent
kernels and spectra, without a separate matrix commutator. The reduced
DC model makes the conversion explicit because the unresolved layer has
been replaced by a rate and an operator on the retained coefficients.
The commutator describes that operator's spectral response; $a_T$ describes
its local rate response.

For a frequency shift of the mean distortion,
$\delta\bm y=\Theta M_{\rm B}\bm y^{(0)}$, the identity
\begin{equation*}
 \mathsf D(\Theta M_{\rm B}\bm y^{(0)})
 +\Theta[M_{\rm B},\mathsf D]\bm y^{(0)}
 =\Theta M_{\rm B}\mathsf D\bm y^{(0)}
\end{equation*}
recovers the frequency shift of the mean collision spectrum. The
commutator and scalar rate response describe different parts of the
variation of one local operator. Changing its rate alone does not supply
the frequency shift of its input and output.

Together with the electron-density and lapse modulation, this gives
\begin{equation}
 \frac{\bm c_{{\rm em},0}^{\rm N}}{a}
 =\mathsf D\bm y_0^{\rm N}
  +\left[(\delta_e^{\rm N}+\Psi+a_T\Theta_{\rm N})\mathsf D
          +\Theta_{\rm N}[M_{\rm B},\mathsf D]\right]\bm y^{(0)}.
 \label{eq:th-reduced-emission-source}
\end{equation}
Here $\Theta_{\rm N}$ is the ordinary photon-temperature input of
Sec.~\ref{app:thermal-variables}; $\mathsf D\bm e_G=0$, so the first term
is unchanged if the ordinary blackbody perturbation is removed from
$\bm y_0^{\rm N}$. Relative to the original effective prescription in
Eq.~\eqref{eq:th-effective-dcbr-sector}, the addition is
\begin{equation}
 \frac{\Delta\bm c_{{\rm em},0}^{\rm N}}a
 =\Theta_{\rm N}\left([M_{\rm B},\mathsf D]
                         +(a_T-1)\mathsf D\right)\bm y^{(0)}.
 \label{eq:th-reduced-emission-addition}
\end{equation}
The factor $a_T-1$ supplies the rate response beyond the explicit unit
prefactor already present in the old prescription; the commutator determines
the frequency-coordinate response.

The derivation in Ref.~\cite{Chluba:2026nmu} includes absorption and
Compton redistribution [Sec.~4.5, Eqs.~(4.28)--(4.32), and Appendix~B].
Restoring the missing $(T_\gamma/T_z)^3$ in its fixed-reference DC
emissivity changes the derivative of $\ln\Lambda$ with respect to
$\ln T_\gamma$ at fixed $x,T_z$ from $2$ to $5$. The local-frequency
reduction above then gives the leading rate response $a_T=3/2$.
The frequency shifts of the input and output are represented separately
by the commutator. Both are required for the same mean emission matrix.

To incorporate this result in App.~\ref{app:thermal-fh}, use the same
matched $x_{c,\rm eff}$ in the mean and perturbed emission matrix, so
$a\mathsf D=\tau'\theta_z(M_{\rm T}-M_{\rm K})$ at the retained order.
Add $\Delta\bm c_{{\rm em},0}^{\rm N}$ to the Newtonian monopole in
Eq.~\eqref{eq:th-fh-monopole}. The background and kinematic-dipole equations,
Eqs.~\eqref{eq:th-fh-background} and \eqref{eq:th-fh-dipole}, are unchanged. Equation~\eqref{eq:th-fh-general} reconstructs the general-gauge
source from this completed monopole. The addition vanishes when emission
is switched off; Sec.~\ref{app:br-extension} gives its extension with BR.

\subsubsection{General-gauge consistency}
\label{app:dc-gauge-check}

On an arbitrary slicing, the local temperature and electron density
perturb the rate, $S_\Theta$ converts frequency coordinates, and the lapse
$A$ converts proper time to conformal time. This independent local
construction gives
\begin{equation}
 \frac{\bm c_{{\rm em},0}}a
 =\mathsf D\bm y_0
 +\left[(\delta_e+A+a_T\Theta)\mathsf D
       +\Theta[M_{\rm B},\mathsf D]\right]\bm y^{(0)}.
 \label{eq:th-reduced-general-source}
\end{equation}
Here $\Theta$ is the common leading photon/electron temperature
perturbation. To compare with the reconstructed Newtonian source, use
the photon reconstruction of Sec.~\ref{sec:gauge-reconstruction},
Eq.~\eqref{eq:invariant-metric} for $\Psi$, and
\begin{equation*}
 \Theta_{\rm N}=\Theta+\Hc s,\qquad
 \delta_e^{\rm N}=\delta_e-\frac{n_{e0}'}{n_{e0}}s.
\end{equation*}
The temperature relation uses $T_z\propto a^{-1}$ at the retained order.
The definition of $a_T$ in Eq.~\eqref{eq:th-dc-rate-response} gives
\begin{equation*}
 \mathsf D'=\left(\frac{n_{e0}'}{n_{e0}}-a_T\Hc\right)\mathsf D.
\end{equation*}
Substituting these inputs in Eq.~\eqref{eq:th-reduced-emission-source}
and transforming the collision term as in Eq.~\eqref{eq:th-fh-general},
with $\bm c_{\rm em}^{(0)}=a\mathsf D\bm y^{(0)}$, yields
\begin{align}
 &\bm c_{{\rm em},0}^{\rm N}
 -\Hc sM_{\rm B}\bm c_{\rm em}^{(0)}
 +(s\bm c_{\rm em}^{(0)})'-\bm c_{{\rm em},0}\nonumber\\
 &\quad=as\left[\mathsf D'
 -\left(\frac{n_{e0}'}{n_{e0}}-a_T\Hc\right)\mathsf D\right]
                     \bm y^{(0)}=0.
 \label{eq:th-reduced-general-check}
\end{align}
The commutator supplies the frequency-shift part of this equality, while
the local rate derivative supplies its evolving-background part. Thus the
completed Newtonian prescription reconstructs the independently specified
local source for arbitrary $s$ and evolving $\bm y^{(0)}$, without an
adiabatic or superhorizon assumption. The dipole follows the local boost
of Eq.~\eqref{eq:th-local-boost}. Together with the Compton comparison in
Sec.~\ref{app:compton}, this establishes the agreement of the FH
prescription with the specified local thermalization model.

\subsubsection{Injection thermalization history}
\label{app:dcbr-adiabatic}

DC thermalization continues to relax the mean spectrum after the
injection of Sec.~\ref{sec:pulse}. In the long-wavelength adiabatic
limit, patches on source-comoving synchronous slices share the same
homogeneous injection and thermalization history. Their common spectrum can evolve
without generating an intrinsic difference between patches. We compare
the original and corrected perturbative prescriptions using this same
DC-only mean history.

After external injection ends, the perturbation $\delta\tau_X^{\rm N}$
of the freely falling source clock obeys
$\delta\dot{\tau}_X^{\rm N}=\Psi$. Define
$\bm d_0^{\rm N}=\bm y_0^{\rm N}-\Theta_{\rm N}\bm e_G$.
The reconstructed intrinsic synchronous monopole is
\begin{equation}
 \bm d_0^{\rm S}=\bm d_0^{\rm N}
       +H\delta\tau_X^{\rm N} M_{\rm B}\bm y^{(0)}-\delta\tau_X^{\rm N}\dot{\bm y}^{(0)}.
 \label{eq:th-reduced-synchronous-null}
\end{equation}
The clock subtraction includes the continuing thermalization of the
mean. In the long-wavelength adiabatic limit,
$\Theta_{\rm N}=-H\delta\tau_X^{\rm N}$ and $\delta_e^{\rm N}=-3H\delta\tau_X^{\rm N}$, so the
common local history requires $\bm d_0^{\rm S}=0$. In the fully ionized
scaling $C\propto a^{-4}$, the original source instead leaves
\begin{equation}
 \left.\dot{\bm d}_0^{\rm S}\right|_{\rm baseline}
 =H\delta\tau_X^{\rm N}\left([M_{\rm B},\mathsf D]+\alpha\mathsf D\right)\bm y^{(0)},
 \qquad
 \alpha=\frac{\dd\ln x_{c,\rm eff}}{\dd\ln(1+z)}=a_T-1.
 \label{eq:th-reduced-unmatched-response}
\end{equation}
This unsuppressed forcing measures the departure of the reconstructed
Newtonian evolution from the direct synchronous history. Evaluating
Eq.~\eqref{eq:th-reduced-emission-addition} on this mode cancels it.
The correction therefore preserves the common adiabatic thermalization
history while the mean evolves; it does not require a static mean.
This is a physical realization of the general source equality in
Eq.~\eqref{eq:th-reduced-general-check}.

The original source assigns different amounts of chemical-potential
relaxation to patches that should follow the same proper-time history.
It thereby produces a false inhomogeneity in photon number relative to
energy, and hence a local chemical-potential fluctuation. Subtracting
the common frequency shift and the clock displacement does not remove
this intrinsic component. Energy-conserving Compton redistribution can
damp higher spectral transients without erasing the chemical potential.
After photon production freezes out, Thomson scattering transports the
fluctuation and supplies diffusion damping at finite $k$ rather than
local chemical-potential relaxation. The unsuppressed intrinsic mode thus persists.

For a numerical example, we evolve the continuous pulse centered at
$z_{\mathrm p}=10^6$ with the source parameters of
Sec.~\ref{sec:cosmological-propagation}, an $N=15$ hierarchy and BR
omitted in both background and perturbations. Both source prescriptions
use the same DC-only profile, mean history and proper-age cutoff
corresponding to $z=2\times10^4$. The original prescription gives
\begin{equation}
 \delta\mu_{\rm intr}\simeq0.08\,\bar\mu_{\rm f}\mathcal R,
 \label{eq:th-spurious-mu-dc}
\end{equation}
where $\bar\mu_{\rm f}$ is the surviving mean chemical potential and
$\mathcal R$ is the primordial curvature fluctuation. The energy/number
moment construction of Eq.~\eqref{eq:mu-moment-diagnostic}, evaluated at
equal source age, and the $M$ component of the independent FH reconstruction
in Eq.~\eqref{eq:th-reduced-synchronous-null} give this same approximate
amplitude after prompt
Compton equilibration. At the cutoff, changing $k$ from $10^{-4}$ to
$10^{-5}\,{\rm Mpc}^{-1}$ leaves the coefficient $0.0797$ unchanged
to the quoted precision. The original response coefficient remains
finite as $k\to0$. The corrected source passes the adiabatic
separate-universe test, leaving only gradient-suppressed departures.
The coefficient in Eq.~\eqref{eq:th-spurious-mu-dc} belongs to this
pulse and local rate model.

Because this spurious mode is proportional to $\mathcal R$ without a
factor of $k^2$, it generates a leading intrinsic $\mu T$ cross-correlation
under subsequent transport. Its spectral response is distinct from the
stationary frequency shift of the common mean and survives the dipole
calibration of Sec.~\ref{sec:calibration}.
This intrinsic mode is absent in the adiabatic separate-universe limit. The test therefore concerns
the appearance of an intrinsic degree of freedom, rather than just a
fractional change in a distortion-basis cross spectrum that already
contains the stationary response. It establishes consistency of the
completed perturbation within the retained DC model; the accuracy of
its frequency-profile reduction remains a separate question.

\subsection{Extension to bremsstrahlung}
\label{app:br-extension}

BR can be included by extending the same local collision model in both
the background and perturbed equations. We give its frequency-dependent
kernel and the corresponding effective source for completeness. The
accuracy of reducing its low-frequency photons to a single chemical-
potential relaxation rate requires a separate qualification.

\subsubsection{Local kernel and shared low-frequency profile}

For BR the local frequency is $u_e=E_b/T_e$. In the nonrelativistic
thermal kernel the prefactor per Thomson depth is
$P_{\rm BR}=\Lambda_{\rm BR}^{\rm loc}/u_e^3$, with
$\Lambda_{\rm BR}^{\rm loc}\propto
\theta_e^{-7/2}\sum_i Z_i^2 n_i g_i(u_e,\theta_e)$
\cite{ChlubaSunyaev:2012}. Here $n_i$ and $Z_i$ are the density and charge
number of ion species $i$, and $g_i$ is its free--free Gaunt factor.
The electron density is already in the Thomson rate. In reference units
$P_{\rm BR}=(T_e/T_z)^3\Lambda_{\rm BR}^{\rm loc}/x_b^3$,
with the Gaunt factors still evaluated at $u_e$. The reference factors
cancel, leaving a kernel that depends on physical electron-frame energy,
temperature and ion densities. Its balance factor is that of
Eq.~\eqref{eq:th-emission-balance}. The scalar and frequency variations
then obey Eq.~\eqref{eq:th-emission-covariance}; the local boost supplies
the same dipole transformation as for DC.

When $T_e=T_\gamma=T$, both processes act on the same low-frequency
profile. Replace $\mathcal A$ in Eq.~\eqref{eq:th-reduced-profile} by
\begin{equation*}
 \mathcal A(u)=\frac{\Lambda_{\rm DC}(u)+\Lambda_{\rm BR}(u)}{\theta}.
\end{equation*}
The emission coefficients are added before solving for $m(u)$ and matching
the integrated rate. Independently reduced DC and BR rates cannot in
general be added, since each process changes the profile on which the
other acts. The resulting $x_{c,\rm eff}$ defines the same form of
$\mathsf D$ and $\Gamma_\mu=\gamma_N Cx_{c,\rm eff}$ as in
Sec.~\ref{app:dcbr-rate}, now with the joint profile.

When Compton redistribution becomes inefficient, BR can still change
the low-frequency spectrum while contributing little to relaxation near
the spectral peak \cite{Chluba2014,Chluba:2015hma,Evangelista2025}.
The single-profile conversion between $\mu$ and temperature assumes that
these frequency ranges remain coupled, and residual spectral modes also
become important outside the $\mu$ era. These limitations concern the
effective $G/M$ reduction, not the existence of a local BR kernel. They
motivate the DC-only numerical benchmark in
Sec.~\ref{app:dcbr-adiabatic}. Including BR consistently satisfies the
same gauge test, even where that reduction is quantitatively inaccurate.

\subsubsection{Local rate responses and perturbed source}

With BR, the rate depends independently on temperature and ion density.
Define
\begin{equation}
 a_T=\left.\frac{\partial\ln\Gamma_\mu}{\partial\ln T}\right|_{n_e,n_i},
 \qquad
 a_I=\left.\frac{\partial\ln\Gamma_\mu}{\partial\ln n_i}\right|_{T,n_e}.
 \label{eq:th-reduced-rate-responses}
\end{equation}
Here $n_i$ denotes the charge-weighted ion density, varied independently
of $n_e$ at fixed ion fractions. The electron-density derivative is
unity. The explicit $C\propto T$ contributes one to $a_T$, while the
remaining temperature and ion responses include readjustment of the
shared profile. The ion term therefore describes the dependence beyond
the electron-density prefactor.

In the leading nonrelativistic limit with a constant BR Gaunt factor,
$\Lambda_{\rm BR}\propto n_iT^{-7/2}$. For a BR-dominated asymptotic
profile, $x_c=\sqrt{\Lambda_{\rm BR}/\theta}$ gives
\begin{equation*}
 \Gamma_\mu\propto n_en_i^{1/2}T^{-5/4},\qquad
 (a_T,a_I)=(-5/4,1/2).
\end{equation*}
The square-root density response includes profile readjustment. For fully
ionized matter at fixed composition, $\delta_i=\delta n_i/n_i=\delta_e
=\delta_b$, and the total density modulation is $(1+a_I)\delta_b$.
With frequency-dependent Gaunt factors and both processes present, the
partials vary through the DC/BR transition. The illustrative joint
closure gives $(a_T,a_I)\simeq(1.28,0.037)$ at $z=10^6$ and
$(-0.87,0.43)$ at $z=10^5$, obtained by independent local variations of
temperature and ion density in the profile calculation.

Use this same joint $\mathsf D$ in the mean and perturbed equations.
The frequency-coordinate conversion of
Eq.~\eqref{eq:th-reduced-operator-response}, with the rate factor replaced
by $1+a_T\Theta+a_I\delta_i$, gives on an arbitrary slicing
\begin{equation}
 \frac{\bm c_{{\rm em},0}}a
 =\mathsf D\bm y_0
 +\left[(\delta_e+A+a_T\Theta+a_I\delta_i)\mathsf D
       +\Theta[M_{\rm B},\mathsf D]\right]\bm y^{(0)}.
 \label{eq:th-br-emission-source}
\end{equation}
On Newtonian slicing, set $A=\Psi$ and use the corresponding reconstructed
temperature and density perturbations. Relative to
Eq.~\eqref{eq:th-effective-dcbr-sector}, the addition is
\begin{equation}
 \frac{\Delta\bm c_{{\rm em},0}^{\rm N}}a
 =\left\{\Theta_{\rm N}
       \left([M_{\rm B},\mathsf D]+(a_T-1)\mathsf D\right)
       +a_I\delta_i^{\rm N}\mathsf D\right\}\bm y^{(0)}.
 \label{eq:th-br-emission-addition}
\end{equation}
In addition to the frequency shift, both the temperature derivative and
the mean matrix must therefore use the joint profile. Omitting BR from
only the perturbed response leaves its contribution to the mean relaxation
unmatched.

The general-gauge comparison proceeds as in
Sec.~\ref{app:dc-gauge-check}, now also using
$\delta_i^{\rm N}=\delta_i-(n_{i0}'/n_{i0})s$ and
\begin{equation*}
 \mathsf D'
 =\left(\frac{n_{e0}'}{n_{e0}}-a_T\Hc
                 +a_I\frac{n_{i0}'}{n_{i0}}\right)\mathsf D.
\end{equation*}
The difference of the transformed Newtonian and direct local sources is
\begin{align}
 &\bm c_{{\rm em},0}^{\rm N}
 -\Hc sM_{\rm B}\bm c_{\rm em}^{(0)}
 +(s\bm c_{\rm em}^{(0)})'-\bm c_{{\rm em},0}\nonumber\\
 &\quad=as\left[\mathsf D'
 -\left(\frac{n_{e0}'}{n_{e0}}-a_T\Hc
                   +a_I\frac{n_{i0}'}{n_{i0}}\right)\mathsf D\right]
                     \bm y^{(0)}=0.
 \label{eq:th-br-general-check}
\end{align}
The adiabatic comparison of Sec.~\ref{app:dcbr-adiabatic} has
$\delta_i^{\rm N}=-3H\delta\tau_X^{\rm N}$ and
$\alpha=a_T+3a_I-1$. Equation~\eqref{eq:th-br-emission-addition} cancels
the corresponding unmatched forcing, preserving the common local history.

A mean thermalization history alone does not specify all independent
local responses. The adiabatic comparison constrains $a_T+3a_I$.
A superhorizon baryon-isocurvature perturbation supplies another
direction: neglecting the small baryon contribution to radiation-era
gravity, patches with different baryon-to-photon ratios can be compared
at the same temperature and expansion. Both $n_e$ and $n_i$ vary, giving
the independent rate response $(1+a_I)\delta_b$. Together these comparisons
test the local partials of the specified model.

Independent homogeneous full-frequency evolution with varied temperature
and densities corroborates the scalar-response structure, including the
change of sign of $a_T$ across the DC/BR transition. These frozen-bath
comparisons assess scalar-rate accuracy within the adopted microscopic
approximations; agreement is approximate, especially near cancellation
of the two temperature responses. More accurate homogeneous rates and
their independent local derivatives could replace the illustrative
profile while retaining the same source construction.

\section{Direct photon injection and fluid energy-momentum transfer}
\label{app:numerics}

Here we connect the treatment in Section~\ref{sec:source} for direct injection of photons into the FH with the gauge-covariant photon and source fluid equations through the energy moments.  The energy moments evolve under energy-momentum conservation with the source and in App.~\ref{app:ks-moments} we provide the gauge-covariant photon conservation equations in the fluid description of KS \cite{Kodama:1984ziu}.  
We present a detailed discussion of the choice of time slicing and source frame for the injection.  In App.~\ref{app:emission-moments}, we specialize the treatment to a frame where the emission is isotropic and in App.~\ref{app:source-moments} complete this system with the source fluid equations.  This energy moment description, when supplemented by an emitted spectrum or number moment transfer in Section~\ref{sec:source}, determines the post-injection chemical potential in the number-conserving Compton-equilibrium example. 

\subsection{Energy and momentum conservation}
\label{app:ks-moments}

The energy moment of the monopole and dipole FH equations  recovers the
photon continuity and momentum equations for an arbitrary local
energy--momentum transfer. The collision moments give the energy and
momentum exchanged with
other components, independently of their spectral representation.
In the general normal frame, define
\begin{equation}
 R_\gamma\equiv\frac{\rho_\gamma}{\rho_z},\qquad
 \mathcal E_\ell\equiv\mathcal E[n_\ell^{(1)}]
 \quad(\ell\geq0).
 \label{eq:app-ks-moment-definitions}
\end{equation}
Here $\rho_\gamma$ is the homogeneous photon energy density and
$\rho_z\propto a^{-4}$ is the reference blackbody energy density.
For the other components of the photon stress energy tensor, we give their relation to the energy moment multipoles in the KS notational convention. As with a fluid, $v_\gamma$ is the photon scalar velocity amplitude and
$V_\gamma\equiv v_\gamma-B$ its velocity relative to the normal observers,
$p_\gamma=\rho_\gamma/3$ its pressure, and $\pi_{T\gamma}$ its
dimensionless scalar anisotropic stress. The trace-free spatial stress
is $\delta T^i{}_{\gamma j}|_{\rm TF}
=p_\gamma\pi_{T\gamma}{\mathbb Y}^i{}_j$. Angular integration then gives
$\delta\rho_\gamma=\rho_z\mathcal E_0$,
$(\rho_\gamma+p_\gamma)V_\gamma=\rho_z\mathcal E_1$, and
$p_\gamma\pi_{T\gamma}=\rho_z\mathcal E_2$.
The corresponding isotropic pressure perturbation is
$\delta p_\gamma=\delta\rho_\gamma/3$.

The total photon collision operator is $\sum_j C_j$,
including Thomson scattering. Its $\ell$th normal-frame multipole has
energy moment $\mathcal E[\sum_j C_{j,\ell}^{(1)}]$; multiplying by
$\rho_z$ gives the energy-density transfer per unit conformal time.
Since
$\mathcal E[U]=4R_\gamma$, Eqs.~\eqref{eq:general-fh-monopole} and
\eqref{eq:general-fh-dipole} imply
\begin{align}
 \mathcal E_0'&=-k\mathcal E_1-4R_\gamma\left(H_L'+\frac{kB}{3}\right)
                   +\mathcal E\!\left[\sum_j C_{j,0}^{(1)}\right],\nonumber\\
 \mathcal E_1'&=\frac{k}{3}(\mathcal E_0-2\mathcal E_2)
          +\frac{4k}{3}R_\gamma A
          +\mathcal E\!\left[\sum_j C_{j,1}^{(1)}\right].
 \label{eq:app-ks-hierarchy-moments}
\end{align}
Using $\rho_z'=-4\Hc\rho_z$, these become
\begin{align}
 (\delta\rho_\gamma)' +4\Hc\delta\rho_\gamma
   +(\rho_\gamma+p_\gamma)(kv_\gamma+3H_L')
   &=\rho_z\mathcal E\!\left[\sum_j C_{j,0}^{(1)}\right],\nonumber\\
 [ (\rho_\gamma+p_\gamma)V_\gamma ]'
  +4\Hc(\rho_\gamma+p_\gamma)V_\gamma
  -k\delta p_\gamma-k(\rho_\gamma+p_\gamma)A
  +\frac{2k}{3}p_\gamma\pi_{T\gamma}
   &=\rho_z\mathcal E\!\left[\sum_j C_{j,1}^{(1)}\right].
 \label{eq:app-ks-stress-moments}
\end{align}
The shift contribution in the monopole combines with $kV_\gamma$
to give $kv_\gamma$ in the continuity equation; the momentum density
uses the normal-frame velocity $V_\gamma$. These equations give the
photon specialization in three spatial dimensions and a flat background
of the component conservation laws
in KS, Sec.~II.5.  They are gauge-covariant and apply to any general scalar gauge.  

To make contact with the energy-based exchange description of KS, we can express the collision moments in terms of energy momentum transfer
\begin{equation}
 Q^{\nu}_\gamma =\nabla_\mu T_\gamma^{\mu\nu} .
 \label{eq:energy-momentum-transfer}
\end{equation}
Following KS Eqs.~(5.18)--(5.19), we split the transfer into 
$Q_\gamma^\mu=Q_\gamma u^\mu+f_\gamma^\mu$, where $u^\mu$ is the
total-fluid 4-velocity; $v$ is its scalar amplitude and
$V\equiv v-B$ its normal-frame velocity.
The proper rate $Q_\gamma=-u_\mu Q_\gamma^\mu$ is measured in the total-fluid rest frame,
and $u_\mu f_\gamma^\mu=0$. The momentum-transfer scalar $f_\gamma$
is normalized by
$f_{\gamma i}=\Hc(\rho_\gamma+p_\gamma)f_\gamma
{\mathbb Y}_i$ at first order.
The background energy moment of the Boltzmann equation gives
\begin{equation}
 \rho_z\mathcal E\!\left[\sum_j C_j^{(0)}\right]
 =\rho_\gamma'+4\Hc\rho_\gamma=aQ_\gamma.
 \label{eq:app-source-normalization}
\end{equation}
The collision moment supplies the photon energy gained per unit conformal
time; the factor $a$ converts the proper transfer rate $Q_\gamma$ to that
time variable. This relation uses the total collision operator and can
also be applied to an individual transfer contribution.

Denote the perturbation of the proper scalar rate by $\delta Q_\gamma$
on the chosen slicing. Then
the collision moments required by Eq.~(\ref{eq:app-ks-stress-moments}) are
\begin{align}
 \rho_z\mathcal E\!\left[\sum_j C_{j,0}^{(1)}\right]
 &=a(\delta Q_\gamma+Q_\gamma A),\nonumber\\
 \rho_z\mathcal E\!\left[\sum_j C_{j,1}^{(1)}\right]
 &=aQ_\gamma V
       +\Hc(\rho_\gamma+p_\gamma)f_\gamma.
 \label{eq:app-ks-transfer-moments}
\end{align}
The second expression separates the momentum
advected with the total velocity from momentum transfer orthogonal to
that velocity. 
When the rate is nonzero one can write the fractional perturbation as 
$\delta Q_\gamma=Q_\gamma\varepsilon_\gamma$.
These forms follow directly from  Eqs.~\eqref{eq:energy-momentum-transfer} and \eqref{eq:app-ks-stress-moments}.\footnote{For the KS equations used here, the following corrections are needed. In the published KS Eq.~(5.46a), the rate perturbation should instead appear as
$\varepsilon_\alpha+A$, consistent with their covariant transfer
component in KS Eq.~(5.22a). The stress coefficients multiplying
$\pi_{L\alpha}$ and $\pi_{T\alpha}$ in KS Eq.~(5.46b) are
$p_\alpha$, as in KS Eq.~(5.16d). With the force normalization of
KS Eq.~(5.21b), its contribution to KS Eq.~(5.46b) is
$a H(\rho_\alpha+p_\alpha)f_\alpha$.  %
Here $\alpha$ labels a component; its density, pressure, rate
perturbation, and force are defined as in the photon equations above.
The symbols $\pi_{L\alpha}$ and $\pi_{T\alpha}$ denote pressure-normalized
isotropic and trace-free stress perturbations of component $\alpha$.
Additional typos occur in the transfer terms of KS Eq.~(5.48b), which we do
not use here.}  Total energy-momentum conservation requires that the transfers to all species satisfy $\sum_\alpha Q_\alpha^\mu=0$.   In the finite spectral basis, the projection $\mathcal P$ in Sec.~\ref{sec:frequency} must preserve the energy moment conservation law through an appropriate choice of frequency weights $\mathcal L$.

The KS invariant rate variables reconstruct the same proper scalar on
different slicings [KS Eqs.~(5.41a) and (5.42a)--(5.42c)]. Where $Q_\gamma\ne0$,
$E_{c\gamma}=(\delta Q_\gamma-Q_\gamma'V/k)/Q_\gamma$ uses total-fluid
comoving slicing, while
$E_\gamma=(\delta Q_\gamma-Q_\gamma'V_\gamma/k)/Q_\gamma$ uses photon-comoving
slicing. The shear-free/Newtonian value is
$E_{s\gamma}=(\delta Q_\gamma-Q_\gamma's)/Q_\gamma$; the spatially flat value
$E_{g\gamma}$ uses the time shift $(H_L+H_T/3)/\Hc$ instead.
These are the scalar reconstructions of Eqs.~\eqref{eq:app-scalar} and
\eqref{eq:app-rest-rate}. This slicing choice is distinct from the physical
frame used to decompose $Q_\gamma^\mu$. 

\subsection{Emission frame and isotropy}
\label{app:emission-moments}

The KS split into an energy-transfer rate $Q_\gamma$  and a 
force $f_\gamma^\mu$ that is orthogonal to the total fluid velocity provides a general description. For a specific source,
it is often convenient to specify the emission in its own physical frame
and then project the transfer onto the total-fluid frame used by KS.
For the direct-emission contribution, let $u_Q^\mu$ be the unit future-directed
4-velocity parallel to $Q_\gamma^\mu$. Thus $Q_\gamma^\mu=Qu_Q^\mu$,
where $Q=-u_{Q\mu}Q_\gamma^\mu$ is the proper emitted energy per unit
volume and time. The emitted photons carry zero net spatial momentum in
this transfer rest frame, which generally differs from the photon-fluid
frame $u_\gamma^\mu$ and the total-fluid frame $u^\mu$.
The subscript $Q$ identifies the transfer that defines this frame.

\pagebreak[3]
Projecting this emission transfer onto the total-fluid frame at the same spacetime
event gives
\begin{align}
 \gamma_{\rm rel}&\equiv-u_\mu u_Q^\mu,\qquad
 Q_\gamma=\gamma_{\rm rel}Q,\nonumber\\
 f_\gamma^\mu&=Q\bigl(u_Q^\mu-\gamma_{\rm rel}u^\mu\bigr),
 \label{eq:app-emission-frame-projection}
\end{align}
where $\gamma_{\rm rel}$ is the relative Lorentz factor. These relations
are exact on a fixed slicing. In the isotropic cosmological background velocities are first order perturbations, so $\gamma_{\rm rel}-1$ is second order and  $Q_\gamma=Q$ and $\delta Q_\gamma=\delta Q$
through linear order, while the relative force is first order and evaluated with $\gamma_{\rm rel}=1$.
Its gauge-covariant form is given below in Eq.~\eqref{eq:app-emission-ks-force}.

For the gauge-covariant FH, we additionally assume that emission is
isotropic in the $u_Q^\mu$ frame at every photon energy
$q_Q\equiv-p_\mu u_Q^\mu$. Isotropy makes the spatial momentum transfer
vanish separately at each photon energy in this frame. Alone the frequency-integrated
zero-momentum condition above leaves energy-dependent dipoles and higher
angular moments unspecified.
The decay model
of Sec.~\ref{sec:source} sets $u_Q^\mu=u_X^\mu$, the 4-velocity of the source, but we retain the $Q$ notation here for generality. 

Isolating this photon-injection piece of the FH collision term gives
$C_Q^{(0)}(\eta,x)$ as the homogeneous photon occupation-number
injection per conformal time, isotropic in $u_Q^\mu$. Define the source
velocity relative to the normal observers by $V_Q=v_Q-B$,
where $v_Q$ is the scalar velocity amplitude of $u_Q^\mu$ as in
Eq.~\eqref{eq:fluidvelocity_vector}.
For this isolated injection contribution, $Q_\gamma=Q$ in the background.
Applying Eq.~\eqref{eq:app-source-normalization} to this sector gives
$\mathcal E[C_Q^{(0)}]=aQ/\rho_z$, the energy normalization used in the
main text. Its
photon-number moment, $\mathcal N[C_Q^{(0)}]$, requires a choice for the spectral shape of the injection $F(\xi)$ in Eq.~(\ref{eq:source-spectrum}).

Boosting a source isotropic in $u_Q^\mu$ into the normal frame gives
the angular dipole and the $\ell=1$ collision multipole. The Doppler shift of
the spectral argument contributes $-i\chi V_Q\Ox C_Q^{(0)}$;
the transformation of the rate per unit time contributes
$+i\chi V_Q C_Q^{(0)}$, giving
\begin{equation}
 C_{Q,\mathrm{dip}}^{(1)}
 =-i\chi V_Q(\Ox-1)C_Q^{(0)},\qquad
 C_{Q,1}^{(1)}=\frac{V_Q}{3}(\Ox-1)C_Q^{(0)}.
 \label{eq:app-source-dipole}
\end{equation}
Using the moment definition in Eq.~\eqref{eq:spectral-moments}, integration
by parts gives the energy moment of this dipole.
On the full physical frequency domain, provided $x^4f\rightarrow0$
at both endpoints,
$\mathcal E[(\Ox-1)f]=3\mathcal E[f]$ and therefore
\begin{equation}
 \mathcal E[C_{Q,1}^{(1)}]
 =V_Q\mathcal E[C_Q^{(0)}]
 =\frac{aQ}{\rho_z}V_Q .
 \label{eq:app-source-energy-dipole}
\end{equation}
Matching the dipole term of Eq.~\eqref{eq:app-ks-transfer-moments} to  \eqref{eq:app-source-energy-dipole} gives the KS relative force in terms of the difference of velocities:
\begin{align}
 \Hc(\rho_\gamma+p_\gamma)f_\gamma&=aQ(V_Q-V),%
 \label{eq:app-emission-ks-force}
\end{align}
and one can verify that Eq.~\eqref{eq:app-emission-frame-projection} gives the same result.
Zero spatial transfer in the transfer rest frame therefore need not mean zero
momentum transfer in the normal frame. The latter separates into advection
with the total fluid and force relative to it.
Changing the slicing shifts $V_Q$ by $-k\mathcal{T}$ and $\delta Q$ by $-Q'\mathcal{T}$,
while the relative velocities $V_Q-V$ and $v_Q-v$ are unchanged.

At linear order, the boost of emission isotropic in the $Q$ frame generates
only a dipole, so $C_{Q,\ell}^{(1)}=0$ for $\ell\ge2$ in the normal frame.
For the monopole, the perturbed proper rate is
 $Q(\eta,\bm X)=Q(\eta)+\delta Q(\eta){\mathbb Y}$, with
$\delta Q=Q\varepsilon_Q$ where $Q\ne0$.  
Multiplication by the lapse $\alpha=a(1+A{\mathbb Y})$ gives the
perturbed conformal-time transfer $a(\delta Q+QA)$, so
\begin{equation}
\mathcal E[C_{Q,0}^{(1)}]=\frac{a}{\rho_z}(\delta Q+QA).
\end{equation}
Along with a choice of the spectral shape of the emission, this completes
the first-order collision multipoles of the isotropic $Q$-frame source
in the normal frame. For the fixed source-frame spectrum of
Eq.~\eqref{eq:source-spectrum}, the full monopole is
$C_{Q,0}^{(1)}=(A+\delta Q/Q)C_Q^{(0)}$. Together with
Eq.~\eqref{eq:app-source-dipole} and $u_Q^\mu=u_X^\mu$, this reproduces
the main-text collision term in Eq.~\eqref{eq:source-injection}.
Inserting the emission moments
into the conservation equations of Sec.~\ref{app:ks-moments}  gives
the photon continuity and Euler equations.

\subsection{Source fluid equations}
\label{app:source-moments}

The source loses the energy and momentum supplied to the photons.
For a cold component $X$, the opposite transfer $Q_X^\mu=-Qu_Q^\mu$
gives the source continuity and Euler equations. With $p_X=0$,
$\delta_X=\delta\rho_X/\rho_X$, and
$\rho_X'+3\Hc\rho_X=-aQ$, these are
\begin{align}
 \delta_X'
 &=-kv_X-3H_L'
   -\frac{a}{\rho_X}\left[\delta Q+Q(A-\delta_X)\right],
\nonumber\\
 V_X'+\Hc V_X
 &=kA+\frac{aQ}{\rho_X}(V_X-V_Q).
 \label{eq:app-donor-fluid}
\end{align}
The force density is controlled by $V_X-V_Q$, the relative velocity between
the $X$ fluid frame and the $Q$ frame of zero net momentum transfer. 

To make contact with the model in the main text, we adopt the decay prescription in Sec.~\ref{sec:source}, with
$u_Q^\mu=u_X^\mu$ and $Q=\Gamma(\tau_X)\rho_X$.   
With
$\delta\Gamma=\Gamma_{,\tau_X}\delta\tau_X$, the rate perturbation is
$\delta Q=\Gamma\delta\rho_X+\rho_X\delta\Gamma$ in Eq.~\eqref{eq:app-donor-fluid}. The
$\delta_X$ term on the right hand side of the density equation also cancels, leaving the lapse
and proper-clock modulation.   Since the $X$-frame and the $Q$ frame coincide,  $V_X=V_Q$ and an isotropic decay in the $Q$ frame provides no force: the loss of momentum density from emission
is purely due to the depletion of $\rho_X$, leaving $V_X'+\Hc V_X=kA$. In the main text, we referred to this property as the source remaining in free fall
after the transfer.  This completes the derivation of Eq.~\eqref{eq:sourceperturbation} of the main text.

\bibliography{methods_paper}

\end{document}